\documentclass[%
floatfix,
showpacs,
superscriptaddress,
aps,
prc,
nofootinbib,
reprint]{revtex4-1}

\usepackage{hyperref}
\usepackage{color}
\usepackage[T1]{fontenc}
\usepackage{comment}
\usepackage{tgtermes}
\usepackage{graphicx}
\usepackage{graphicx}
\usepackage[dvipsnames]{xcolor}
\usepackage{dsfont}
\usepackage{bm}
\usepackage{slashed}
\usepackage{changepage} 
\usepackage{mathtools}
\usepackage{amsmath}
\usepackage{subfigure}
\usepackage{indentfirst}
\usepackage{hyperref}
\usepackage{booktabs, tabularx}
\usepackage{enumitem}
\usepackage[utf8]{inputenc}
\usepackage{ulem}

\hypersetup{
    colorlinks=true,
    linkcolor=blue,
    filecolor=magenta,      
    urlcolor=blue,
    citecolor=blue
}

\usepackage{amssymb}
\usepackage{multirow,array}
\usepackage{lipsum}
\usepackage{enumitem}

\newcommand{\be}{\begin{equation}}
\newcommand{\ee}{\end{equation}}
\newcommand{\bs}{\begin{subequations}}
\newcommand{\es}{\end{subequations}}
\newcommand{\beal}{\begin{align}}

\newcommand{\trento}{\texttt{T$_\mathrm{R}$ENTo}}
\newcommand{\sqrts}{\sqrt{s_\textrm{NN}}}

\newcommand{\xperp}{\mathbf{x}_\perp}

\newcommand{\VAH}{\texttt{VAH}}
\newcommand{\SMASH}{\texttt{SMASH}}
\newcommand{\CORNELIUS}{\texttt{CORNELIUS}}
\newcommand{\isd}{\texttt{iS3D}}

\newcommand{\variance}{{\mathbf{C}}}
\newcommand{\mean}{\pmb{{\mu}}}

\DeclareMathOperator{\sign}{sign}
\DeclareMathOperator*{\argmax}{arg\,max}

\begin{document}

\title{Bayesian calibration of viscous anisotropic hydrodynamic simulations of heavy-ion collisions}

\author{Dananjaya Liyanage}
\affiliation{Department of Physics, The Ohio State University, Columbus, OH 43210, USA}
\author{\"{O}zge S\"{u}rer}
\affiliation{Department of Information Systems and Analytics, Miami University, Oxford, OH 45056, USA}
\affiliation{Northwestern-Argonne Institute of Science and Engineering (NAISE),
Northwestern University, Evanston, IL 60208, USA}
\author{Matthew Plumlee}
\affiliation{Northwestern-Argonne Institute of Science and Engineering (NAISE),
Northwestern University, Evanston, IL 60208, USA}
\affiliation{Industrial Engineering and Management Sciences Department,
Northwestern University, Evanston, IL 60208, USA}
\author{Stefan M.\ Wild}
\affiliation{Northwestern-Argonne Institute of Science and Engineering (NAISE),
Northwestern University, Evanston, IL 60208, USA}
\affiliation{Applied Mathematics \& Computational Research Division, 
Lawrence Berkeley National Laboratory, Berkeley, California 94720, USA}
\author{Ulrich Heinz}
\affiliation{Department of Physics, The Ohio State University, Columbus, OH 43210, USA}

\date{\today}

\begin{abstract}
Due to large pressure gradients at early times, standard hydrodynamic model simulations of relativistic heavy-ion collisions do not become reliable until $O(1)$\,fm/$c$ after the collision. To address this one often introduces a pre-hydrodynamic stage that models the early evolution microscopically, typically as a conformal, weakly interacting gas. In such an approach the transition from the pre-hydrodynamic to the hydrodynamic stage is discontinuous, introducing considerable theoretical model ambiguity. Alternatively, fluids with large anisotropic pressure gradients can be handled macroscopically using the recently developed Viscous Anisotropic Hydrodynamics (VAH). In high-energy heavy-ion collisions VAH is applicable already at very early times, and at later times transitions smoothly into conventional second-order viscous hydrodynamics (VH). We present a Bayesian calibration of the VAH model with experimental data for Pb--Pb collisions at the LHC at $\sqrts{}=2.76$\,TeV. We find that the VAH model has the unique capability of constraining the specific viscosities of the QGP at higher temperatures than other previously used models.
\end{abstract}

\maketitle

\section{Introduction}
\label{sec1}
\vspace*{-2mm}

Atomic nuclei are made of fundamental particles called quarks and gluons (a.k.a.\ partons). Due to the property of color confinement, partons cannot be studied in isolation. Confinement can be broken, however, in many-parton systems of very high densities \cite{Linde_1979, Collins1975}. Quark-Gluon Plasma (QGP) is one of several such emergent phases in chunks of strongly interacting matter. It is a hot and dense soup of quarks and gluons with liquid properties that can be created in nucleus--nucleus collisions at very high energies \cite{Gyulassy:2004zy} and has strong similarity with the matter that filled the universe right after the Big Bang before it cooled down to produce the hadronic matter that we observe today \cite{Yagi:2005yb}.  

The largest experimental heavy-ion facility in the world, the Large Hadron Collider (LHC) at CERN, collides nuclei at center-of-mass energies of several TeV per nucleon pair. At these high energies all of the baryon number carriers from the incoming nuclei pass through each other without being stopped, while a fraction of the total energy is deposited in the mid-rapidity region of the collision, creating new matter with approximately zero net baryon density~\cite{Muller:2012zq}. Strong interactions among its constituents quickly turn this matter into a baryon-neutral QGP that rapidly cools via collective expansion, converting into color-neutral hadrons when the temperature drops below the hadronization temperature. These hadrons continue to interact until their density becomes too low, after which unstable hadronic resonances decay and the stable decay products fly out toward the detectors. 

The QGP phase has an extremely short lifetime (${\sim\,}10^{-23}$\,s) and size (${\sim\,}10^{-14}$\,m). It can only be studied by recording the distributions of, and correlations among, the energies and momenta of the thousands of finally emitted hadrons. A quantitative theoretical analysis of these distributions requires their simulation using sophisticated dynamical evolution models and advanced statistical techniques~\cite{big_picture}. 

Many large-scale heavy-ion simulation studies using Bayesian statistical techniques have been carried out in recent years to infer from the experimental measurements the properties of the QGP  \cite{Petersen:2010zt, Novak:2013bqa, Sangaline:2015isa, Bernhard:2015hxa, Bernhard:2016tnd, Moreland:2018gsh, Bernhard:2019bmu, JETSCAPE:2020shq, Nijs:2020ors, Nijs:2020roc,  JETSCAPE:2020mzn, Heffernan_CGC}. One of the major limitations of the current simulations is the breakdown of the hydrodynamic approach that describes the evolution of the QGP phase at early times when the QGP is being formed. This breakdown is caused by extremely large pressure gradients in the incipient stage of the newly created matter. To circumvent this issue, almost all previous Bayesian model calibrations invoked a weakly-coupled, gaseous pre-hydrodynamic stage.\footnote{%
    A very recent study \cite{Heffernan_CGC} employed the Color-Glass Condensate based IP-Glasma model to dynamically evolve the pre-hydrodynamic stage. While this is a significant conceptual improvement over free-streaming partons, it shares with the latter approach that, being rooted in Classical Yang-Mills dynamics for the interacting gluon fields, it keeps the system from naturally approaching local thermal equilibrium.}
In this stage the fireball medium evolves initially as a conformal, weakly interacting gas. After a ``switching time'' of order 1\,fm/$c$ the density and pressure gradients have decreased to a level where second-order viscous hydrodynamics, the framework used to evolve the QGP fluid, becomes applicable. Strong interactions in the QGP break its conformal symmetry; the matching between a conformally symmetric pre-hydrodynamic stage to a non-conformal QGP fluid stage, with a realistic equation of state (EoS) $p(e)$ taken from lattice QCD simulations, introduces considerable theoretical ambiguity. The discontinuity of the EoS gives rise to a number of unphysical effects, including a large positive bulk viscous pressure at the start of the hydrodynamic stage \cite{Liu:2015nwa, NunesdaSilva:2020bfs}. The second-order viscous hydrodynamic equations must then rapidly evolve the bulk pressure toward its first-order Navier-Stokes value, which is of more moderate magnitude but has the opposite sign \cite{McNelis_2021}. 

In this work we perform the first Bayesian model calibration of a novel dynamical evolution model based on viscous {\it anisotropic} hydrodynamics (\VAH) as described in Refs.~\cite{mcnelis20183+, McNelis_2021} instead of standard second-order viscous hydrodynamics.\footnote{%
    Our \VAH{} approach differs from that used in Refs.~\cite{Martinez:2010sc, florkowski2011highly, alqahtani20173+, alqahtani2017anisotropic,  almaalol2019anisotropic, alqahtani2021bulk} by including in the hydrodynamic evolution equations all dissipative terms arising from the residual viscous correction $\delta\tilde f$
    \cite{bazow2014second}.
}
We use the JETSCAPE simulation and model calibration framework \cite{JETSCAPE:2020mzn} for relativistic heavy-ion collisions but modify it by eliminating the free-streaming pre-equilibrium module and replacing the second-order viscous hydrodynamic (VH) stage with the \VAH{} module described in \cite{McNelis_2021}. \VAH{} can handle the large anisotropy between the longitudinal and transverse pressure gradients that characterizes the early expansion stage in heavy-ion collisions much better than VH; this allows us to start the hydrodynamic stage at such an early time that neglecting the pre-hydrodynamic evolution completely becomes a good approximation. \VAH{} transitions automatically into standard second-order viscous hydrodynamics (although with an algorithm based on a different decomposition of the fluid's energy-momentum tensor) at later times \cite{McNelis_2021}. All the simulation data and code for the Bayesian parameter inference in this work can be found at \cite{gitcode}.

The rest of this paper is organized as follows. An overview of the \VAH{} hybrid evolution model for ultra-relativistic heavy-ion collisions, including a description of its model parameters, is presented in Sec.~\ref{sec2}. In Sec.~\ref{sec3} we briefly describe the Bayesian model calibration workflow, including several innovations introduced and tested in the present study. Section~\ref{sec4} describes the construction and training of fast emulators for the \VAH{} model. Closure test results for these emulators are reported in Sec.~\ref{sec5}. The actual model calibration process is described in Sec.~\ref{sec6}, and the posterior probability distributions for the inferred parameters are described and discussed. A model sensitivity analysis for the observables used in the calibration procedure is presented in Sec.~\ref{sec7}. The performance of the calibrated model in describing the calibration and other experimental data is discussed in Sec.~\ref{sec8}. Our conclusions are presented in Sec.~\ref{sec9}. Appendices \ref{app:datacollection}--\ref{app:Sobol} provide technical details related to procedures described in the main body of the text.

\section{Overview of the hybrid model for relativistic heavy-ion collisions}
\label{sec2}
\vspace*{-2mm}

The dynamical evolution of relativistic heavy-ion collisions involves physics ranging from small to large length scales and demands a multistage modeling approach, with different modules simulating different stages of the evolution. In this section, we briefly discuss each evolution stage and its simulation. 

\subsection{Initial conditions}
\label{trento}
\vspace*{-2mm}

Our simulation of a relativistic heavy-ion collision starts right after the two incoming nuclei collide. In the lab frame both nuclei move initially close to the speed of light in opposite directions and, for an observer on the beam axis, appear as strongly Lorentz-contracted pancakes. As the nuclei hit each other, their valence quarks (which carry their baryon number) pass through each other without being fully stopped in the center-of-momentum frame. Due to interactions between their gluon clouds they lose, however, a large fraction of their energy, some of which is deposited in the form of newly created matter near mid-rapidity (i.e., with low longitudinal momenta in the lab frame) \cite{big_picture}. The spatial distribution of this matter is described phenomenologically with the stochastic model \trento{} \cite{Moreland:2014oya, trento_code}.\footnote{%
    For the default values of the \trento{} model parameters described in the following please consult \cite{Moreland:2014oya, trento_code}.}

\trento{} models the incoming nuclei as a conglomerate of nucleons whose positions in the plane perpendicular to the beam are held fixed during the extremely short nuclear interpenetration time. The transverse density distribution of a nucleon is parameterized by a three-dimensional Gaussian with width parameter $w$, integrated over the beam direction $z$:
\be
\label{eq:fluctuated_thickness}
    \rho(\xperp)=\int_{-\infty}^\infty \frac{dz}{(2\pi w^2)^{3/2}}\exp\left(-\frac{\xperp^2+z^2}{2w^2}\right).
\ee
The positions of the nucleons inside each incoming nucleus are  sampled with a minimum pairwise separation $d_\mathrm{min}$ to simulate their repulsive hard core, but otherwise independently, from Woods-Saxon distributions whose radius $R$ and surface diffusion parameter $\alpha$ are adjusted such that, after folding with the above nucleon density, the measured nuclear charge density distributions are reproduced.\footnote{%
    We do not account for the possibility of a neutron skin.
}
Next, assuming the experiment measures collisions with minimum bias, an impact parameter vector $\mathbf{b}$ in the transverse plane is sampled from a uniform distribution, and each nucleon in nucleus $A$ ($B$) is shifted by $+\mathbf{b}/2$\ \ ($-\mathbf{b}/2$). For each pair of colliding nucleons their collision probability is then sampled by taking into account the proton-proton collision cross section at the specified center-of-mass energy ($\sqrts{}=2.76$\,TeV for this work). Nucleons that do not undergo any collisions and thus do not contribute to mid-rapidity energy deposition are then thrown away. The remaining nucleons in each nucleus are labeled as participants, and their density distributions, integrated along the $z$-direction, are added to get the areal densities in the transverse plane of the participants in each of the two nuclei (their so-called nuclear thickness functions):
\be
T_{A,B}(\xperp) = \sum_{i=1}^{N^{A,B}_\mathrm{part}} \gamma_i \int_{-\infty}^\infty dz \, \rho(\mathbf{x}-\mathbf{x}_i).
\ee
Here $\mathbf{x}_i$ are the nucleon positions, and the $\gamma_i$ are independent random weights, sampled from a Gamma distribution with unit mean and standard deviation $\sigma_k$. These parameterize the measured large multiplicity fluctuations observed in minimum-bias proton--proton collisions. Note that the nuclear thickness functions fluctuate from event to event, due to the fluctuating nucleon positions $\mathbf{x}_i$ and their fluctuating ``interaction strengths'' $\gamma_i$. They describe how much longitudinally integrated matter each nucleus contributes to the collision at each transverse position $\xperp$.

The implementation chosen here then sets the initially deposited energy density at transverse position $\xperp$ to
\be
\label{eq:edens}
  \epsilon(\xperp)=\frac{1}{\tau_0} \frac{dE}{d\eta \, d^2\xperp}=\frac{1}{\tau_0} N\,T_R(\xperp;p),
\ee
where the ``reduced thickness function'' $T_R$ is defined in terms of the two nuclear thickness functions by their ``generalized mean'' \cite{Moreland:2014oya}
\be
\label{eq:harmonic_mean}
  T_{R}(\xperp;p)=\left(\frac{T_A^p(\xperp)+T_B^p(\xperp)}{2}\right)^{1/p}\,.
\ee
Here the longitudinal proper time $\tau_0$ marks the end of the energy deposition process and is taken as small as technically possible (see below). The normalization $N$ depends on $\sqrts$ and allows one to describe the growth of the energy deposited at mid-rapidity with increasing collision energy.

In this work, we assume longitudinal boost invariance for the collision system (i.e., $\eta$-independence of all spatial distributions where $\eta=\frac{1}{2}\ln\bigl[(t{+}z)/(t{-}z)\bigr]$ is the space-time rapidity, and $y$-independence of all momentum distributions where $y=\frac{1}{2}\ln\bigl[(E{+}p_z)/(E{-}p_z)\bigr]$ is the momentum rapidity) and use only observables measured at mid-rapidity as also done in previous Bayesian calibrations of relativistic heavy-ion collision models \cite{JETSCAPE:2020mzn, Nijs:2020ors, Heffernan_CGC}. We start the viscous anisotropic hydrodynamic evolution at $\tau_0=0.05$\,fm/$c$. We use the above \trento{} initial condition model to simulate Pb--Pb collisions at $\sqrts{}=2.76$\,TeV. The Woods-Saxon parameters used for Pb are $R=6.62$\,fm and $\alpha=0.546$\,fm. The nucleon width $w$, their minimum distance $d_\mathrm{min}$, the normalization $N$, the harmonic mean parameter $p$, and the standard deviation of the Gamma distribution $\sigma_k$ are model parameters of interest that we infer from the experimental data using Bayesian parameter estimation. 

\subsection{Viscous Anisotropic Hydrodynamics (\VAH{})}
\label{vah}
\vspace*{-2mm}

Hydrodynamics is an effective theory that can accurately model the space-time evolution of macroscopic properties of many dynamical systems found in nature, ranging from large-scale galaxy formation to small-scale systems such as cold atomic gases \cite{big_picture, Romatschke:2017ejr}. At even smaller scales, hydrodynamic modeling has also been extensively used for decades in describing relativistic heavy-ion collisions \cite{Kolb:2003dz, Heinz:2009xj, Heinz:2013th, Gale, Jeon:2015dfa}. Within the framework of Bayesian parameter estimation it is the workhorse for modeling the QGP stage of the collision fireball \cite{Nijs:2020ors, Bernhard:2015hxa, JETSCAPE:2020mzn, Heffernan_CGC}.

Taking the macroscopic degrees of freedom to be the energy density $\epsilon$ and the fluid four velocity $u_\mu$ (i.e., the four-velocity of the local rest frame (LRF) relative to the global frame), the energy momentum tensor for an ideal fluid (i.e., at zeroth-order in gradients of the macroscopic degrees of freedom) can be decomposed as
\be
    T^{\mu\nu} = \epsilon u^\mu u^\nu - p\Delta^{\mu\nu}.
\label{eq:T_munuideal}
\ee
Here, $\Delta^{\mu\nu}=g^{\mu\nu}-u^\mu u^\nu$ projects on the space-like part in the LRF. The signature of the metric $g^{\mu\nu}$ is taken to be ``mostly minus'' $(+,-,-,-)$. $p(\epsilon)$ is the isotropic equilibrium pressure in the LRF, given by the EoS which is a property of the medium that reflects it microscopic degrees of freedom and their interactions. 

Ideal fluid dynamics embodies the unrealistic assumption of instantaneous local equilibration of the medium, i.e., zero mean free path for the microscopic constituents. For finite microscopic relaxation times (i.e., non-zero mean free path) an expanding fluid is always somewhat out of local equilibrium. These non-equilibrium effects can be written as first- and higher-order gradient corrections to the energy momentum tensor,
\be
    T^{\mu\nu} = \epsilon u^\mu u^\nu - (p+\Pi)\Delta^{\mu\nu} + \pi^{\mu\nu},
\label{eq:T_munuviscous}
\ee
whose evolution is then described by viscous hydrodynamic equations of motion. In second-order viscous hydrodynamics one considers dissipative corrections $\Pi, \pi^{\mu\nu}$ given by a sum of terms containing up to two derivatives of the macroscopic degrees of freedom $\epsilon$ and $u^\mu$, multiplied by so-called transport coefficients that reflect the microscopic transport properties of the fluid medium. To ensure causality, these so-called constituent relations can, however, not be imposed instantaneously (as done in Navier-Stokes theory), but must be implemented via additional equations of motion describing their dynamical evolution toward their Navier-Stokes limits, on time scales related to the microscopic relaxation time. The resulting ``M\"uller-Israel-Stewart (MIS) type'' theories \cite{Muller:1967, Israel:1976tn, Israel:1979wp} introduce additional non-hydrodynamic modes of higher frequencies and shorter wavelengths \cite{Romatschke:2017ejr} which dominate the macroscopic dynamics when the medium develops  large spatial gradients, leading to a breakdown of the gradient expansion and eventually rendering the hydrodynamic approach invalid. In particular, second-order viscous hydrodynamics is not applicable for situations with very large pressure anisotropies such as those encountered in the earliest stage of ultra-relativistic heavy-ion collisions directly after nuclear impact. 

State-of-the-art simulation models for relativistic heavy-ion collisions circumvent this issue by introducing a pre-hydrodynamic stage that evolves the out-of-equilibrium quark-gluon system microscopically until the gradients have decayed sufficiently that the medium can be fed into the hydrodynamic evolution module. One of these pre-hydrodynamic models is based on the simplifying assumption of free-streaming massless partons for which the microscopic kinetic theory can be solved exactly (see, for example, Refs.~\cite{Baym:1984np, Broniowski:2008qk, Liu:2015nwa, Romatschke:2015dha, Liu-thesis}). In this work we avoid this overly simplified picture by using a different formulation of hydrodynamics, viscous anisotropic hydrodynamics (\VAH) \cite{Martinez:2010sc, florkowski2011highly, alqahtani20173+, alqahtani2017anisotropic,  almaalol2019anisotropic, alqahtani2021bulk, bazow2014second, mcnelis20183+, McNelis_2021}, that significantly extends the range of validity of hydrodynamics in the presence of large pressure anisotropies like those encountered in the early stage of heavy-ion collisions \cite{McNelis_2021}, and can even describe the far-off-equilibrium free-streaming limit in that situation, for both massless \cite{bazow2014second} and massive partons \cite{Chattopadhyay:2021ive, Jaiswal:2021uvv}.

The development of viscous anisotropic hydrodynamics was motivated by the fact that the momentum distribution of the QGP is highly anisotropic right after the relativistic heavy-ion collision~\cite{PhysRevD.22.2793}. Initially, the medium has a large expansion rate along the longitudinal beam direction and can be approximated by a boost-invariant longitudinal velocity profile (Bjorken flow \cite{PhysRevD.27.140}). The transverse expansion rate of the medium is initially small, building up only gradually in response to the transverse pressure gradients. The initially highly anisotropic expansion rate engenders a large shear pressure, resulting in a high degree of anisotropy between the pressures in the transverse and longitudinal (beam) directions. 

To formulate viscous anisotropic hydrodynamic one starts by decomposing the spatial projector $\Delta^{\mu\nu}$ further, splitting it into projectors along the beam direction ($z^\mu$) and a transverse projector $\Xi^{\mu\nu}$) as $\Delta^{\mu\nu}=\Xi^{\mu\nu}-z^\mu z^\nu$. Multiplying these projectors by different longitudinal and transverse pressures, the energy-momentum tensor is now decomposed as \cite{Molnar:2016vvu}
\be
\label{eq:vah}
T^{\mu\nu} = \epsilon u^\mu u^\nu + P_L z^\mu z^\nu - P_\perp \Xi^{\mu\nu} + 2 W_{\perp z}^{(\mu} z^{\nu)} + \pi_{\perp}^{\mu \nu}.
\ee
The decompositions (\ref{eq:T_munuviscous}) and (\ref{eq:vah}) are mathematically equivalent  \cite{Molnar:2016vvu}, but the \VAH{} equations evolve the longitudinal and transverse pressures $P_L$ and $P_\perp$ separately, treating them on par with the thermal pressure in standard viscous hydrodynamics and not by assuming that their differences from the thermal pressure and from each other are small viscous corrections.

The evolution equations for the energy density and the flow velocity are obtained from the conservation laws for energy and momentum: 
\be
\partial_\mu T^{\mu \nu} = 0.
\ee
The equilibrium pressure $p(\epsilon)$ is taken from lattice QCD calculations by the HotQCD collaboration \cite{PhysRevD.97.014510}. The dynamical evolution equations for the non-equilibrium flows $P_L, P_\perp, W_{\perp z}^{(\mu} z^{\nu)}, \pi_\perp^{\mu \nu}$ are derived assuming that the fluid's microscopic physics can be described by the relativistic Boltzmann equation with a medium-dependent mass \cite{Alqahtani:2016rth, PhysRevD.95.054007, mcnelis20183+}. The relaxation times for $P_L$ and $P_\perp$ are written in terms of those for the bulk and shear viscous pressures \cite{mcnelis20183+}, which are parameterized as 
\begin{align}
\tau_\pi = \frac{\eta}{s\beta_\pi}, && \tau_\Pi = \frac{\zeta}{s\beta_\Pi}.
\end{align}
$\beta_\pi$, $\beta_\Pi$, as well as all of the anisotropic transport coefficients, are computed within the quasiparticle kinetic theory model discussed in Ref.~\cite{mcnelis20183+}.\footnote{%
    Specifically, $\beta_\pi$ and $\beta_\Pi$ are given by the temperature-dependent isotropic thermodynamic integrals defined in Eq.~(82) in Ref.~\cite{mcnelis20183+}.}

For the temperature-dependent specific shear and bulk viscosities, $\bigl(\eta/s\bigr)(T)$ and $\bigl(\zeta/s\bigr)(T)$, we use the same parameterizations as the JETSCAPE Collaboration \cite{JETSCAPE:2020mzn}:
\begin{eqnarray}
\label{positivity}
    \Bigl(\frac{\eta}{s}\Bigr)(T) &=& \max\left[\left.\frac{\eta}{s}\right\vert_{\rm lin}\!\!\!\!(T), \; 0\right],
\\    
    \text{with}\quad\left.\frac{\eta}{s}\right\vert_{\rm lin}\!\!\!(T) &=& a_{\rm low}\, (T{-}T_{\eta})\, \Theta(T_{\eta}{-}T)+ (\eta/s)_{\rm kink}
\nonumber\\
    &+& a_{\rm high}\, (T{-}T_{\eta})\, \Theta(T{-}T_{\eta}),
\end{eqnarray}
and
\begin{eqnarray}
    \Bigl(\frac{\zeta}{s}\Bigr)(T) &=& \frac{(\zeta/s )_{\max}\Lambda^2}{\Lambda^2+ \left( T-T_\zeta\right)^2},
\\
    \text{with}\qquad\Lambda &=& w_{\zeta} \Bigl(1 + \lambda_{\zeta} \sign \left(T{-}T_\zeta\right) \Bigr)\nonumber.
\end{eqnarray}
A figure illustrating these parameterizations can be found in \cite{JETSCAPE:2020mzn}. 

Apart from the eight parameters related to the viscosities, there is one additional parameter that we will infer from the experimental data: the initial ratio $R=(P_L/P_\perp)_0$ at the time $\tau_0$ when \VAH{} is initialized:
\be
    P_{\perp 0}=\frac{3}{2{+}R}\, p_0\,, \qquad
    P_{L 0}=\frac{3R}{2+R}\, p_0\,.
\label{eq:R}
\ee
Here $p_0{\,\equiv\,}p(\epsilon_0)$ is the equilibrium pressure at time $\tau_0$. The allowed range for $R$ is restricted to $R \in (0.3, 1)$ for technical reasons: For $R<0.3$ the inversion of the relations expressing the macroscopic densities in terms of the microscopic parameters of the distribution function (needed for the calculation of the transport coefficients) fails to converge.\footnote{%
    This can be avoided (Kevin Ingles, private communication) when using the modified Romatschke-Strickland distribution introduced in \cite{Chattopadhyay:2021ive, Jaiswal:2021uvv} but this modification has not yet been implemented in the Bayesian calibration code used here.}
By using this parameterization we also assume that the initial bulk viscous pressure is $\Pi=(2P_\perp{+}P_L)/3-p=0$. The initial flow profile is assumed to be static in Milne coordinates, $u^\mu=(1,0,0,0)$, and the residual shear stresses $W_{\perp z}^\mu$ 
and $\pi_\perp^{\mu \nu}$ are initially set to zero.

\subsection{Particlization}
\label{is3d}
\vspace*{-2mm}

As the QGP expands and cools, it eventually reaches the  critical temperature for hadronization. Below that temperature, the fireball medium can be described as a hadron resonance gas. Its constituents are color neutral and thus interact much less strongly with each other than do the quarks and gluons in the QGP just before hadronization. Correspondingly, the mean free path increases rapidly below this transition, and fluid dynamics quickly becomes inadequate \cite{Song:2011hk}. This breakdown forces a transition back to a microscopic description, for which we use the Boltzmann transport code \SMASH{} \cite{smash_code} briefly described in the following subsection.

This so-called particlization transition is not a physical phase transition but simply a change of language, from fluid dynamical degrees of freedom in \VAH{} to particle degrees of freedom in \SMASH. Following earlier calibration efforts \cite{Bernhard:2016tnd, Moreland:2018gsh, Bernhard:2019bmu, JETSCAPE:2020mzn, JETSCAPE:2020shq, Nijs:2020ors, Nijs:2020roc, Heffernan_CGC}, we keep the particlization temperature as a model parameter to be inferred from the experimental data and denote it by the switching temperature $T_\mathrm{sw}$. Fluid cells that reach the switching temperature and their surface normal vector $s^3\sigma_\mu$ are identified using the code \CORNELIUS{} \cite{Huovinen:2012is}. The particlization hypersurface is passed to the \isd{} hadron sampler \cite{McNelis:2019auj,is3d_code}. There, the Cooper-Frye prescription is used to convert all energy and momentum of the fluid into different hadron types and momenta emitted from the switching hypersurface. 

According to the Cooper-Frye formula  \cite{Cooper:1974mv}, the Lorentz-invariant particle momentum distribution is given by
\be
    p^0 \frac{dN_i}{d^3p} = \frac{g_i}{(2\pi)^3} \int_{\Sigma} d^3\sigma_{\mu}(x) p^{\mu} f_i(x;p).
\ee
Here $p$ is particle four-momentum, $x$ is the position four-vector of the fluid cell, $f_i(x,p)$ is the phase-space distribution function for the particle species $i$, and $\Sigma$ is the switching hypersurface, with volume element $d^3\sigma_\mu(x)$ at point $x\in\Sigma$.

For a locally equilibrated hadron resonance gas the distribution function takes the form 
\be
    f_{\mathrm{eq},i}(x,p)=\frac{g_i}{\exp\bigl[p{\cdot}u(x)/T(x)-\alpha_i(x)\bigr]+\Theta_i}.
\ee
Here $g_i{\,=\,}2s_i{+}1$ is the spin degeneracy for species $i$, $u(x)$ is the 4-velocity of the fluid, $T(x)$ is the temperature, $\alpha_i(x)=b_i\mu_B(x)/T(x)$ is the baryon chemical potential-to-temperature ratio for a hadron with baryon number $b_i$, and $\Theta_i \in [-1, 1]$ accounts for the quantum statistics of fermions and bosons, respectively. In the present work $\mu_B{\,=\,}0$ since at midrapidity the fireball is taken to have zero net baryon number.

On the switching hypersurface $\Sigma$ the QGP liquid is still characterized by large dissipative flows and thus cannot be modeled as a locally equilibrated hadron resonance gas. The distribution functions must be modified such that the momentum moment $\sum_i\langle p^\mu p^\nu\rangle_i=T^{\mu\nu}$ matches the energy-momentum tensor (\ref{eq:vah}) on the particlization surface, including the dissipative flows. There are many types of modifications that fulfill this constraint -- here we choose the Pratt-Torrieri-McNelis-Anisotropic (PTMA) prescription \cite{PhysRevC.103.064903}:\footnote{%
    The theoretical uncertainty introduced by different assumptions for the form of the momentum dependence of $f_i(x,p)$ on the particlization surface and its importance in Bayesian model parameter inference is discussed in Refs.~\cite{PhysRevC.103.064903, McNelis:2019auj, JETSCAPE:2020shq}. In the absence of a full microscopic kinetic description of the evolution preceding particlization it cannot be avoided. A maximum entropy (minimum information) approach to defining $f_i(x,p)$ at particlization was proposed in \cite{Everett:2021ulz} but has not yet been implemented in the Bayesian parameter inference framework. A full Bayesian assessment of the contribution to the error budget of the model parameters arising from the particlization and other model uncertainties will have to await the development of suitably adapted Bayesian model mixing tools \cite{Phillips:2020dmw}.}
\be
\label{eq:PTMA}
  f_{a,i}^{\mathrm{PTMA}}(x,p)=\frac{\mathcal{Z}(x)\, g_i}{\exp\Bigl[\frac{ \sqrt{{p'}^2(x)+m_i^2}}{\Lambda(x)}\Bigr]+\Theta_i}\,.
\ee
The normalization $\mathcal{Z}$, effective temperature $\Lambda(x)$, and the transformation relating $p'(x)$ with $p$ depend on the dissipative terms in the \VAH-decomposition (\ref{eq:vah}) of the energy-momentum tensor at point $x$; the interested reader is referred to Ref.~\cite{PhysRevC.103.064903} for details. Compared to some other prescriptions (see discussion in \cite{PhysRevC.103.064903, McNelis:2019auj}) the PTMA distribution (\ref{eq:PTMA}) has the advantage of being positive definite by construction. 

\subsection{Hadronic afterburner}
\label{smash}
\vspace*{-2mm}

For the final decoupling stage of the fireball evolution, the hadrons sampled from the particlization surface are fed into the hadronic Boltzmann transport code \SMASH{} \cite{smash_code}, which allows the hadrons to rescatter and the unstable resonances to decay and to be recreated in hadronic interactions, until the system becomes so dilute and collisions so rare that first the chemical composition and then the momentum distributions fall out of equilibrium and eventually freeze out \cite{Nonaka:2006yn, Hirano:2007ei, Petersen:2008dd, Song:2010aq, Heinz:2011kt, Song:2013qma, Zhu:2015dfa, Ryu:2017qzn}. To obtain sufficient particle statistics at limited computational cost, we limit the full hydrodynamic runs to a representative sample of the initial-state quantum fluctuations between 200 and 1600 hydro events, distributed over all collision centralities, per design point in parameter space (see App.~\ref{app:datacollection}), but oversample the switching hypersurface for each hydro event many times until a sufficient number of emitted hadrons has been generated for good statistical precision of all observables of interest \cite{Petersen:2010zt, Novak:2013bqa, Sangaline:2015isa, Bernhard:2015hxa, Bernhard:2016tnd, Moreland:2018gsh, Bernhard:2019bmu, JETSCAPE:2020shq, Nijs:2020ors, Nijs:2020roc,  JETSCAPE:2020mzn, Heffernan_CGC}. In practice we oversample each hypersurface until a total of $10^5$ hadrons has been generated per hydrodynamic event, but not more than 1000 times. Finally, the experimental observables measured by the \texttt{ALICE} detector \cite{Aamodt:2010cz, Adam:2016thv, Abelev:2013vea, ALICE:2011ab} are calculated from the final ensemble of hadrons generated by the \SMASH{} output, following the experimental procedures. 

\section{Bayesian calibration}
\label{sec3}
\vspace*{-2mm}

This section provides a brief conceptual summary of Bayesian model calibration; details will be fleshed out in later sections. Let us represent a generic simulation model by a mathematical function $\mathbf{y}_\mathrm{sim}(\cdot)$ that takes values of parameters\footnote{%
    The reader is asked to distinguish this use of $\mathbf{x}$ to describe sets of model parameters from its use as a spatial position vector elsewhere in the text. Its meaning should be clear from the context.
}
$\mathbf{x}{\,=\,}(x_1, \ldots, x_q) {\,\in\,}\mathcal{X}$ and returns output with mean $\mathbf{y}_\mathrm{sim}(\mathbf{x}){\,\in\,}\mathbb{R}^d$. A Bayesian parameter estimation process uses experimental observations, denoted by a vector $\mathbf{y_{\rm exp}}{\,=\,}(y_{{\rm exp}, 1}, \ldots, y_{{\rm exp}, d})$, to infer the simulation parameters $\mathbf{x}$~\cite{Sivia}. In this paper we carry out a Bayesian calibration of the \VAH{} simulation for relativistic heavy-ion collisions, to align the simulation outputs $\mathbf{y}_\mathrm{sim}(\mathbf{x})$ with the experimental data $\mathbf{y_{\rm exp}}$. To do so we write down a statistical model of the form
\begin{equation}\label{eq:generalstatmodel}
    \mathbf{y_{\rm exp}} = \mathbf{y}_\mathrm{sim}(\mathbf{x}) + \pmb{\epsilon}, 
\end{equation}
where the residual error, $\pmb{\epsilon}$, follows a multivariate normal (MVN) distribution with mean $\mathbf{0}$ and covariance matrix $\mathbf{\Sigma}$. In this work, we consider $d=98$ experimental observables for Pb--Pb collisions at a center-of-mass energy of $\sqrts{}=2.76$\,TeV (see Sec~\ref{sec6A} for details). The experimental measurements have experimental uncertainties due to finite measurement statistics and other instrumental effects. Our simulations are also stochastic, and thus the simulation outputs also have uncertainties (included in the error $\pmb{\epsilon}$ in (\ref{eq:generalstatmodel})). Hence, finding the model parameter values that would fit the experimental data (i.e., solving ``the inverse problem'' of determining the model parameters by analyzing the model output), taking into account {\it all} of the uncertainties, requires an advanced probabilistic framework.

In a Bayesian viewpoint, probability is defined as the degree of belief about an hypothesis considering all available information \cite{Sivia}. Parameter inference for physical model simulations is based on Bayes rule:
\begin{equation}
\label{eq:bayes}
    \mathcal{P}(\mathbf{x}|\mathbf{y}_{\rm exp}) =  \frac{\mathcal{P}(\mathbf{y_{\rm exp}|x})\mathcal{P}(\mathbf{x})} 
         {\mathcal{P}(\mathbf{y_{\rm exp}})}.
\end{equation}
The term $\mathcal{P}(\mathbf{x}|\mathbf{y}_{\rm exp})$ on the left-hand side of Eq.~\eqref{eq:bayes} is called the posterior (short for ``the posterior probability density"), which is the probability of the model parameter $\mathbf{x}$ to take a particular value given the experimental data $\mathbf{y_{\rm exp}}$. It is the primary focus of Bayesian parameter inference. On the right-hand side of Eq.~\eqref{eq:bayes}, $\mathcal{P}(\mathbf{x})$ represents the prior probability density for the parameters to take values $\mathbf{x}$, given information from previous experiments and/or independent theoretical input. $\mathcal{P}(\mathbf{y_{\rm exp}|x})$ is the likelihood function, describing the probability density that the model output for a given set of model parameters $\mathbf{x}$ agrees with the experimental data $\mathbf{y_{\rm exp}}$. Given the distribution of $\pmb{\epsilon}$, the likelihood $\mathcal{P}(\mathbf{y_{\rm exp}|x})$ is
\begin{equation}
    \label{eq:likelihood}
     \frac{1}{\sqrt{|2\pi\mathbf{\Sigma}|}}
    \exp\Bigl[-\frac{1}{2}(\mathbf{y}_\mathrm{sim}(\mathbf{x}) - \mathbf{y}_\mathrm{exp})^\top \mathbf{\Sigma}^{-1} (\mathbf{y}_\mathrm{sim}(\mathbf{x}) - \mathbf{y}_\mathrm{exp})\Bigr],
\end{equation}
where the $d \times d$ matrix $\mathbf \Sigma$ represents the total uncertainty, obtained by adding the experimental and simulation uncertainties.  

The posterior generally does not have a closed-form expression, in particular not for heavy-ion collisions. To find the posterior for the parameters $\mathbf{x}$ and to quantify their uncertainty therefore in practice requires numerical techniques that directly produce samples from the posterior. This is typically achieved by using Markov Chain Monte Carlo (MCMC) techniques \cite{Gelman2004, Trotta:2008qt}. These techniques require only relative probabilities, so the normalization $\mathcal{P}(\mathbf{y_{\rm exp}})$ in the denominator on the right of Eq.~\eqref{eq:bayes} (which is independent of the parameters to be inferred) does not need to be calculated.

To produce each sample, MCMC techniques have to evaluate the right-hand side of Eq.~\eqref{eq:bayes} many times for different values of $\mathbf{x}$. When the simulation model is computationally intensive, evaluating the likelihood $\mathcal{P}(\mathbf{y_{\rm exp}|x})$ becomes computationally expensive. Typically, MCMC methods require very many evaluations of the likelihood; this can make Bayesian parameter estimation computationally prohibitive for expensive simulation models such as the one described in the preceding section.

A popular solution is to build a computationally cheaper emulator to be used in place of the simulation model \citep{BMC, Higdon2004}. Gaussian Process (GP) emulators can serve as computationally cheap surrogates to replace an expensive simulation \cite{gramacy2020surrogates, Rasmussen2004}. GP emulators have to be trained on simulation data before they can be used to predict the simulation outputs at any other parameter set $\mathbf{x}$. Once the emulator is built, it returns the prediction mean $\mean(\mathbf{x})$ and the covariance $\variance(\mathbf{x})$ to represent the simulation output $\mathbf{y}_\mathrm{sim}(\mathbf{x})$. In this case, the likelihood in Eq.~(\ref{eq:likelihood}) is approximated as \cite{BMC}
\begin{equation}
    \label{eq:emulikelihood}
     \frac{1}{\sqrt{|2\pi\mathbf{V}(\mathbf{x})|}}
    \exp\Bigl[-\frac{1}{2}(\mean(\mathbf{x}) - \mathbf{y}_\mathrm{exp})^\top \mathbf{V}(\mathbf{x})^{-1} (\mean(\mathbf{x}) - \mathbf{y}_\mathrm{exp})\Bigr],
\end{equation}
where $\mathbf{V}(\mathbf{x}) = \variance(\mathbf{x}) + \mathbf{\Sigma}$. MCMC techniques can then use Eq.~(\ref{eq:emulikelihood}) to draw samples from the posterior in order to avoid a sampling process that depends on extensive evaluation of the expensive computer simulation.

\section{Emulators for the \VAH{} model}
\label{sec4}
\vspace*{-2mm}

Relativistic heavy-ion collision simulations are computationally expensive. Due to irreducible quantum fluctuations in the initial state, for each set of model parameters the simulation has to run multiple times, with stochastically fluctuating initial conditions, in order to produce output that can be meaningfully compared with the experimental measurements. For example, in the JETSCAPE framework for a single set of model parameters, the full-model simulations for 2000 fluctuating initial conditions take approximately $1000$ core hours \cite{ JETSCAPE:2020mzn}. A majority ($\approx$80\%) of the core hours is spent on the hadron transport stage after particlization; the remaining fraction of core hours is mostly utilized by the hydrodynamic QGP evolution code. The MCMC techniques employed in Bayesian parameter estimation require many evaluations of the simulation that, without the use of efficient emulators, renders the inference for heavy-ion collision models computationally infeasible. 

Getting the training data necessary for the GP emulators from the full heavy-ion model is the most computationally demanding step in Bayesian parameter estimation. In this work, we employ several novel methods to significantly reduce this computational cost: a) we use a novel Minimum Energy Design (MED) \cite{MED} for selecting the model parameter values at which we run the full simulation, replacing the Min-Max Latin Hypercube Design \cite{LHS} used in most previous works \cite{Petersen:2010zt, Novak:2013bqa, Sangaline:2015isa, Bernhard:2015hxa, Bernhard:2016tnd, Moreland:2018gsh, Bernhard:2019bmu, JETSCAPE:2020shq, Nijs:2020ors, Nijs:2020roc,  JETSCAPE:2020mzn}; (b) instead of getting all the simulation data at once, we follow a sequential process to obtain batches of simulation data with increasing accuracy in the most probable parameter region; (c) we test several GP emulation methods for relativistic heavy-ion collisions and use a thorough validation to select the best one. All of these features are explained in detail below. 

\subsection{Generating the simulation data for emulator training}
\label{sec4b}

The popular Latin Hypercube Design (LHD) fills the multidimensional model parameter input space in a way that ensures that each parameter is broadly distributed within its full range \cite{LHS}. The emulators that are built using LHD effectively treat all regions of the parameter space as important.

Building emulators for Bayesian parameter estimation is a unique application of emulators. In theory the emulators built for this purpose need to accurately predict the simulation output only in the high posterior regions of the model parameter space. Following this idea naturally leads to a sampling method where some regions of the input parameter space have a higher density of design/training points than others, deviating strongly from a LHD. This targeted sampling approach saves computational cost by requiring a smaller overall number of high-precision simulations, by sacrificing emulation accuracy in uninteresting (low posterior probability) regions of the model parameter space for higher precision simulations in regions of interest (high posterior probability). This idea of building emulators by sequentially sampling with a bias towards the interesting regions while still exploring, albeit with less model precision, the full parameter space domain specified by the prior $\mathcal{P}(\mathbf{x})$ is known as {\it active learning} in the machine learning and statistics literature (see Chapter~6 in \cite{gramacy2020surrogates} for a detailed review).

In this work, we adopt an active learning approach to generate the VAH{} simulation data for training our emulators. Our simulation is inherently stochastic due to the random initial energy depositions and the probabilistic conversion of QGP fluid to hadrons at particlization. Our final aggregated observables are calculated from the stochastic simulation by evaluating it multiple times with random initial conditions at any given model parameter set. We refer to each of these evaluations as an event. As the number $N_\mathrm{events}$ of events simulated for a given design point increases, the statistical accuracy of the final simulation output observables increases roughly proportional to $\sqrt{N_\mathrm{events}}$ while the computational expense increases linearly. In the following we describe the sequential design approach to simulate design points with varying precision (from 200 events per design to 1600 events per design, which we refer to as low-precision and high-precision simulations, respectively) to build emulators for Bayesian parameter estimation. 

First, we use low-precision simulations on a very sparse set of initial design points to estimate an intermediate posterior for potential additional intermediate design point. Informed by the initial design, we decide on which regions in the parameter space we should put more weight when sampling the next batch of design points, and use maximum energy design (MED) to do intelligent sequential sampling. MED selects the points that minimize the total ``potential energy'' as defined  in \cite{Joseph2015}, and fast procedures for generating MED samples are provided in \cite{MED}. Suppose that we start with $j$ samples of the $q$-dimensional parameter space and their corresponding simulation outputs. According to the MED selection criterion, the $(j{+}1)$th point is selected at
\begin{equation}
\label{eq:medcriterion}
    \mathbf{x}_{j+1} = \argmax_{\mathbf{x} \in \mathcal{L}} \min_{i=1:j} \mathcal{P}^{1/2q}(\mathbf{y_{\rm exp}|x}) \mathcal{P}^{1/2q}(\mathbf{y_{\rm exp}}|\mathbf{x}_i) d(\mathbf{x}, \mathbf{x}_i)
\end{equation}
where $\mathcal{L}$ represents a candidate list of design parameters that we generate from a LHD and $d(\mathbf{x}, \mathbf{x}_i)$ represents the Euclidean distance between $\mathbf{x}$ and $\mathbf{x}_i$. 
Since model simulations have not yet been performed for any of the points $\mathbf{x}{\,\in\,}\mathcal{L}$, the likelihood $\mathcal{P}(\mathbf{y_{\rm exp}|x})$ at such an ``unseen'' point $\mathbf{x}$ must be estimated using Eq.~(\ref{eq:emulikelihood}) with the emulator built using the simulation data retrieved at design points $\{\mathbf{x}_i|i=1,\dots,j\}$. After generating a batch of design points in regions of large estimated values for the posterior via Eq.~(\ref{eq:medcriterion}), high-precision full-model simulations with increased event statistics per design are performed for this batch, and their output is added to the previously simulated events and used to build new emulators with improved precision. The process is iterated until the desired emulator precision has been reached. Full details are given in App.~\ref{app:datacollection}.

\subsection{Gaussian process emulation}
\label{sec4a}

In this study, we leverage a GP-based emulator using the basis vector approach \cite{Higdon2008} since the simulation returns a length $d{\,=\,}98$ vector. Principal component analysis (PCA) \cite{Ramsay97functionaldata} is used to project the high-dimensional outputs into a low-dimensional space where the projection is a collection of latent outputs. Then, each latent output is modeled using an independent GP model. We have tested different emulation strategies, and we provide the results for stochastic kriging \cite{Ankenman2009} since it returns the best prediction accuracy on the test data. Relativistic heavy-ion collision simulations are stochastic, meaning that each time the simulation model is evaluated with  the same parameter setting the simulation output is different. The stochastic kriging approach incorporates both the intrinsic uncertainty inherent in a stochastic simulation and the extrinsic uncertainty about the unknown simulation output. Additional information on other emulators we explored are presented in App.~\ref{app:additionalemu}.

For the rest of this section, let $\{\mathbf{x}_1^{\rm tr}, \ldots, \mathbf{x}_n^{\rm tr}\}$ denote the ${n}$ unique parameter samples used to train each of the emulators. Let $\bar{\mathbf{y}}_\mathrm{sim}(\mathbf{x}^{\rm tr}_i)$ be a $d$-dimensional aggregated output vector of observables from the simulation model evaluated at the model parameter value $\mathbf{x}^{\rm tr}_i$ across a number of random initial conditions. The standardized outputs are stored in a $d \times n$ matrix $\Xi$ where the $i$th column is $(\bar{\mathbf{y}}_\mathrm{sim}(\mathbf{x}^{\rm tr}_i) - \mathbf{h})/\mathbf{c}$ (computed element-wise); and $\mathbf{h}$ and $\mathbf{c}$ are the centering and scaling vectors. Principal component analysis finds a $d{\,\times\,}p$ linear transformation matrix $\mathbf{A} = [\mathbf{a}_1, \ldots, \mathbf{a}_p]$, which projects $d$-dimensional simulation outputs $\Xi$ to a collection of latent outputs $\mathbf{Z} = [\bm{z}_1, \ldots, \bm{z}_p] = \mathbf{A}^{\!\top}\Xi$ in a $p$-dimensional space. In this work, we find that keeping twelve principal components ($p{\,=\,}12$) explains 98\% of the variance in the original simulation data set.  After transformation, we build an independent GP for each latent output $z_t(\cdot) = \mathbf{a}_t^\top \mathbf{G}^{-1}(\bar{\mathbf{y}}_\mathrm{sim}(\cdot) - \mathbf{h})$ where $\mathbf{G} = {\rm diag}(\mathbf{c})$. The GP model results in a prediction with mean $m_t(\mathbf{x})$ and variance $s^2_t(\mathbf{x})$ such that
\begin{equation}
    z_t(\mathbf{x})|\bm{z}_t \sim  {\rm N}(m_t(\mathbf{x}), s^2_t(\mathbf{x})), \quad \text{for} \quad t = 1, \ldots, p.
    \label{eq:gppost}
\end{equation}
Following the properties of GPs \cite{santner2003design}, the mean $m_t(\mathbf{x})$ and variance $s^2_t(\mathbf{x})$ are given by
\begin{align}
\begin{split}
    m_t(\mathbf{x}) &= \mathbf{k}_t^\top \mathbf{K}_t^{-1} \bm{z}_t\\
    s^2_t(\mathbf{x}) &= k_t(\mathbf{x},\mathbf{x}) - \mathbf{k}_t^\top \mathbf{K}_t^{-1} \mathbf{k}_t,
    \label{eq:gpeqns}
\end{split}
\end{align}
where $k_t(\cdot,\cdot)$ is the covariance function, $\mathbf{k}_t = \bigl[k_t(\mathbf{x}, \mathbf{x}^{\rm tr}_i)\bigr]_{i=1}^n$ is the covariance vector between $n$ parameters $\{\mathbf{x}_1^{\rm tr}, \ldots, \mathbf{x}_n^{\rm tr}\}$ and any parameter $\mathbf{x}$, and $\mathbf{K}_t = \bigl[k_t(\mathbf{x}_i^{\rm tr},\mathbf{x}^{\rm tr}_j) + \delta_{ij} \frac{r_{t,i}}{a_i}\bigr]_{i,j=1}^n$ is the covariance matrix between $n$ parameters used to train the GP (see App.~\ref{gpfitting} for additional details). Here, $r_{t,i}$ represents the intrinsic uncertainty due to the stochastic nature of the simulation output across different events (i.e., different random initializations), $a_i$ is the number of events, and $\delta_{ij}$ is the Kronecker-$\delta$. For the covariance function, there are several popular choices including Gaussian,\footnote{%
        In the statistics literature the Gaussian function is often called a ``squared-exponential", indicated here by the superscript SE.}
Mat\'ern, and cubic covariances \cite{Rasmussen2004}.

Once the GP is fitted, the goal is to predict the simulation output at an unseen point $\mathbf{x}$ as follows. Equations (\ref{eq:gppost})--(\ref{eq:gpeqns}) provide the basis for emulator modeling: the posterior mean $m_t(\mathbf{x})$ serves as the emulator model prediction at a new point $\mathbf{x}$, and the posterior variance $s_t^2(\mathbf{x})$ quantifies the emulator model uncertainty. A key appeal of GP emulators is that both their prediction and uncertainty can be efficiently computed via such closed-form expressions. For a prediction of the model output at any test point $\mathbf{x}$, we first obtain the mean $m_t(\mathbf{x})$ and variance $s^2_t(\mathbf{x})$ in Eq.~(\ref{eq:gpeqns}) for each of the corresponding latent outputs $t = 1, \ldots, p$. These are then transformed back to the original $d$-dimensional space through the inverse PCA transformation as follows: Define the $p$-dimensional vector $\mathbf{m}(\mathbf{x}) = (m_1(\mathbf{x}), \ldots, m_p(\mathbf{x}))$ and a $p \times p$ diagonal matrix $ \mathbf{S}(\mathbf{x})$ with diagonal elements $s^2_t(\mathbf{x})$ ($t = 1, \ldots, p$). Then the inverse PCA transformation yields
\begin{equation}
    \mathbf{y}_\mathrm{sim}(\mathbf{x}) \sim \text{MVN}(\mean(\mathbf{x}), \variance(\mathbf{x})),
\label{emu_final}
\end{equation}
where $\mean(\mathbf{x}) = \mathbf{h} + \mathbf{G} \mathbf{A} \mathbf{m}(\mathbf{x})$ is the emulator predictive mean and $\variance(\mathbf{x}) = \mathbf{G} \mathbf{A} \mathbf{S}(\mathbf{x})\mathbf{A}^\top \mathbf{G}$ is the covariance matrix. By plugging $\mean(\mathbf{x})$ and $\variance(\mathbf{x})$ into Eq.~(\ref{eq:emulikelihood}) we obtain the approximate likelihood at parameter point $\mathbf{x}$. We then use Eq.~(\ref{eq:bayes}) to compute the posterior $\mathcal{P}(\mathbf{x|y_{\rm exp}})$.

In this way, by employing the MCMC method on the sequentially updated emulators, we extract posterior distributions for the parameters of the \VAH{} simulations of heavy-ion collisions without running the computationally expensive simulation model when doing the MCMC sampling.

\begin{figure}[b]
\hspace*{-5mm}
\includegraphics[width=1.1\linewidth]
{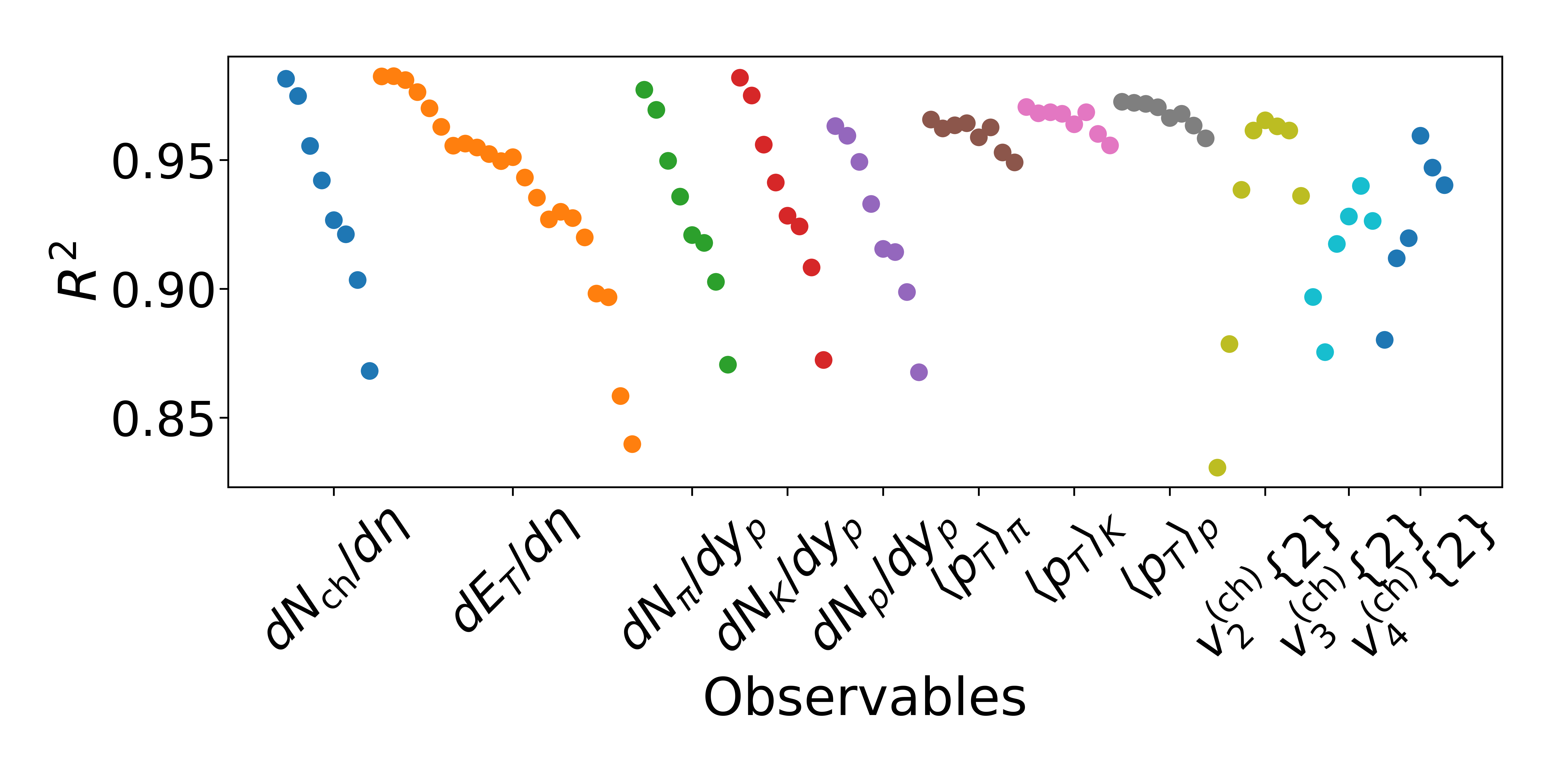}
\caption{$\mathrm{R}^2$ scores calculated using the test simulations and emulator predictions for principal component stochastic kriging (PCSK) for eleven observable types. Each observable type is represented by a unique color, and each dot corresponds to measurements of that observable type at increasing centrality when going from left to right.
}
\label{R2_pcgpr_sub_groups}
\end{figure}

Once the emulators are fitted with training data, and before using them for Bayesian parameter inference, we test the accuracy of the emulators. To do that, we use the model simulation output from a batch of $m$ design points that was put aside and not used for emulator training (specifically batch (b) described in App.~\ref{app:datacollection}). The emulator accuracy is measured using the coefficient of determination ($R^2$ score) for the $l^ \mathrm{th}$ observable ($l = 1, \ldots, d$) defined by 
\begin{eqnarray}
\label{eq:R2}
    && R^2_l = 1 - \frac{\sum_{i=1}^m(\bar{\mathbf{y}}_{{\rm sim}, l}(\mathbf{x}_i^{\rm test})-\mean_l(\mathbf{x}_i^{\rm test}))^2}
    {\sum_{i=1}^m(\bar{\mathbf{y}}_{{\rm sim},l}(\mathbf{x}_i^{\rm test})-\bar{\bar{\mathbf{y}}}_{{\rm sim},l}(\mathbf{x}_i^{\rm test}))^2}\,, 
\\\nonumber
    && {\rm where} \quad \bar{\bar{\mathbf{y}}}_{{\rm sim},l}(\mathbf{x}_i^{\rm test}) = \frac{\sum_{i=1}^{m} \bar{\mathbf{y}}_{{\rm sim},l}(\mathbf{x}_i^{\rm test})}{m}.
\end{eqnarray}
The maximum possible value for $R_l^2$ is $R_l^2{\,=\,}1$, which occurs only when the emulation predictions for observable $l$ are identical to the simulation output for all $m$ test designs (i.e., the emulator is perfect on the test designs). The $R^2$ score can assume negative values when the emulator predictions deviate strongly from the simulation outputs for the test designs. 

The $R^2$ scores we obtain for the \VAH{} emulators trained by using the principal component stochastic kriging (PCSK) method\footnote{%
    Two other types of emulators, obtained with the {\it principal component Gaussian process regression} (PCGP) and the {\it principal component Gaussian process regression with grouped observables} (PGPRG) methods, respectively, were also studied (see App.~\ref{app:additionalemu} for a description and the corresponding $R^2$ values). The $R^2$ values exhibit a general tendency to decrease as we switch from PCSK to PCGPR emulators, and then again as we move to PCGPRG emulators. For the analysis in the rest of the paper we have therefore chosen the PCSK emulators as the most accurate ones.}
%
%
\begin{figure*}[!t]
\noindent\makebox[1\textwidth]{%
  \centering
  \begin{minipage}{0.5\textwidth}
    \includegraphics[width=\linewidth]{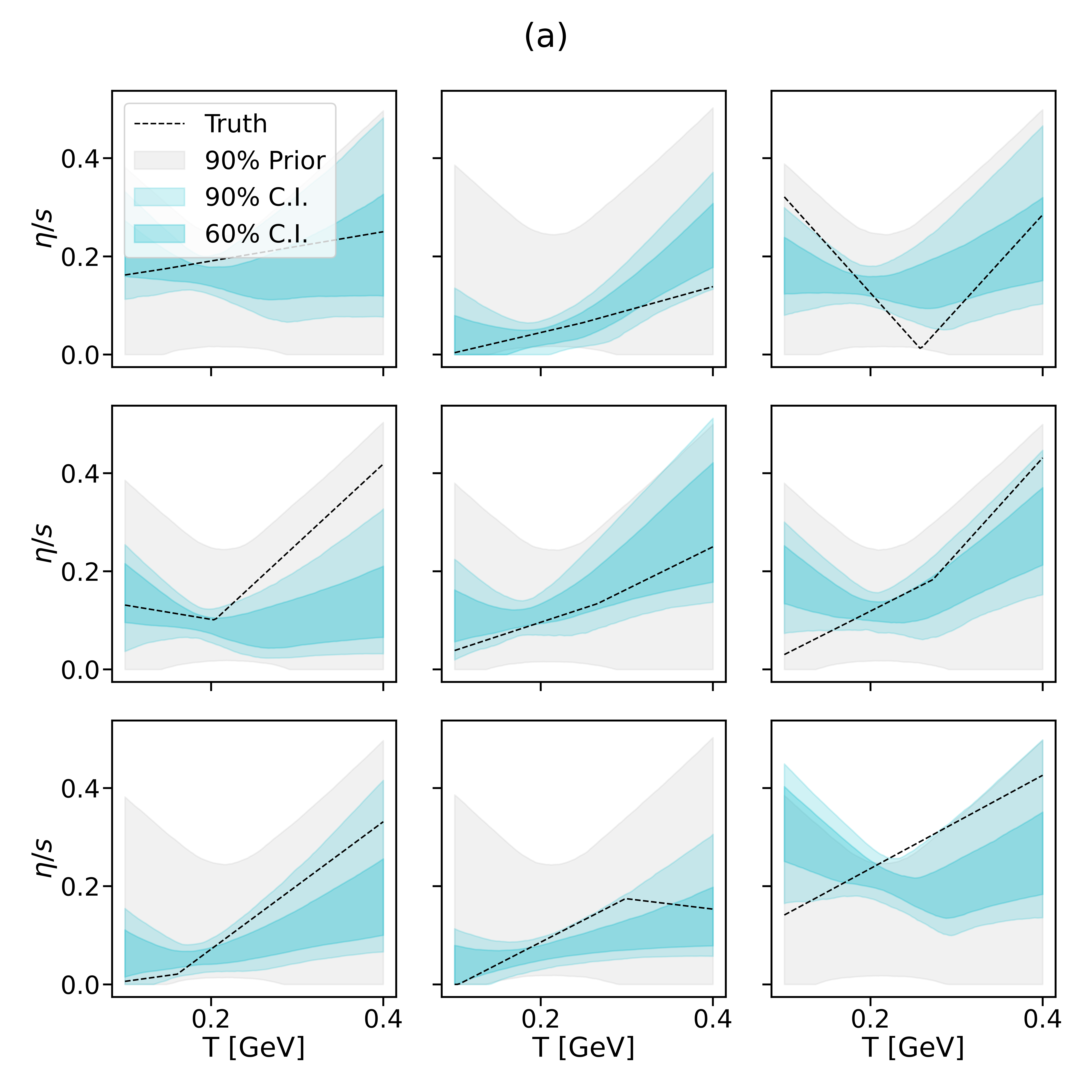}
  \end{minipage}
  \begin{minipage}{0.5\textwidth}
    \includegraphics[width=\linewidth]{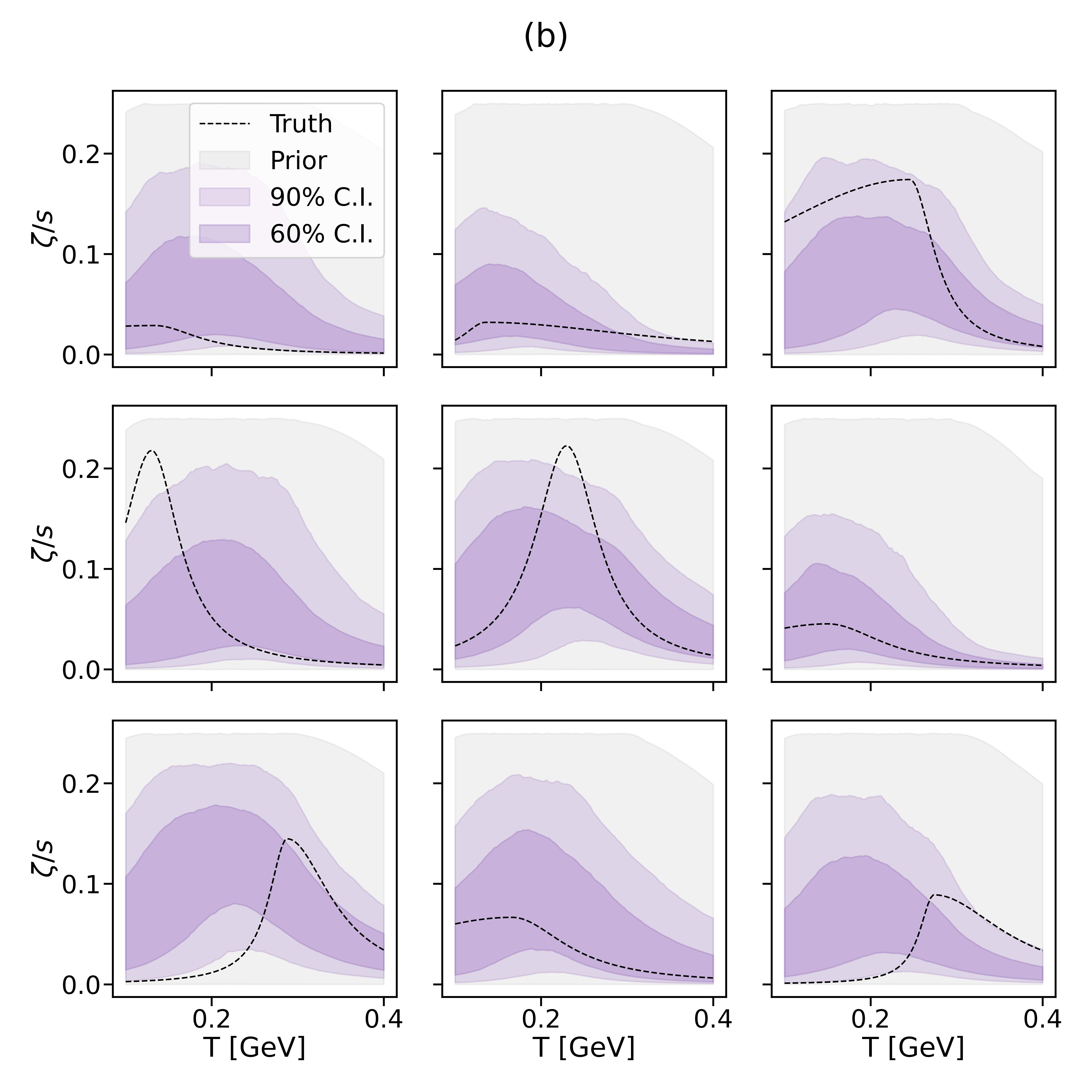}
  \end{minipage}
  }
  \caption{Closure tests for the temperature-dependent specific (a) shear and (b) bulk  viscosities. The colored bands show different confidence intervals (C.I.).}
  \label{fig:closure_tests}
\end{figure*}%
%
are shown in Fig.~\ref{R2_pcgpr_sub_groups}. Each color corresponds to a different type of observable, and each dot corresponds to a different collision centrality, ordered from left to right by increasing centrality. In total there are 98 dots, corresponding to the 98 observables considered in this work.  

\section{Closure tests}
\label{sec5}
\vspace*{-2mm}

Following Ref.~\cite{JETSCAPE:2020mzn}, we perform closure tests before proceeding to the model calibration stage. By using Bayesian inference to reconstruct the model parameters from simulated (``mock'') data generated by model runs for a known point in model parameter space, closure tests are an important check of the ability of the inference framework to correctly infer model parameters from real experimental measurements, and they also provide a deeper understanding of the behavior of the uncertainties associated with the inference process.
 
We randomly pick nine design points from our most accurate simulation data set (batch (e) in App.~\ref{app:datacollection}), which has 1600 events per design. We train our emulators without including these nine design points in our training data. Then, the simulation outputs of these nine design points are taken as pseudo-experimental data and we perform the Bayesian parameter inference to find the most probable values of the model parameters that can reproduce the pseudo-experimental data. Finally, we compare the inferred model parameter values with the known truth values that generated these pseudo-experimental data and validate our Bayesian parameter estimation framework. 

Figure~\ref{fig:closure_tests} shows the closure test results for the temperature-dependent specific bulk and shear viscosities. One sees that in most of the nine cases the true model parameter values (giving rise to the dotted lines) are well captured by the 90\% inferred posteriors for these model parameters (indicated by lighter and darker colored regions for the 90\% ad 60\% confidence intervals (C.I.), respectively). We expect the true value (black dashed line) to lie inside the 90\% confidence interval 90\% of the time. We also observe that for some pseudo-experimental data values (e.g., left column, middle row in both panels (a) and (b)) the closure can be poor. There are several possible reasons for this to happen: a) emulators are not well trained to capture the simulation behavior around some of the parameter values used to generate pseudo-experimental data; b) in some regions of the parameter space the temperature-dependent viscosities are not sensitive enough to be accurately inferred using the currently used experimental observables; a similar issue was reported in Section~V.B of Ref.~\cite{Heffernan_CGC}. The observation of such instances of poor closure highlights the importance of further validation tests after completing the model calibration. For this purpose we perform posterior predictive tests for the experimental data as discussed in Sec.~\ref{sec8}. For the closure tests shown here, in addition to the viscosities we also checked the consistency of the posteriors for the remaining model parameters with their true values; we found all of them to be inferred accurately. 

\section{Bayesian calibration of \VAH{} with LHC data}
\label{sec6}
\vspace*{-2mm}
\subsection{Calibration procedure}
\label{sec6A}
\vspace*{-2mm}

With our Bayesian tool set in place we perform Bayesian parameter inference using experimental data for Pb--Pb collisions at center-of-mass energy $\sqrts{}=2.76$\,TeV collected at the LHC. For ease of comparison, we use (almost\footnote{%
   We omit the fluctuations in the mean transverse momentum, $\delta p_T /\langle p_T\rangle$ \cite{Abelev:2014ckr}, for centrality bins ranging from 0 to 70\% centrality in our analysis. We found that excluding this observable increases the validation scores $R^2$ for all observables. We suspect this is due to the current number of events per design being insufficient to calculate the mean transverse momentum accurately.}%
) the same set of measurements as the recent JETSCAPE analysis \cite{JETSCAPE:2020mzn} (98 measurements in total):
\begin{itemize}
    \item the charged particle multiplicity $dN_{\text{ch}}/d\eta$ \cite{Aamodt:2010cz} for centrality bins covering $0{-}70$\% centrality;
    \item the transverse energy $dE_T/d\eta$ \cite{Adam:2016thv} for centrality bins covering $0{-}70$\% centrality;
    \item the multiplicity $dN/dy$ and mean transverse momenta $\langle p_T \rangle $ of pions, kaons, and protons \cite{Abelev:2013vea} for centrality bins covering $0{-}70$\% centrality;
    \item the two-particle cumulant harmonic flows $v_n\{2\}$ $(n{\,=\,}2,3,4)$ for centrality bins covering $0{-}70$\% centrality for $n{\,=\,}2$ and $0{-}50$\% centrality for $n{\,=\,}3,\,4$ \cite{ALICE:2011ab}.
\end{itemize}

The 15 model parameters that we infer and their priors are listed in Table~\ref{table:prior}. Since \VAH{} has no free-streaming stage, the associated JETSCAPE parameters are missing from the table. The additional parameter $R{\,\in\,}[0.3,\,1]$, defined in Eq.~(\ref{eq:R}), controls the initial pressure anisotropy $(\mathcal{P}_L/\mathcal{P}_L)_0$ and is unique to \VAH. 

\begin{table}
\begin{tabular}{l|l}
parameter & prior range\\ \hline \hline
$N$ & $[10, 30]$  \\ \hline
$p$ & $[-0.7, 0.7]$  \\ \hline
$w$ [fm] & $[0.5, 1.5]$ \\ \hline
$d_{\mathrm{min}}$ [fm] & $[0.0, 1.7]$ \\ \hline
$\sigma_k$ & $[0.3, 2.0]$ \\ \hline
$T_\mathrm{sw}$ [GeV] & $[0.135, 0.165]$ \\ \hline
$R$ & $[0.3, 1]$ \\ \hline
$T_{\eta/s, \mathrm{kink}}$\ [GeV] & $[0.13, 0.3]$  \\ \hline
$(\eta/s)_{\rm kink}$ & $[0.01, 0.2]$  \\ \hline
$a_{\mathrm{high}}$ [GeV$^{-1}$] & $[-1, 2]$ \\ \hline
$a_{\mathrm{low}}$ [GeV$^{-1}$] & $[-2, 1]$ \\ \hline
$(\zeta/s)_{\mathrm{max}}$ & $[0.01, 0.25]$ \\ \hline
$T_{\zeta}$ [GeV] & $[0.12, 0.3]$ \\ \hline
$w_{\zeta}$ [GeV] & $[0.025, 0.15]$ \\ \hline
$\lambda_\zeta$ & $[-0.8, 0.8]$   
\end{tabular}
\caption{
List of \VAH{} model parameters. We use uniform priors throughout, with prior ranges (specified in the right column) that agree with the ones used in Ref.~\cite{JETSCAPE:2020mzn}.}
\label{table:prior}
\end{table}

For the likelihood function we use the multivariate normal distribution (\ref{eq:likelihood}). The variances characterizing the experimental and simulation uncertainties are added together to obtain the total uncertainty appearing in the likelihood.

The posterior is found by combining the (multivariate Gaussian) likelihood with the (multivariate uniform) prior according to Bayes'\ rule (\ref{eq:bayes}). The resulting 15-dimensional posterior probability distribution is analyzed by sampling it using MCMC \cite{MCMC, Trotta_2008}. Specifically, we use the Parallel Tempering MCMC technique \cite{Vousden_2015, Foreman_Mackey_2013, ptemcee_code}, on account of its robustness in sampling multi-modal posterior distributions. For the temperature ladder we choose 500 values for the tempering temperature, evenly distributed between 0 and 1000. For each temperature in the ladder we run 100 randomly initialized chains (see \cite{ptemcee_code} for technical details of the parallel tempering algorithm). In each chain we discard the first 1000 steps as burn-in. The final posterior samples are obtained from the next 5000 steps, after thinning the chains by a factor 10, by saving only every 10th sample.  

\subsection{Posterior for the model parameters}
\label{sec6B}

\begin{figure*}[!h]
    \noindent\makebox[1\textwidth]{%
    \centering
    \includegraphics[width=1.03\linewidth]{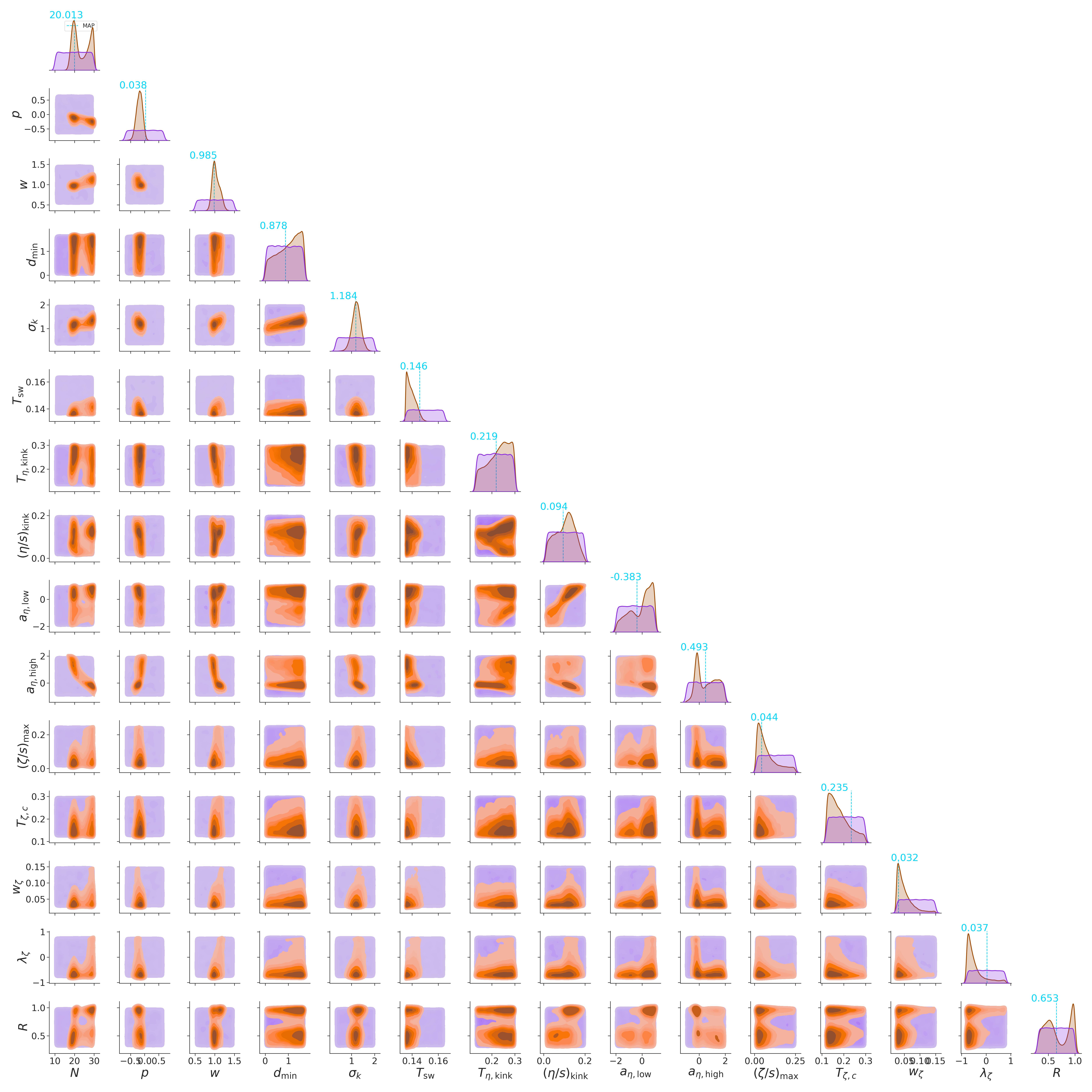}
    }
\caption{Joint marginal distributions of the posterior for all model parameters, for \VAH{} with PTMA viscous corrections at particlization and using experimental data from Pb--Pb collisions at $\sqrts{}=2.76$\,TeV. Posterior distributions are represented by shades of red while the (uniform) prior distributions are shown in lilac. Cyan numbers and vertical dotted lines in the diagonal panels indicate the MAP values for each parameter. (See Table~\ref{table:prior} for the units of the model parameters.)}
\label{fig:full_posterior}
\end{figure*}

The joint marginal posterior distributions (obtained by projecting the posterior on two dimensions in all possible ways) for all the model parameters are shown in Fig.~\ref{fig:full_posterior}; the diagonal shows the marginal posterior distributions for each model parameter. These marginal distributions are obtained by integrating the posterior over all other model parameters, except the chosen one or two. The model parameter values that maximize the high-dimensional posterior (the ``mode'' of the distribution) are called Maximum a Posteriori (MAP) parameters --- they are indicated by blue dotted vertical lines in the diagonal panels. 

In each panel the lilac color shades the uniform prior distribution density; shades of red color indicate the projected density of the posterior distribution. In cases where the available experimental data have insufficient constraining power on a parameter one expects the prior (blue) and posterior (red) marginal distributions to largely agree. In these cases the posterior  essentially returns the information already contained in the prior. Figure~\ref{fig:full_posterior} shows that two of our model parameters related to the very early stage of the fireball, the initial pressure ratio $R$
and the minimum distance $d_\mathrm{min}$ between nucleons, are not well constrained by the experimental data.\footnote{%
    To the extent that the very early fireball expansion stage respects Bjorken symmetry \cite{PhysRevD.22.2793} (an approximation expected to hold well in collisions between large nuclei at LHC energies), the dynamical evolution of $R$ is controlled by a far-from-equilibrium hydrodynamic attractor to which it decays rapidly on a time scale ${\sim\,}\tau_0$, irrespective of its initial value, following a power law decay \cite{Florkowski:2017olj, Jaiswal:2019cju, Kurkela:2019set, Chattopadhyay:2021ive, Jaiswal:2021uvv}, and which is well described by viscous anisotropic hydrodynamics \cite{Chattopadhyay:2021ive, Jaiswal:2021uvv}. This may be the main reason for our inability to constrain it well using final-state experimental measurements.}
%
%
The shape of the probability density contours in the off-diagonal panels conveys information about correlations among the inferred model parameters. When considering parameters unrelated to the viscosities one observes slight positive correlations between the nucleon width parameter $w$, as well as the multiplicity fluctuation parameter $\sigma_k$, and the \trento{} normalization $N$, and between $\sigma_k$ and the minimum nucleon distance $d_\mathrm{min}$; the power $p$ in the \trento{} parameterization (\ref{eq:harmonic_mean}) of the reduced thickness function $T_R$ is slightly anti-correlated with $N$ and $R$. More striking is the bimodal structure of the \trento{} normalization parameter $N$. In all previous Bayesian parameter inferences for heavy-ion collision simulations for Pb--Pb at $\sqrts{}=2.76$\,TeV energy, the marginal distribution of $N$ was well constrained to a Gaussian-like distribution \cite{Petersen:2010zt, Novak:2013bqa, Sangaline:2015isa, Bernhard:2015hxa, Bernhard:2016tnd, Moreland:2018gsh, Bernhard:2019bmu, JETSCAPE:2020shq, Nijs:2020ors, Nijs:2020roc,  JETSCAPE:2020mzn}. Since $N$ determines the initial energy deposited after the collision, it is also closely related to the initial entropy. This causes the final particle yields to directly scale with $N$. Using primarily the final measured particle yields and transverse energy, all previous Bayesian studies were able to tightly constrain $N$. The deviation from this in the \VAH{} model warrants further investigation into the sub-dominant peak that appears at higher $N$ values closer to the prior bound.

As a first step towards understanding the bimodal structure, we look at all the correlations seen in the joint marginal distributions between $N$ and the other model parameters, shown in the leftmost column of Figure~\ref{fig:full_posterior}.
Five parameters are seen to correlate most strongly with the two peaks in the marginal posterior for $N$: $p$, $w$, $\sigma_k$, $a_{\mathrm{high}}$, and $R$ all prefer different values for the two peaks of $N$. This suggests that one of these parameters, or some combination of them, acts to compensate the effect that a large $N$ value has on the final observables.\footnote{%
    We note that increasing $N$ causes all the final observables considered in this work to increase.}

To narrow things down further requires going beyond the pairwise joint marginal posterior distributions. We therefore jump ahead to the model sensitivity shown in Fig.~\ref{fig:sensitivity} which will be discussed in greater detail in Sec.~\ref{sec7} below. Figure~\ref{fig:sensitivity} shows that the charged particle yields for central collisions show significant sensitivities only to $N$ and the specific shear viscosity $\eta/s$. Further checking along these lines points to the possibility that the two-peak structure in $N$ might be related specifically to the high-temperature slope of the specific shear viscosity, $a_{\mathrm{high}}$. 

Closer inspection of the joint marginal distribution between $a_{\mathrm{high}}$ and $N$ in Figure~\ref{fig:full_posterior} shows that these two parameters are anti-correlated. The peak at large $N$-values correlates with negative slope $a_{\mathrm{high}}<0$ (i.e. a specific shear viscosity that decreases with temperature) whereas $a_{\mathrm{high}}>0$ (i.e. $\eta/s$ increases with temperature) for smaller values of $N$. 

The \VAH{} model starts the hydrodynamic evolution at the very early time of $\tau_0=0.05$\,fm/$c$, thus probing much higher temperatures than any other hydrodynamic model for heavy-ion collisions used so far. Smaller (but still positive) high-temperature values of $\eta/s$ result in reduced viscous heating during the earliest evolution stage and, therefore, in lower final particle yields as well as smaller mean transverse momenta and flow anisotropies. To hold these observables constant as $N$ increases, the Bayesian fit compensates by reducing the specific shear viscosity in particular at early times (i.e. at high temperature) when the rate of viscous entropy production is largest. This is accomplished by selecting a negative slope parameter $a_{\mathrm{high}}$.\footnote{\label{widget}%
    The interested reader is invited to confirm these claims with the help of the emulator-based ``\VAH{} widget'' which can be found at \url{https://danosu-visualization-vah-streamlit-widget-wq49dw.streamlit.app/} \cite{vah_tool}}
This unique characteristic of the \VAH{} model explains the bimodal structure of the marginal posterior for $N$ which has not been observed in any other model for heavy-ion collisions. We note that a similar bimodal structure is also seen in the joint marginal distributions of several other parameter pairs but hidden in their individual marginal distributions.

For the \VAH{} model the MAP value for the harmonic mean parameter $p$ in \trento{} (see Eq.~(\ref{eq:harmonic_mean})) is 0.038, close to the preferred value of zero found in all previous Bayesian parameter inferences using a free-streaming pre-hydrodynamic stage \cite{Petersen:2010zt, Novak:2013bqa, Sangaline:2015isa, Bernhard:2015hxa, Bernhard:2016tnd, Moreland:2018gsh, Bernhard:2019bmu, JETSCAPE:2020shq, Nijs:2020ors, Nijs:2020roc,  JETSCAPE:2020mzn}. However, its marginal posterior distribution is shifted toward slightly negative values. Its correlation with $N$ suggests that this shift is associated with the large posterior density around negative values of $a_{\mathrm{high}}$, which we argued above corresponds to small specific shear viscosity values at early times. The small specific shear viscosity causes the fluid to behave more ideal, quickly erasing all initial momentum anisotropies and thereby reducing all finally measured flow harmonics. To compensate for this effect the harmonic mean parameter $p$ tends negative, which, as observed in \cite{Moreland:2014oya} and confirmed with the \VAH{} widget in \cite{vah_tool}, increases all flow harmonics. Decreasing the specific shear viscosity also correlates with a growth of the multiplicity fluctuation parameter $\sigma_k$, compensating for the faster decay of initial momentum anisotropies with larger initial density fluctuations to keep the final anisotropic flow response strong.

We note that the marginal posterior peak at the larger $N$ values correlates not only with a reduced specific shear viscosity at high temperatures, but also with a large value $R\approx 1$ of the initial pressure ratio $P_L/P_\perp$ (i.e. a small value for the initial shear stress). We point out that the degeneracy in probability between the two peaks seen in the marginal posterior for $R$ can be broken, at the level of the prior distribution for $R$, by invoking additional ``prior theoretical'' input: In the widely considered Color Glass Condensate (CGC) model \cite{Gelis:2010nm}, which is expected to apply to heavy-ion collisions at LHC energies, the longitudinal pressure $P_L$ is predicted to be initially negative, rising very quickly (on a time scale ${\sim\,}1/Q_s$ where $Q_s{\,\sim\,}1{\,-\,}3$\,GeV is the saturation momentum of the CGC) to positive values and approaching the transverse pressure $P_\perp$ from below at late times as the system moves closer to thermal equilibrium \cite{Epelbaum:2013ekf}. This picture would definitely tilt the prior for $R$ in the lower right panel of Fig.~\ref{fig:full_posterior} strongly toward the left, assigning very low prior (and therefore also posterior) probability to initial pressure ratios near $R{\,=\,}1$.

Similar to earlier Bayesian model calibrations \cite{Moreland:2018gsh, Bernhard:2019bmu, Nijs:2020ors, JETSCAPE:2020mzn, Nijs:2021clz} we find the nucleon width parameter $w$ constrained to a value around 1\,fm. Since this value is larger than the nucleon width needed to match the recently measured total hadronic cross sections for $p$--Pb \cite{CMS:2015nfb} and Pb--Pb \cite{ALICE:2022xir} collisions at $\sqrts{\,=\,}5.02$\,GeV at the LHC \cite{Nijs:2022rme}, this suggests that a successful description of the experimental data used in our model calibration requires an initial density profile (\ref{eq:edens}) for the energy density deposited near mid-rapidity that fluctuates on a larger length scale than the strong interaction radius of the proton. The \VAH{} widget \cite{vah_tool} (see footnote \ref{widget}) shows that bumpier initial conditions whose density fluctuates on shorter length scales increase the radial and anisotropic flows (reflected in the average momenta $\langle p_T\rangle$ and harmonic flow coefficients $v_n$ of the emitted hadrons) which must be compensated by larger values for the shear and bulk viscosities. Thus, the choice of $w$ has direct consequences for the viscosity coefficients inferred from the experimental data.

The fluctuation length scale of the initially deposited matter should be larger than that of the nuclear thickness functions of the colliding nuclei makes immediate sense once one realizes that the matter created by a collision of two colored partons at transverse position $\mathbf{x}_\perp$, which is characterized by a typical transverse momentum scale $p_\perp{\,\lesssim\,}1$\,GeV, cannot be deposited at precisely the same point $\mathbf{x}_\perp$: the uncertainty relation requires it to be distributed around $\mathbf{x}_\perp$ in a cloud of transverse radius $\Delta r_\perp{\,\sim\,}\mathcal{O}(1/p_\perp)$. Unfortunately, the \trento{} ansatz (\ref{eq:edens},\ref{eq:harmonic_mean}) does not
%
\begin{figure*}[!htb]
\noindent\makebox[\textwidth]{%
  \centering
  \begin{minipage}{0.5\textwidth}
    \includegraphics[width=8cm]{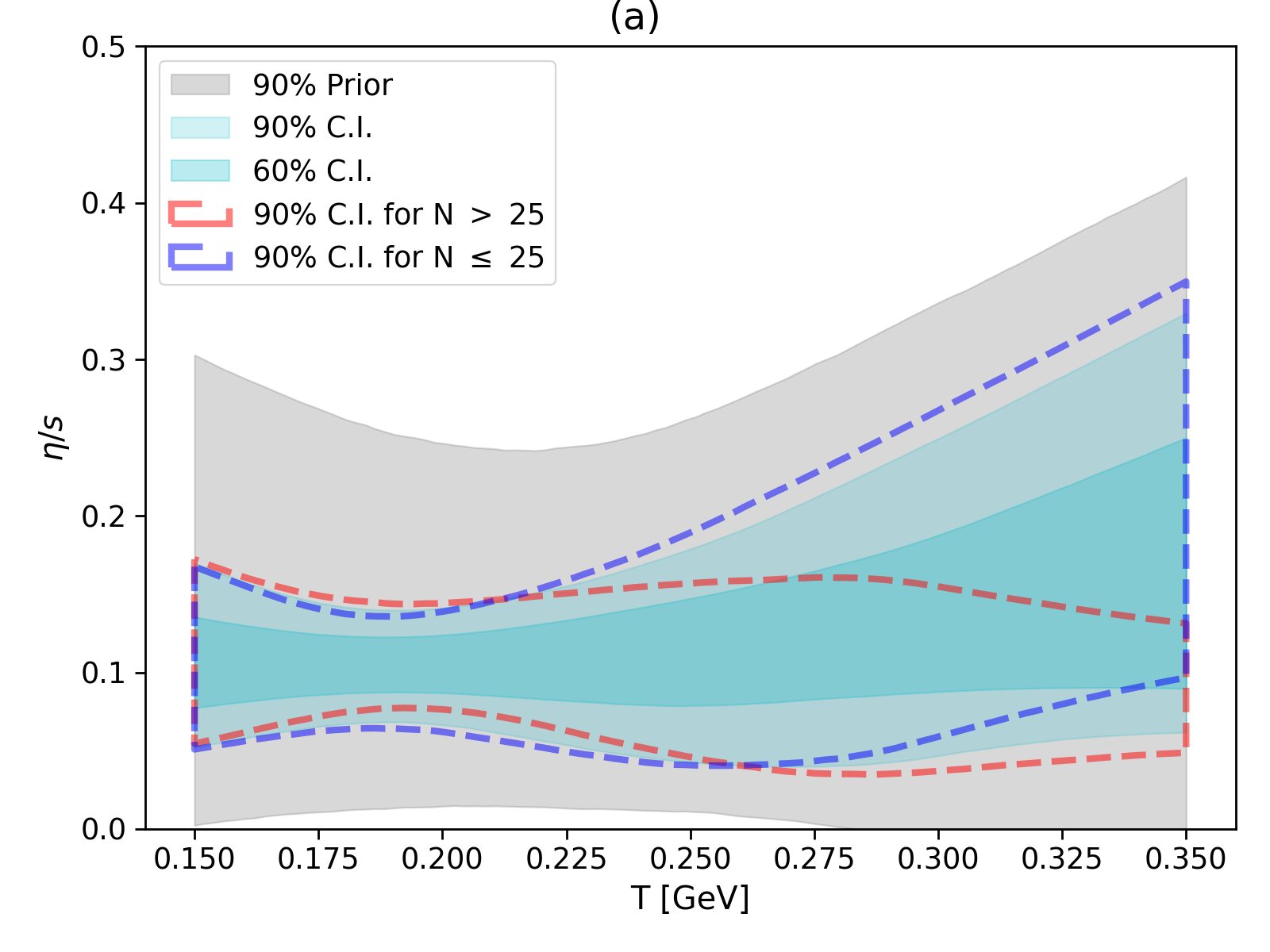}
  \end{minipage}
  \begin{minipage}{0.5\textwidth}
    \includegraphics[width=8cm]{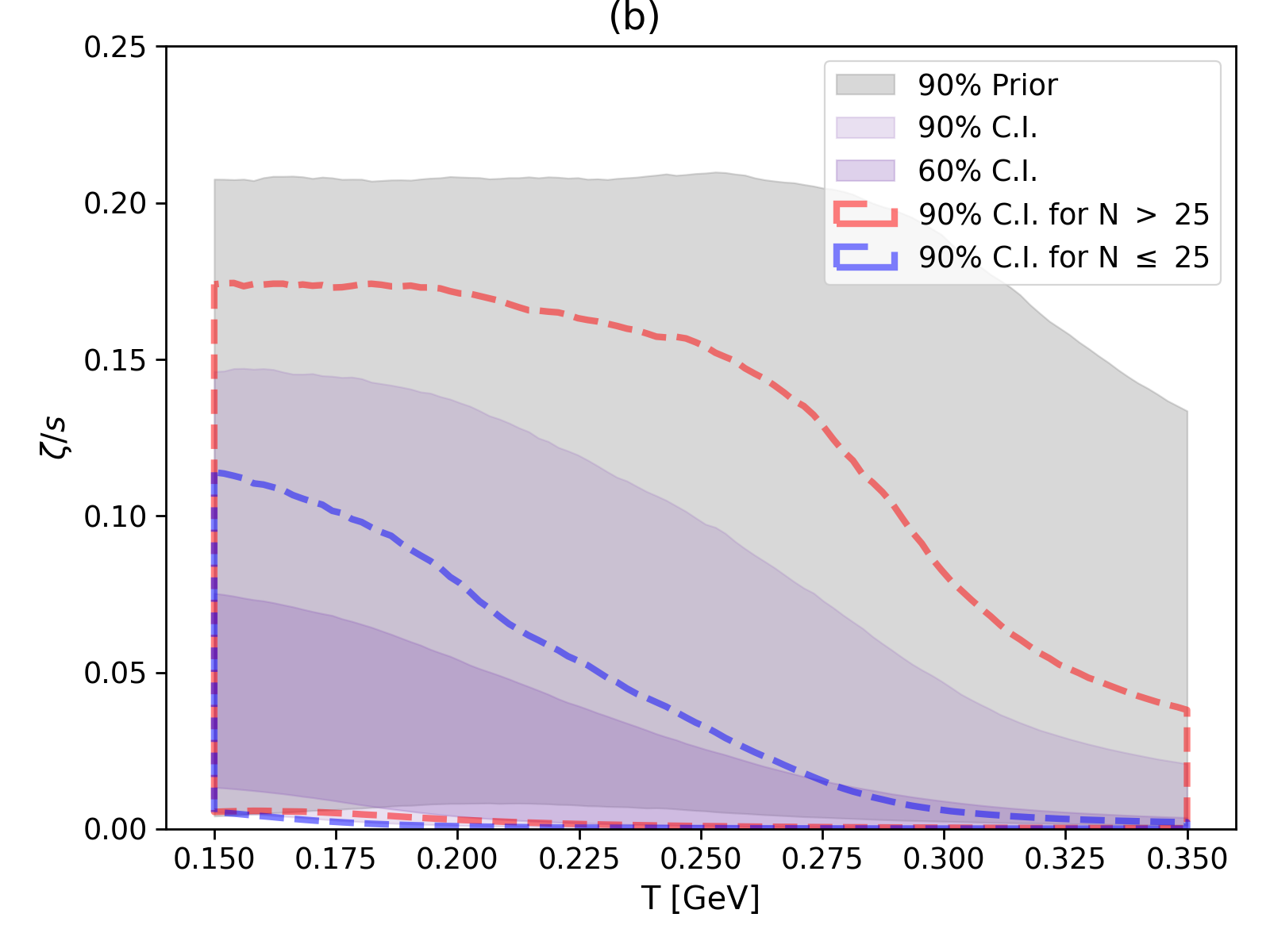}
  \end{minipage}
  }
  \caption{
  Posteriors for the temperature-dependent specific (a) shear and (b) bulk viscosities in the \VAH{} model with PTMA viscous corrections at particlization, using experimental data from Pb--Pb collisions at $\sqrts{}=2.76$\,TeV for model calibration. Grey bands show the 90\% prior interval. The colored regions corresponds to 90\% (light) and 60\% (dark) posterior credible intervals. Dashed lines indicate the 90\% posterior credible intervals within the parameter subspaces corresponding to \trento{} normalization $N>25$ (red) and $N\leq25$ (blue), respectively. See text for discussion.
  }
  \label{fig:posterior_viscosities}
\end{figure*}%
%
allow to change the width parameter $w$ independently in the nuclear thickness functions $T_{A,B}$ (which control, among other observables, the total inelastic $p$--Pb and Pb--Pb cross sections \cite{CMS:2015nfb, ALICE:2022xir, Nijs:2022rme}) and in the reduced thickness function $T_R$ (\ref{eq:harmonic_mean}) which controls the fluctuation length scale of the initially deposited transverse density profile. We suggest that future implementations of the \trento{} model should allow for an additional Gaussian smearing in Eq.~(\ref{eq:edens}),
\be
\label{eq:edens_new}
  \epsilon(\xperp)=\frac{N}{\tau_0} \int d^2r_\perp T_R(\mathbf{r}_\perp;p,w)\, \frac{e^{-\frac{1}{2} \bigl(\frac{\xperp{-}\mathbf{r}_\perp}{\Delta r_\perp}\bigr)^2}}{2\pi(\Delta r_\perp)^2},
\ee
where $T_R$ is characterized by a nucleon width $w$ matched to reproduce the total inelastic $p$--Pb and Pb--Pb cross sections \cite{Nijs:2022rme} while the additional smearing scale $\Delta r_\perp$, setting the length scale of the initial energy density fluctuations, is taken as a model parameter to be inferred from the heavy-ion collision data by Bayesian inference.\footnote{%
    As suggested by Weiyao Ke (private communication) the above uncertainty argument $\Delta r_\perp{\,\sim\,}\mathcal{O}(1/p_\perp)$ implies that the smearing radius $\Delta r_\perp$ could be smaller for the production of hard particles than for the soft matter considered here which makes up the QGP medium.}
    
Our Bayesian model calibration leaves the minimum distance $d_{\mathrm{min}}$ between nucleons unconstrained. This has also been observed in other Bayesian parameter estimation studies done with the \trento{} model \cite{JETSCAPE:2020mzn}. The switching temperature $T_{\mathrm{sw}}$ where the QGP is converted into hadrons is constrained to a somewhat lower temperature than in previous studies \cite{JETSCAPE:2020mzn, Nijs:2020ors}. While its MAP value $T_{\mathrm{sw}}^\mathrm{MAP}=146$\,MeV is compatible with earlier analyses, its marginal posterior is tilted toward lower temperatures and abuts the lower edge of its prior interval. This shift toward lower particlization temperatures suggests that the very tight constraints for $T_\mathrm{sw}$ observed in earlier Bayesian analyses using different evolution models are affected by significant model uncertainties, and that in future Bayesian parameter studies with the \VAH{} model the prior for this parameter should be extended toward lower temperature values.  

\subsection{Posteriors for the temperature-dependent specific shear and bulk viscosities}
\label{sec6C}

The posterior probability distributions for the temperature-dependent specific shear and bulk viscosities of the QGP, inferred from experimental data for Pb--Pb collisions at $\sqrts{}=2.76$\,TeV, are shown in Fig.~\ref{fig:posterior_viscosities}. 

Comparing with two other recent Bayesian parameter studies \cite{JETSCAPE:2020mzn, Nijs:2020ors} we observe improved posterior constraints especially in the upper temperature range. The specific bulk viscosity is constrained to very low values ($\zeta/s{\,<\,}0.01$ with 60\% confidence) at temperatures above 350\,MeV; at temperatures below 220\,MeV the constraints are similar to those in Ref.~\cite{JETSCAPE:2020mzn}. For the specific shear viscosity $\eta/s$ we find posterior constraints that again are consistent with previous Bayesian parameter inference studies at temperatures below 250 MeV, but are much tighter and weighted toward lower $\eta/s$ values at higher temperatures when using \VAH{} than found with the earlier models that assumed a free-streaming pre-hydrodynamic stage.

The better constraints at higher temperatures make sense when noting that in our \VAH{} model the specific bulk and shear viscosities enter as model parameters into the description of the dynamical evolution at much earlier times, when the energy density (which controls the associated equilibrium temperature by Landau matching) is much higher. In the JETSCAPE \cite{JETSCAPE:2020mzn} and {\sl Trajectum} \cite{Nijs:2020ors} models $\eta/s$ and $\zeta/s$ do not enter until the beginning of the viscous hydrodynamic stage after about 1\,fm/$c$ or later, when the energy density has already dropped by a factor 20 or more, corresponding to a decrease by more than factor 2 in temperature. The initial free-streaming stage assumed in Refs.~\cite{JETSCAPE:2020mzn, Nijs:2020ors} can in fact be thought of as a fluid with infinite shear and bulk viscosities. The present work shows that by replacing the unphysical free-streaming stage by viscous anisotropic hydrodynamics we can achieve a description of the experimental measurements that is at least as good as achieved in the previous model calibrations (see further discussion below) and which allows us to probe the temperature dependence of the QGP viscosities to much higher temperatures than possible before. An additional benefit of the \VAH{} approach is that it eliminates the unphysical switching time from free-streaming to hydrodynamics as a model parameter, and also avoids the large and positive (i.e., wrong-signed) artificial starting values of the bulk viscous pressure that arise from matching the conformal free-streaming stage to a hydrodynamic fluid with a realistic, non-conformal EoS.

In Section~\ref{sec6B} we found that the entropy generated by viscous heating drives the unique bimodal structure seen in the posterior shown in Fig.~\ref{fig:full_posterior}. The dashed lines in Fig.~\ref{fig:posterior_viscosities} provide additional support for our analysis: In order to identify the contribution from each of the two modes to the final posteriors for the temperature dependent viscosities, we divide the 15-dimensional parameter space into subspaces with \trento{} normalization parameter $N>25$ and $N\leq 25$. The red and blue dashed lines in Fig.~\ref{fig:posterior_viscosities} delineate the corresponding 90\% posterior credible intervals for the specific viscosities when the range of $N$ is restricted in this way to capture one or the other of these two modes.

For large initial entropy ($N{\,>\,}25$) the specific shear viscosity $\eta/s$ is seen to most likely remain small (close to $1/4\pi$) as the temperature increases, thereby suppressing additional shear viscous entropy production during the earliest and hottest expansion stage. For small initial entropy deposition ($N{\,\leq\,}25$), $\eta/s$ rises with increasing temperature, entailing significant shear viscous heating during the hottest earliest stage. Panel (b) shows that complex constraints provided by the experimental data correlate an increase of the shear viscosity $\eta/s$ at high $T$ with a concomitant decrease of the bulk viscosity $\zeta/s$, and vice versa. The model sensitivity analysis in the next section reveals an overall rather weak sensitivity of the observables used for model calibration to the bulk viscosity $\zeta/s$: only the proton yield and the pion and proton $\langle p_T\rangle$ respond significantly to changes in $\zeta/s$, but at the same time they respond much more strongly to several other model parameters. The anti-correlation between $\eta/s$ and $\zeta/s$ at high temperature seen in Fig.~\ref{fig:posterior_viscosities} when comparing the regions encircled by the dashed red and blue lines is therefore likely caused by a combination of several small effects that requires 
retuning multiple parameters simultaneously in the \VAH{} widget \cite{vah_tool}.

\section{Model sensitivity}
\label{sec7}
\vspace*{-2mm}

\begin{figure}[t]
\includegraphics[width=\linewidth]{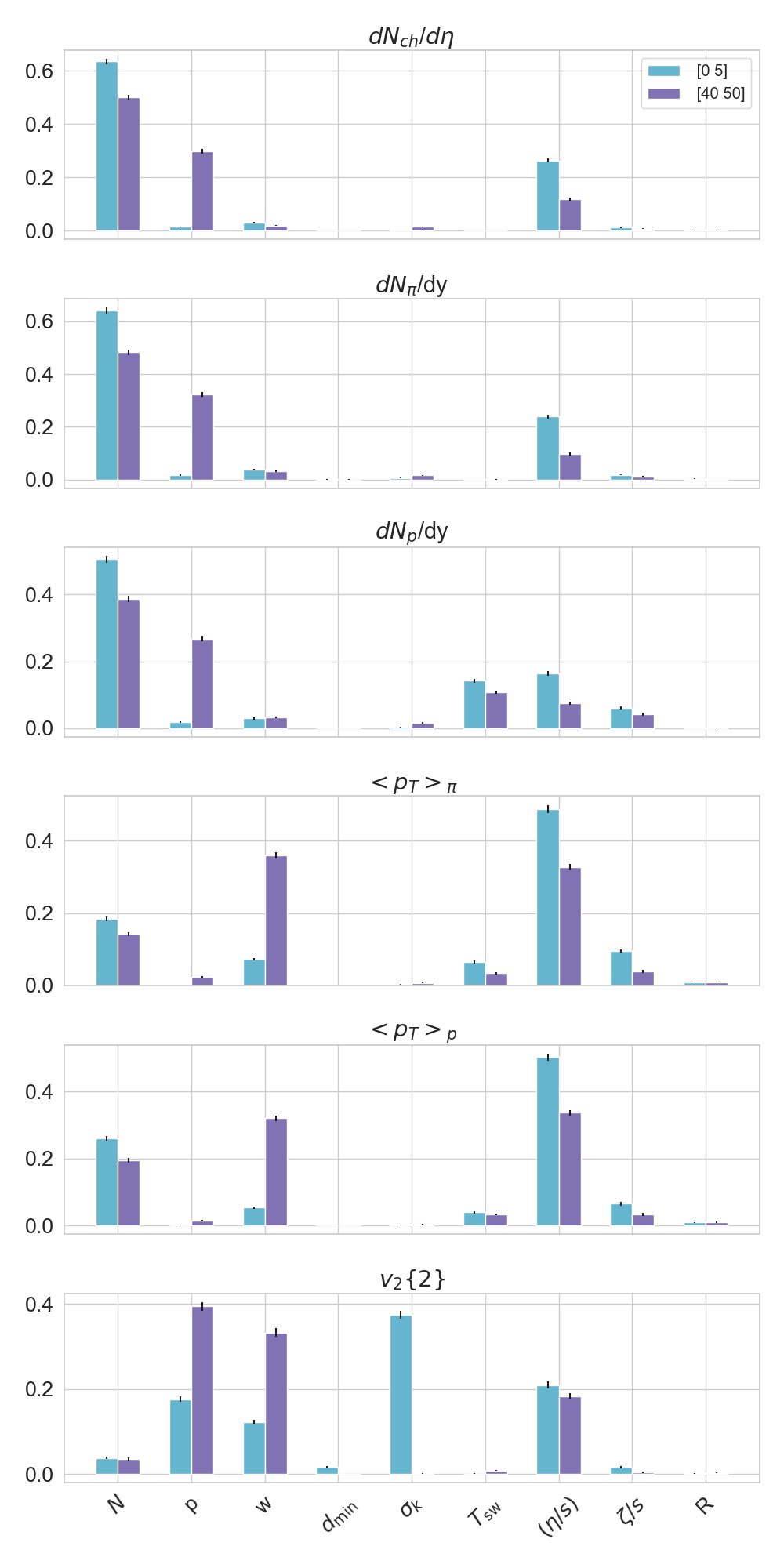}
\caption{First-order Sobol' sensitivities (App.~\ref{app:Sobol}) for the \VAH{} model.}
\label{fig:sensitivity}
\end{figure}

In this section, we perform a sensitivity analysis on our model emulators to understand how the \VAH{} model observables respond to changes in the model parameters. The first-order Sobol' sensitivity analysis performed here measures the global sensitivity of the model to its parameters \cite{sobol1990sensitivity, dan_tl}. Here we have grouped the parameters related to shear and bulk viscosity into two separate groups and measure the overall sensitivity of the model to these grouped sets for the ease of presentation \cite{jacques2006sensitivity}; see App.~\ref{app:Sobol} for details.

Figure~\ref{fig:sensitivity} shows the first-order Sobol' sensitivity indices calculated for 6 types of observables for the \VAH{} model. The blue color is for observables measured in the most central collisions (0-5\% centrality) while the purple color represents observables in the mid-centrality range (40-50\% centrality). 

We observe that charged and identified particle yields for pions and protons are mostly sensitive to the \trento{} normalization, $N$. The normalization directly scales the magnitude of the initial energy deposition profile from \trento{} and thus controls the number of particles produced at freeze out. Interestingly, we find that the charged particle and pion yields display their second strongest sensitivity to the grouped specific shear viscosity $\eta/s$. Further exploration reveals that this sensitivity is caused by viscous heating effects during the fireball evolution, especially at very early times. We invite the interested reader to further verify this by using the \VAH{} widget link in footnote~\ref{widget}. In the widget, reducing the high temperature slope of the specific shear viscosity, keeping everything else fixed, results in a dramatic decrease of the charged particle and pion yields, showing the effect of viscous heating. This strong sensitivity is a specific feature of \VAH{} which starts the (anisotropic) hydrodynamic evolution very early (from 0.05\,fm/$c$) when the temperature is high, providing a strong lever arm for the slope parameter $a_\mathrm{high}$. 

The same observables measured in peripheral collisions, but not in central collisions, are sensitive to the \trento{} harmonic mean parameter $p$, which controls how the nucleon thickness functions for each of the nucleus are combined to produce the initial energy profile.  

The mean transverse momenta of both pions and protons are most sensitive to the grouped specific shear viscosity parameters, with somewhat weaker sensitivity to the normalization parameter $N$. The same observables also have high sensitivity to the nucleon width $w$ measured in peripheral collisions but not in central collisions.

The elliptic flow observables measured in the most central collisions exhibit the strongest sensitivity to the multiplicity fluctuation parameter $\sigma_k$. The average shape of the nuclear overlap region being perfectly azimuthally symmetric in the most central collisions, flow anisotropies arise only from event-by-event fluctuations in these collisions, explaining their sensitivity to $\sigma_k$. The elliptic flow also shows unsurprisingly significant sensitivities to the grouped specific shear viscosity parameters and the \trento{} parameters $p$ and $w$, in both central and peripheral collisions. The latter again reflects the roles of these parameters in the event-by-event fluctuation spectrum characterizing the initial conditions. 

None of the selected observables exhibit any appreciable sensitivity to the initial pressure anisotropy parameter $R$ or the minimum distance $d_\mathrm{min}$ between nucleons when sampling their positions from the Woods-Saxon distribution. This observation is consistent with the very broad marginal posteriors for these two parameters shown in Fig.~\ref{fig:full_posterior}. These suggest that additional novel observables (to sharpen their marginal posteriors) and/or trustworthy additional theoretical arguments (to better constrain their priors) are needed to better constrain these model parameters.

In Ref.~\cite{McNelis_thesis} the author tried to keep all except two parameters in the \VAH{} model fixed at the MAP values of the recently calibrated JETSCAPE SIMS model \cite{JETSCAPE:2020mzn}, seeking \VAH{} values only for the normalization $N$ and nucleon width $w$. The resulting fit \cite{McNelis_thesis} agreed surprisingly well with the experimental data. The above sensitivity can explain this unexpected success: Fig.~\ref{fig:sensitivity} shows that $N$ and $w$ are the two parameters to which most of the selected observables exhibit significant sensitivity, so adjusting them captures most of the variation of the observables under model parameter change. For this reason we have also included the \VAH{} parameter set proposed in \cite{McNelis_thesis} as their ``best guess" as one of our design points when training emulators. 

Our trained emulators for the \VAH{} model can be accessed by following the link in footnote \ref{widget} and Ref.~\cite{vah_tool}. The ``\VAH{} widget'' produces the centrality dependent observables in real time for any model parameter values within their prior bounds. As long as only model predictions for the set of experimental data used in the \VAH{} calibration are desired, the widget can serve as a fast and quantitatively precise emulator of the full \VAH{} model. We found it very useful for developing intuition for the response of the observables to changes in single or combinations of model parameters. While no substitute for the full solution of the ``inverse problem'' (i.e., the inference of the model parameters from the observables), it can help to develop an in-depth understanding of such a solution.

\vspace*{-2mm}
\section{Predictions from the maximum a posteriori probability}
\label{sec8}
\vspace*{-2mm}

The final result of the Bayesian parameter estimation work presented here are computationally cheap samples of the experimental observables from the most probable region in the multidimensional posterior distribution for the model parameters.  To visualize and understand the full posterior distribution we have used plots of the 1- and 2-dimensional (single or joint) marginal distribution in Figs.~\ref{fig:full_posterior} and \ref{fig:posterior_viscosities}. In this section we will show how to find a point estimate from the posterior, i.e., the set of model parameters that can best fit the experimental data. The comparison between the simulation predictions for this set of model parameter values (the MAP values) and the experimental measurement will be a final test for the validity of the Bayesian parameter inference framework that we have used in this work. We note, however, that model predictions that quantitatively include the full uncertainty range arising from the uncertainties of the inferred model parameters require a posterior-weighted sample from the entire high-probability region in the multidimensional parameter space.

\begin{table}[b]
\begin{tabular}{l|l}
parameter & MAP values\\ \hline \hline
$N$ & $20.013$  \\ \hline
$p$ & $0.038$  \\ \hline
$w$ [fm] & $0.985$ \\ \hline
$d_\mathrm{min}$ [fm] & $0.878$ \\ \hline
$\sigma_k$ & $1.184$ \\ \hline
$T_\mathrm{sw}$ [GeV]& $0.146$ \\ \hline
$R$ & $0.653$ \\ \hline
$T_{\eta/s, \mathrm{kink}}$\ [GeV] & $0.219$  \\ \hline
$(\eta/s)_\mathrm{kink}$ & $0.094$  \\ \hline
$a_{\mathrm{high}}$ [GeV$^{-1}$] & $0.493$ \\ \hline
$a_{\mathrm{low}}$ [GeV$^{-1}$] & $-0.383$ \\ \hline
$(\zeta/s)_{\mathrm{max}}$ & $0.044$ \\ \hline
$T_{\zeta}$ [GeV] & $0.235$ \\ \hline
$w_{\zeta}$ [GeV] & $0.032$ \\ \hline
$\lambda_\zeta$ & $0.037$
\end{tabular}
\caption{
Model parameters corresponding to the mode of the posterior distribution (MAP) for the \VAH{} model. These model parameters provide simulation outputs that agree best with the experimental measurements.}
\label{table:MAP}
\end{table}

\begin{figure}[!h]
\includegraphics[width=8cm]{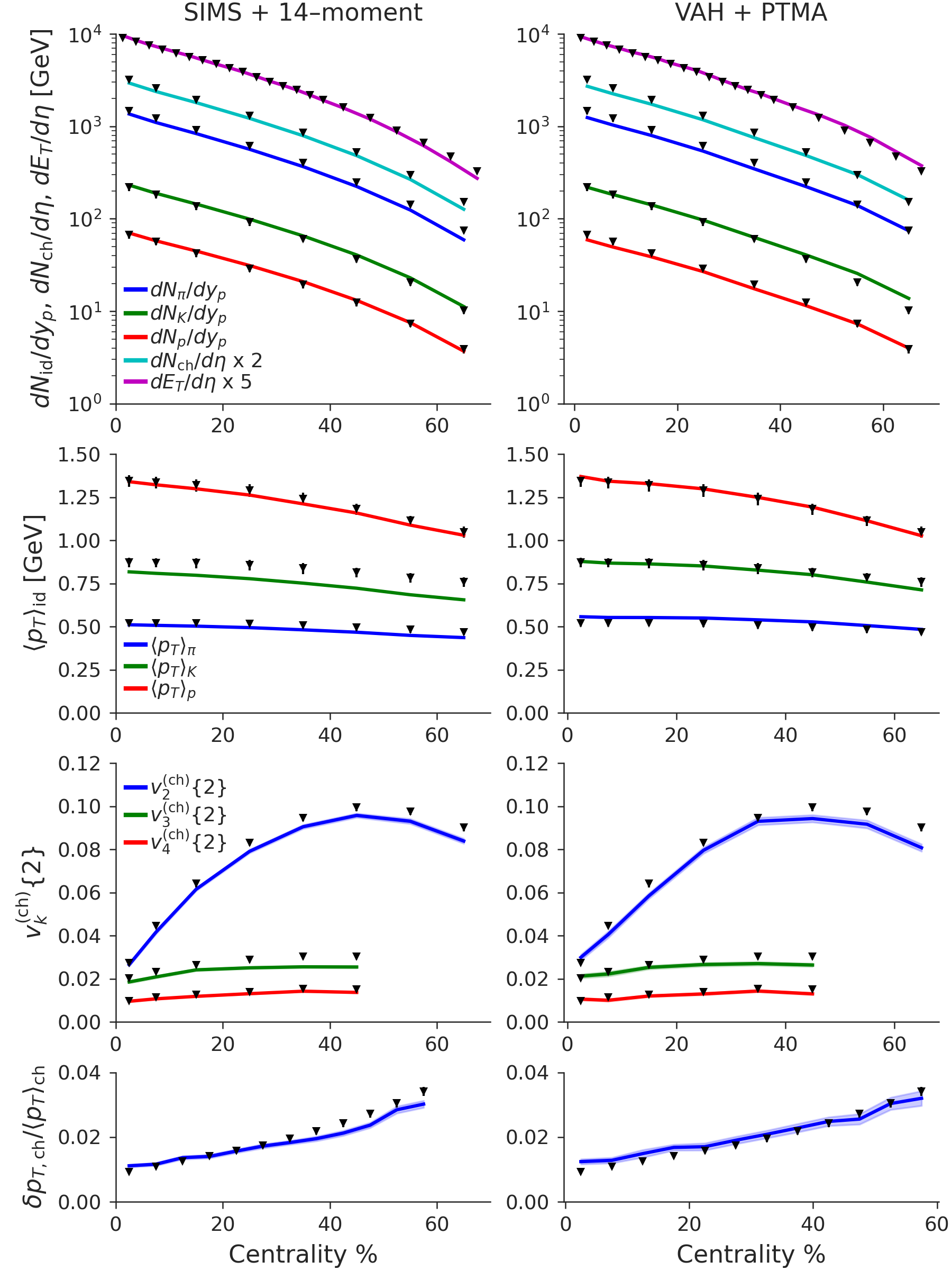}
\caption{MAP predictions for the JETSCAPE SIMS model with Grad 14-moment viscous corrections at particlization from Ref.~\cite{JETSCAPE:2020mzn} (left column) compared with those for the \VAH{} model with PTMA viscous corrections at particlization (right column).}
\label{fig:MAP}
\end{figure}

The model parameters that correspond to the mode of the posterior distribution are called maximum a posterior probability (MAP) values. The MAP values are found using numerical optimization algorithms that find the model parameter values that minimize the negative log posterior distribution \cite{diff_evolution, 2020SciPy-NMeth}. The MAP values for the posterior obtained in this work are listed in Table~\ref{table:MAP}.

In Fig.~\ref{fig:MAP} we compare how the MAP prediction from the \VAH{} model compares with another recent Bayesian parameter estimation study published in Ref.~\cite{JETSCAPE:2020mzn}. The left column of Fig.~\ref{fig:MAP} is obtained by using the JETSCAPE SIMS model \cite{JETSCAPE:2020mzn} with its MAP parameter values and running 3000 events with fluctuating initial conditions. The right column is obtained by the same procedure using the \VAH{} model with its MAP parameter set. In both columns the black triangles show the experimental measurements taken with Pb--Pb collisions at the LHC at $\sqrts{}=2.76$\,TeV. While both fits look good, closer inspection reveals several detailed features in the data that are better described by \VAH{} and none that are significantly better described by JETSCAPE SIMS. It is likely, however, that the two sets of model predictions are statistically consistent with each other\footnote{%
    Perhaps with the exception of the mean $\langle p_T\rangle$ for kaons.}
once a full sample of the posterior probability distribution is generated. We leave this for a future study aimed at a quantitative comparison between the two models and improved observable predictions obtained by combining the models (and variations of them) with Bayesian model mixing techniques \cite{Phillips:2020dmw}.
 
It is important to note that the Bayesian parameter estimation carried out here did not use the experimental data for the transverse momentum fluctuations of charged particles, $\delta p_T^\mathrm{ch}/\langle p_T^\mathrm{ch}\rangle$, in the model calibration, nor any model predictions for this observable when training the model emulators. The fact that the MAP values for the \VAH{} model successfully reproduce this observable (bottom right panel in Fig.~\ref{fig:MAP}) can be interpreted as a successful {\it pre}diction of the calibrated \VAH{} model.

\vspace*{-2mm}
\section{Conclusions}
\label{sec9}
\vspace*{-2mm}

In this work we performed a Bayesian calibration of a novel relativistic heavy-ion collision model, \VAH{}, which treats the early far-from-equilibrium stage of the collision with viscous anisotropic hydrodynamics that is optimized for the particular symmetries that characterize this stage. The experimental input in this study were data taken with Pb--Pb collisions at $\sqrts{}=2.76$\,TeV at the LHC. The \VAH{} model can be used from very early times onward (here we use $\tau_0{\,=\,}0.05$\,fm/$c$), which largely obviates the need for any pre-hydrodynamic evolution at all. This eliminates model parameters and other conceptual uncertainties associated with the free-streaming modeling of the pre-hydrodynamic stage employed in similar analyses that employed standard viscous hydrodynamics to model the QGP.

An important advantage of being able to start the hydrodynamic modeling so much earlier is that transport coefficients enter as parameters of the dynamical description at a stage of much higher energy density and temperature. By using the \VAH{} approach we can therefore constrain the specific shear and bulk viscosities at higher temperatures than accessible in other approaches that invoke an extended pre-hydrodynamic stage modeled microscopically in a way that cannot be meaningfully parameterized by hydrodynamic transport coefficients.  We find that within the \VAH{} approach the specific bulk viscosity is well constrained by the LHC data to very low values, $\zeta/s{\,<\,}0.03$ at 90\% confidence level and ${\,<\,}0.01$ at 60\% confidence level, for temperatures above 350\,MeV. Also the specific shear viscosity is more tightly constrained at temperatures above 250\,MeV than in previous analyses, with the 60\% and 90\% confidence intervals both pushed toward lower $\eta/s$ values than before. At lower temperatures below 220 MeV the \VAH{} constraints agree with those obtained from previous Bayesian parameter estimation studies.

The joint marginal posterior distributions for pairs of model parameters exposed a bimodal posterior which, in the case of the \trento{} normalization $N$, is even visible at the level of the individual marginal posterior. We were able to trace this back to a nontrivial interplay between the {\it initial} entropy deposition controlled by $N$ and the {\it additional} entropy generated by viscous heating during the subsequent evolution. The Bayesian calibration found two different solutions with similar posterior probabilities, corresponding to different trade-offs between these two mechanisms. The use of \VAH{} exposed the possibility of such a trade-off for the first time since previous hybrid evolution models were not able to explore reliably the large viscous heating effects occurring during the very early evolution stage when the longitudinal expansion rate is huge and the fluid is very far from local thermal equilibrium. \VAH{} is optimized to handle this extreme situation within an adapted hydrodynamic approach. We argued that there may be additional theoretical arguments that, in future Bayesian calibration campaigns, can be implemented by appropriate modification of the prior probability distributions for some of the model parameters, in order to resolve the ambiguity between the two modes exposed in this work.

We demonstrated that the primary mode of the full 15-dimensional posterior parameter probability distribution (i.e., its MAP parameter set) leads to a very good description of the data used in the model calibration, with no significant tensions between the calibrated model and the experimental data. The calibrated \VAH{} model was shown to improve on several features where earlier Bayesian model fits exhibited weaknesses. In particular, the MAP parameters of our calibrated \VAH{} model led to a quite successful prediction of the mean-$p_T$ fluctuations, a set of data that was not used in the model calibration on account of its high demand on event statistics but was still correctly predicted by the model after calibration with other, statistically less demanding experimental observables.

The comparison of our results here with those obtained earlier with different variants of the heavy-ion collision evolution model shines a light on model uncertainties and their effects on the uncertainty ranges for the fireball parameters inferred with these models from the experimental data. While state-of-the-art Bayesian analyses provide principled uncertainty quantification given the known experimental uncertainties, there is nothing principled about the way the heavy-ion community presently tries to assess theoretical and modeling uncertainties. The availability of several Bayesian calibrated heavy-ion evolution models now opens the window to, and underlines the urgency for developing new statistical tools that enable principled uncertainty quantification in Bayesian model parameter inference that accounts for experimental and theoretical uncertainties on an equal footing.

\section*{Acknowledgments}

We thank Derek Everett and the JETSCAPE Collaboration for providing the relativistic heavy-ion collision simulation and Bayesian calibration framework which we adapted for this work. The authors express particular gratitude to Mike McNelis for developing and providing the hydrodynamic simulation code for the \VAH{} model and also for continuous support while running the simulations. In addition,
we are grateful for useful discussions with JF. Paquet and M. Heffernan. U.H. acknowledges stimulating discussions with participants of the INT Program INT-23-1a and thanks the Institute for Nuclear Theory for its hospitality. Generous computing resources provided on Bebop, a high-performance computing cluster operated by the Laboratory Computing Resource Center at Argonne National Laboratory, are gratefully acknowledged. This work was supported by the NSF CSSI program under grant \rm{OAC-2004601} (BAND Collaboration). D.L.\ and U.H.\ acknowledge additional partial support within the framework of the JETSCAPE Collaboration under NSF Award No.~\rm{ACI-1550223}, as well as by the DOE Office of Science, Office for Nuclear Physics under Award No.~\rm{DE-SC0004286}. S.M.W\ acknowledges additional support by the U.S.\ Department of
Energy, Office of Science, Office of Advanced Scientific Computing
Research SciDAC program under Contract Nos.\ \rm{DE-AC02-05CH11231} and \rm{DE-AC02-06CH11357}.
\vfil


\appendix

\section{Details of training data collection procedure}
\label{app:datacollection}

The following are, in chronological order, the five batches of design points for which we ran full model simulations. 

\begin{enumerate}[label=\alph*)]

\item We start with a Latin Hypercube Design (LHD) with 300 design points, simulating 200 events per design point. We use this relatively low statistics of simulated events as our starting point to build an initial set of emulators and compute a preliminary posterior. 

\item We then generate another, coarser LHD with 90 design points and run 800 full-model events per design. These simulations are used as test data to check emulation accuracy. 

\item We use the Minimum Energy Design (MED) algorithm to generate a first batch of 90 MED design points at which we run another 800 events per design.

\item Based on revised emulators trained on all the simulation data above, the second batch of 90 MED design points is identified at which another 800 events per design are simulated.

\item Based on revised emulators trained on all the simulation data above, we identify a first batch of 70 MED design points for which we simulate a larger number of 1600 events per design, to increase statistical precision of the training data in the parameter region which the emulators suggest to have a high posterior probability. 

\item The last step is repeated to identify a last batch of 30 MED design points in the high-posterior region for which we again simulate 1600 events per design. A final set of high-precision emulators is trained using all of the simulation data accumulated.

\end{enumerate}

When training and testing emulators we first discard design parameter sets for which more than 5\% of the simulations fail to complete. This ensures that the emulators are not contaminated by large errors in parameter regions where the model starts to break down (i.e., it becomes inapplicable).

\section{Additional emulation strategies}
\label{app:additionalemu}
\subsection{The Principal Component Gaussian Process Regression (PCGPR) emulation method}
\label{sec3d}

PCGPR has been the standard emulation method used in all previous Bayesian parameter studies for relativistic heavy-ion collisions \cite{Petersen:2010zt, Novak:2013bqa, Sangaline:2015isa, Bernhard:2015hxa, Bernhard:2016tnd, Moreland:2018gsh, Bernhard:2019bmu, JETSCAPE:2020shq, Nijs:2020ors, Nijs:2020roc,  JETSCAPE:2020mzn, Heffernan_CGC}. The PCGPR method emulates each of the principal components using independent Gaussian processes. In our study we employed the anisotropic Gaussian covariance function, widely used for computer experiment emulators \cite{santner2003design}. The PCGPR method differs from the PCSK method used in this study mainly due to the fact that PCSK takes into account the intrinsic uncertainty associated with stochastic simulation models. As discussed in the following we find the PCSK method is better suited over PCGPR to build emulators with simulation data of varying accuracies.

\subsection{Principal Component Gaussian Process Regression with Grouping (PCGPRG) emulation method}
\label{sec3e}

PCGPRG is an extension of the PCGPR emulation method, obtained by modifying it as follows: we first separate the observables into two groups, knowing that some of the observables are more prone to noise and might behave very differently when emulated compared to the others. In this study we separate the anisotropic flow observables $\bigl(v^{(\mathrm{ch})}_2\{2\},\  v^{(\mathrm{ch})}_3\{2\},\   v^{(\mathrm{ch})}_4\{2\}\bigr)$ from the rest, on account of the fact that they require many more events for a precise measurement and thus tend to be more noisy when computed from the simulated events. Then, we perform PCGPR emulation for each subgroup separately, i.e., the emulators for each of the groups of observables are trained only with full-model output for the observables in the group. We do so while keeping the total number of principal components at 12, for fair comparison with other methods. The PCGPR emulators for the flow observables produce 8 principal components which cumulatively explain a fraction of 0.965 of their total variance. The PCGPR emulators for the collection of all other observables produce 4 principal components which cumulatively explain a fraction of 0.98 of their total variance. The difference in the number of principal components needed confirms the assumption that the flow observables are more complex and more difficult to model accurately than the rest of the observables. 

\begin{figure}[hb]
\includegraphics[width=\linewidth]{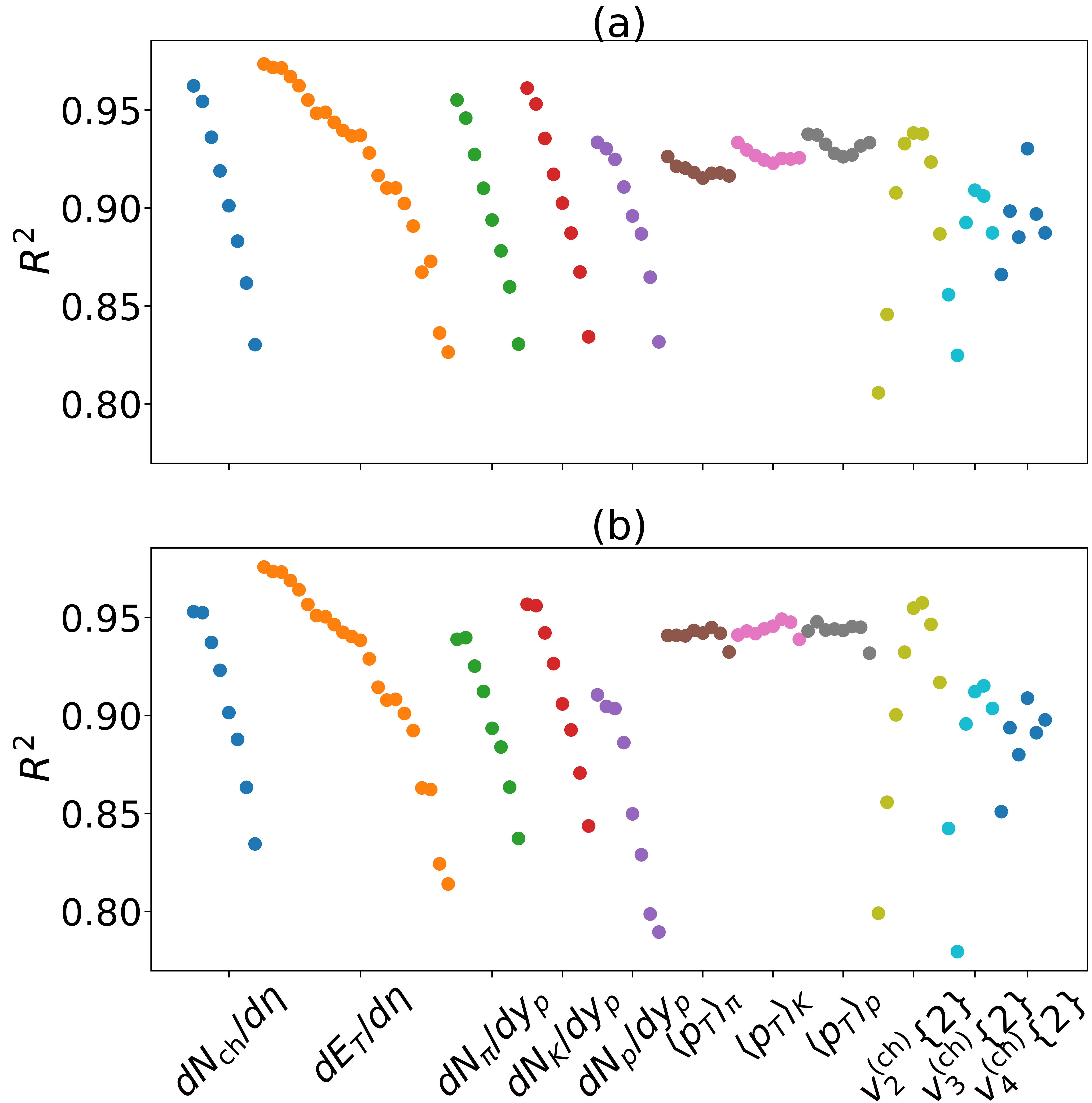}
\caption{$\mathrm{R}^2$ scores calculated using the test simulations and emulator predictions for two different emulation methods: (a) PCGPR, (b) PCGPRG.
}
\label{R2_pcgpr_groups}
\end{figure}

Fig.~\ref{R2_pcgpr_groups} shows the $R^2$ scores for the PCGPR and PCGPRG emulators for the VAH{} model. They are generically smaller than those for the PCSK emulators. In particular one sees that in our situation grouping of the observables does not improve the performance of the emulators -- the disadvantages of using only a fraction of the simulated data for emulator training in each group obviously outweigh the advantages of exploiting the reduced noisiness of the non-flow data when training their emulators.

\section{Gaussian process fitting procedure}
\label{gpfitting}

There exist several choices for the covariance function; here we parameterize it using a Mat\'ern kernel with hyper-parameters $\bm{\theta}$ \cite{Rasmussen2004}. There are many ways to infer the hyper-parameters of the covariance function $k_t(\cdot)$ for latent output $t$. One common way is to maximize the likelihood due to $\bm{z}_t \sim \text{MVN}(\mathbf{0}, \mathbf{K}_t)$. In this study, since the maximum likelihood estimation is computationally efficient, the hyper-parameters are estimated by maximizing the log likelihood 
\begin{align}
\begin{split}
    -\frac{1}{2} (\bm{z}_t^\top \mathbf{K}_t^{-1} \bm{z}_t)- \frac{1}{2} \log(|\mathbf{K}_t|)
    - \frac{n}{2} \log(2\pi)
    \label{eq:log_like}
\end{split}
\end{align}
and then plugged into the predictive Eqs.~\eqref{eq:gpeqns} for emulation (see \cite{santner2003design} for further details on plug-in predictors). 
 
\section{First-order Sobol' sensitivity indices}
\label{app:Sobol}

Here we briefly summarize how to calculate the first-order Sobol' index for a single model output $Y$ of a model that can have multiple outputs (observables). A more detailed description of this approach, as it applies to relativistic heavy-ion collision models, can be found in Ref.~\cite{dan_tl}. The first-order Sobol' index for the sensitivity of observable $Y$ to model parameter $x_j$ is defined as
\begin{equation}
    \frac{\text{Var}_{X_j}(\mathbb{E}_{\bm{X}_{-j}}(Y(\bm{X})|X_j))}
               {\text{Var}_{\bm{X}}(Y(\bm{X}))},\qquad j=1,\dots,q.
               \label{eq:sobind}
\end{equation}
Here, $X_j$ is an independent uniform random variable for parameter $x_j$ over its parameter range, and $\bm{X} = ({X}_1, \cdots, {X}_q)$ is its corresponding random vector for all parameters. The term $\mathbb{E}_{\bm{X}_{-j}}(Y(\bm{X})|x_j)$ is called the \textit{main effect} of parameter $x_j$: given fixed $j$-th parameter $X_j = x_j$, it \textit{averages} the model output $Y$ uniformly over the remaining parameters $\bm{X}_{-j} = \bm{X} \setminus X_j$. It is formally defined as
\begin{eqnarray}
\label{eq:me}
    &&\mathbb{E}_{\bm{X}_{-j}}(Y(\bm{X})|X_j) = 
\\\nonumber
    && \quad \int_{\mathcal{X}_{-j}} \delta(x_1,\dots,x_q) \; dU(x_1,\dots,x_{j-1},x_{j+1},\dots,x_q),
\end{eqnarray}
where $U(x_1,\dots,x_{j-1},x_{j+1},\dots,x_q)$ is the uniform probability measure over $\mathcal{X}_{-j}$, the parameter space $\mathcal{X}$ omitting the $j$-th parameter. The first-order Sobol' index \eqref{eq:sobind} thus quantifies the importance of parameter $x_j$, by taking the ratio of $\text{Var}_{X_j}(\mathbb{E}_{\bm{X}_{-j}}(Y(\bm{X})|X_j))$, the variance accounted for by the \textit{main effects} $\mathbb{E}_{\bm{X}_{-j}}(Y(\bm{X})|X_j)$, over $\text{Var}_{\bm{X}}(Y)$, the \textit{total} variance of model output $Y$ over all parameters. 

One can further modify the Sobol' indices in \eqref{eq:sobind} by grouping together similar model input parameters. The grouped Sobol' indices in \cite{jacques2006sensitivity} accomplish this. The $q$ input parameters $\mathbf{X} = (X_1,\dots,X_q)$ (assumed again to be uniformly distributed) are first divided into $J$ groups $(\mathbb{X}_1,\dots,\mathbb{X}_J)$, given by
\be
\nonumber
(X_1,\dots,X_q) = (\underbrace{X_1,\dots,X_{k_1}}_{\mathbb{X}_1},\dots,\underbrace{X_{k_{J-1}+1},\dots,X_q}_{\mathbb{X}_J}).
\ee
The first-order \textit{grouped} Sobol' indices can then be defined as:
\begin{align}
\begin{split}
    S_j = \frac{\text{Var}_{\mathbb{X}_j}(\mathbb{E}_{\mathbb{X}_{-j}}(Y|\mathbb{X}_j))}
               {\text{Var}_{\mathbf{X}}(Y(X))},
    \quad j=1,\dots,J,
\end{split}
\end{align}
where $\mathbb{X}_{-j} = \mathbf{X} \setminus \mathbb{X}_j$ consists of all parameters except for those in group $j$.\vfil

\bibliography{biblio}

\begin{thebibliography}{110}%
\makeatletter
\providecommand \@ifxundefined [1]{%
 \@ifx{#1\undefined}
}%
\providecommand \@ifnum [1]{%
 \ifnum #1\expandafter \@firstoftwo
 \else \expandafter \@secondoftwo
 \fi
}%
\providecommand \@ifx [1]{%
 \ifx #1\expandafter \@firstoftwo
 \else \expandafter \@secondoftwo
 \fi
}%
\providecommand \natexlab [1]{#1}%
\providecommand \enquote  [1]{``#1''}%
\providecommand \bibnamefont  [1]{#1}%
\providecommand \bibfnamefont [1]{#1}%
\providecommand \citenamefont [1]{#1}%
\providecommand \href@noop [0]{\@secondoftwo}%
\providecommand \href [0]{\begingroup \@sanitize@url \@href}%
\providecommand \@href[1]{\@@startlink{#1}\@@href}%
\providecommand \@@href[1]{\endgroup#1\@@endlink}%
\providecommand \@sanitize@url [0]{\catcode `\\12\catcode `\$12\catcode
  `\&12\catcode `\#12\catcode `\^12\catcode `\_12\catcode `\%12\relax}%
\providecommand \@@startlink[1]{}%
\providecommand \@@endlink[0]{}%
\providecommand \url  [0]{\begingroup\@sanitize@url \@url }%
\providecommand \@url [1]{\endgroup\@href {#1}{\urlprefix }}%
\providecommand \urlprefix  [0]{URL }%
\providecommand \Eprint [0]{\href }%
\providecommand \doibase [0]{http://dx.doi.org/}%
\providecommand \selectlanguage [0]{\@gobble}%
\providecommand \bibinfo  [0]{\@secondoftwo}%
\providecommand \bibfield  [0]{\@secondoftwo}%
\providecommand \translation [1]{[#1]}%
\providecommand \BibitemOpen [0]{}%
\providecommand \bibitemStop [0]{}%
\providecommand \bibitemNoStop [0]{.\EOS\space}%
\providecommand \EOS [0]{\spacefactor3000\relax}%
\providecommand \BibitemShut  [1]{\csname bibitem#1\endcsname}%
\let\auto@bib@innerbib\@empty
\bibitem [{\citenamefont {Linde}(1979)}]{Linde_1979}%
  \BibitemOpen
  \bibfield  {author} {\bibinfo {author} {\bibfnamefont {A.~D.}\ \bibnamefont
  {Linde}},\ }\href {\doibase 10.1088/0034-4885/42/3/001} {\bibfield  {journal}
  {\bibinfo  {journal} {Reports on Progress in Physics}\ }\textbf {\bibinfo
  {volume} {42}},\ \bibinfo {pages} {389} (\bibinfo {year} {1979})}\BibitemShut
  {NoStop}%
\bibitem [{\citenamefont {Collins}\ and\ \citenamefont
  {Perry}(1975)}]{Collins1975}%
  \BibitemOpen
  \bibfield  {author} {\bibinfo {author} {\bibfnamefont {J.~C.}\ \bibnamefont
  {Collins}}\ and\ \bibinfo {author} {\bibfnamefont {M.~J.}\ \bibnamefont
  {Perry}},\ }\href {\doibase 10.1103/PhysRevLett.34.1353} {\bibfield
  {journal} {\bibinfo  {journal} {Phys. Rev. Lett.}\ }\textbf {\bibinfo
  {volume} {34}},\ \bibinfo {pages} {1353} (\bibinfo {year}
  {1975})}\BibitemShut {NoStop}%
\bibitem [{\citenamefont {Gyulassy}\ and\ \citenamefont
  {McLerran}(2005)}]{Gyulassy:2004zy}%
  \BibitemOpen
  \bibfield  {author} {\bibinfo {author} {\bibfnamefont {M.}~\bibnamefont
  {Gyulassy}}\ and\ \bibinfo {author} {\bibfnamefont {L.}~\bibnamefont
  {McLerran}},\ }\href {\doibase 10.1016/j.nuclphysa.2004.10.034} {\bibfield
  {journal} {\bibinfo  {journal} {Nucl. Phys. A}\ }\textbf {\bibinfo {volume}
  {750}},\ \bibinfo {pages} {30} (\bibinfo {year} {2005})},\ \Eprint
  {http://arxiv.org/abs/nucl-th/0405013} {arXiv:nucl-th/0405013} \BibitemShut
  {NoStop}%
\bibitem [{\citenamefont {Yagi}\ \emph {et~al.}(2005)\citenamefont {Yagi},
  \citenamefont {Hatsuda},\ and\ \citenamefont {Miake}}]{Yagi:2005yb}%
  \BibitemOpen
  \bibfield  {author} {\bibinfo {author} {\bibfnamefont {K.}~\bibnamefont
  {Yagi}}, \bibinfo {author} {\bibfnamefont {T.}~\bibnamefont {Hatsuda}}, \
  and\ \bibinfo {author} {\bibfnamefont {Y.}~\bibnamefont {Miake}},\
  }\href@noop {} {\bibfield  {journal} {\bibinfo  {journal} {{Cambridge Monogr.
  Part. Phys. Nucl. Phys. Cosmol.}}\ }\textbf {\bibinfo {volume} {23}},\
  \bibinfo {pages} {1} (\bibinfo {year} {2005})}\BibitemShut {NoStop}%
\bibitem [{\citenamefont {Muller}\ \emph {et~al.}(2012)\citenamefont {Muller},
  \citenamefont {Schukraft},\ and\ \citenamefont {Wyslouch}}]{Muller:2012zq}%
  \BibitemOpen
  \bibfield  {author} {\bibinfo {author} {\bibfnamefont {B.}~\bibnamefont
  {Muller}}, \bibinfo {author} {\bibfnamefont {J.}~\bibnamefont {Schukraft}}, \
  and\ \bibinfo {author} {\bibfnamefont {B.}~\bibnamefont {Wyslouch}},\ }\href
  {\doibase 10.1146/annurev-nucl-102711-094910} {\bibfield  {journal} {\bibinfo
   {journal} {Ann. Rev. Nucl. Part. Sci.}\ }\textbf {\bibinfo {volume} {62}},\
  \bibinfo {pages} {361} (\bibinfo {year} {2012})},\ \Eprint
  {http://arxiv.org/abs/1202.3233} {arXiv:1202.3233 [hep-ex]} \BibitemShut
  {NoStop}%
\bibitem [{\citenamefont {Busza}\ \emph {et~al.}(2018)\citenamefont {Busza},
  \citenamefont {Rajagopal},\ and\ \citenamefont {van~der
  Schee}}]{big_picture}%
  \BibitemOpen
  \bibfield  {author} {\bibinfo {author} {\bibfnamefont {W.}~\bibnamefont
  {Busza}}, \bibinfo {author} {\bibfnamefont {K.}~\bibnamefont {Rajagopal}}, \
  and\ \bibinfo {author} {\bibfnamefont {W.}~\bibnamefont {van~der Schee}},\
  }\href {\doibase 10.1146/annurev-nucl-101917-020852} {\bibfield  {journal}
  {\bibinfo  {journal} {Annual Review of Nuclear and Particle Science}\
  }\textbf {\bibinfo {volume} {68}},\ \bibinfo {pages} {339} (\bibinfo {year}
  {2018})},\ \Eprint
  {http://arxiv.org/abs/https://doi.org/10.1146/annurev-nucl-101917-020852}
  {https://doi.org/10.1146/annurev-nucl-101917-020852} \BibitemShut {NoStop}%
\bibitem [{\citenamefont {Petersen}\ \emph {et~al.}(2011)\citenamefont
  {Petersen}, \citenamefont {Coleman-Smith}, \citenamefont {Bass},\ and\
  \citenamefont {Wolpert}}]{Petersen:2010zt}%
  \BibitemOpen
  \bibfield  {author} {\bibinfo {author} {\bibfnamefont {H.}~\bibnamefont
  {Petersen}}, \bibinfo {author} {\bibfnamefont {C.}~\bibnamefont
  {Coleman-Smith}}, \bibinfo {author} {\bibfnamefont {S.~A.}\ \bibnamefont
  {Bass}}, \ and\ \bibinfo {author} {\bibfnamefont {R.}~\bibnamefont
  {Wolpert}},\ }\href {\doibase 10.1088/0954-3899/38/4/045102} {\bibfield
  {journal} {\bibinfo  {journal} {J. Phys. G}\ }\textbf {\bibinfo {volume}
  {38}},\ \bibinfo {pages} {045102} (\bibinfo {year} {2011})},\ \Eprint
  {http://arxiv.org/abs/1012.4629} {arXiv:1012.4629 [nucl-th]} \BibitemShut
  {NoStop}%
\bibitem [{\citenamefont {Novak}\ \emph {et~al.}(2014)\citenamefont {Novak},
  \citenamefont {Novak}, \citenamefont {Pratt}, \citenamefont {Vredevoogd},
  \citenamefont {Coleman-Smith},\ and\ \citenamefont
  {Wolpert}}]{Novak:2013bqa}%
  \BibitemOpen
  \bibfield  {author} {\bibinfo {author} {\bibfnamefont {J.}~\bibnamefont
  {Novak}}, \bibinfo {author} {\bibfnamefont {K.}~\bibnamefont {Novak}},
  \bibinfo {author} {\bibfnamefont {S.}~\bibnamefont {Pratt}}, \bibinfo
  {author} {\bibfnamefont {J.}~\bibnamefont {Vredevoogd}}, \bibinfo {author}
  {\bibfnamefont {C.~E.}\ \bibnamefont {Coleman-Smith}}, \ and\ \bibinfo
  {author} {\bibfnamefont {R.~L.}\ \bibnamefont {Wolpert}},\ }\href {\doibase
  10.1103/PhysRevC.89.034917} {\bibfield  {journal} {\bibinfo  {journal} {Phys.
  Rev.}\ }\textbf {\bibinfo {volume} {C89}},\ \bibinfo {pages} {034917}
  (\bibinfo {year} {2014})},\ \Eprint {http://arxiv.org/abs/1303.5769}
  {arXiv:1303.5769 [nucl-th]} \BibitemShut {NoStop}%
\bibitem [{\citenamefont {Sangaline}\ and\ \citenamefont
  {Pratt}(2016)}]{Sangaline:2015isa}%
  \BibitemOpen
  \bibfield  {author} {\bibinfo {author} {\bibfnamefont {E.}~\bibnamefont
  {Sangaline}}\ and\ \bibinfo {author} {\bibfnamefont {S.}~\bibnamefont
  {Pratt}},\ }\href {\doibase 10.1103/PhysRevC.93.024908} {\bibfield  {journal}
  {\bibinfo  {journal} {Phys. Rev.}\ }\textbf {\bibinfo {volume} {C93}},\
  \bibinfo {pages} {024908} (\bibinfo {year} {2016})},\ \Eprint
  {http://arxiv.org/abs/1508.07017} {arXiv:1508.07017 [nucl-th]} \BibitemShut
  {NoStop}%
\bibitem [{\citenamefont {Bernhard}\ \emph {et~al.}(2015)\citenamefont
  {Bernhard}, \citenamefont {Marcy}, \citenamefont {Coleman-Smith},
  \citenamefont {Huzurbazar}, \citenamefont {Wolpert},\ and\ \citenamefont
  {Bass}}]{Bernhard:2015hxa}%
  \BibitemOpen
  \bibfield  {author} {\bibinfo {author} {\bibfnamefont {J.~E.}\ \bibnamefont
  {Bernhard}}, \bibinfo {author} {\bibfnamefont {P.~W.}\ \bibnamefont {Marcy}},
  \bibinfo {author} {\bibfnamefont {C.~E.}\ \bibnamefont {Coleman-Smith}},
  \bibinfo {author} {\bibfnamefont {S.}~\bibnamefont {Huzurbazar}}, \bibinfo
  {author} {\bibfnamefont {R.~L.}\ \bibnamefont {Wolpert}}, \ and\ \bibinfo
  {author} {\bibfnamefont {S.~A.}\ \bibnamefont {Bass}},\ }\href {\doibase
  10.1103/PhysRevC.91.054910} {\bibfield  {journal} {\bibinfo  {journal} {Phys.
  Rev. C}\ }\textbf {\bibinfo {volume} {91}},\ \bibinfo {pages} {054910}
  (\bibinfo {year} {2015})},\ \Eprint {http://arxiv.org/abs/1502.00339}
  {arXiv:1502.00339 [nucl-th]} \BibitemShut {NoStop}%
\bibitem [{\citenamefont {Bernhard}\ \emph {et~al.}(2016)\citenamefont
  {Bernhard}, \citenamefont {Moreland}, \citenamefont {Bass}, \citenamefont
  {Liu},\ and\ \citenamefont {Heinz}}]{Bernhard:2016tnd}%
  \BibitemOpen
  \bibfield  {author} {\bibinfo {author} {\bibfnamefont {J.~E.}\ \bibnamefont
  {Bernhard}}, \bibinfo {author} {\bibfnamefont {J.~S.}\ \bibnamefont
  {Moreland}}, \bibinfo {author} {\bibfnamefont {S.~A.}\ \bibnamefont {Bass}},
  \bibinfo {author} {\bibfnamefont {J.}~\bibnamefont {Liu}}, \ and\ \bibinfo
  {author} {\bibfnamefont {U.}~\bibnamefont {Heinz}},\ }\href {\doibase
  10.1103/PhysRevC.94.024907} {\bibfield  {journal} {\bibinfo  {journal} {Phys.
  Rev.}\ }\textbf {\bibinfo {volume} {C94}},\ \bibinfo {pages} {024907}
  (\bibinfo {year} {2016})},\ \Eprint {http://arxiv.org/abs/1605.03954}
  {arXiv:1605.03954 [nucl-th]} \BibitemShut {NoStop}%
\bibitem [{\citenamefont {Moreland}\ \emph {et~al.}(2020)\citenamefont
  {Moreland}, \citenamefont {Bernhard},\ and\ \citenamefont
  {Bass}}]{Moreland:2018gsh}%
  \BibitemOpen
  \bibfield  {author} {\bibinfo {author} {\bibfnamefont {J.~S.}\ \bibnamefont
  {Moreland}}, \bibinfo {author} {\bibfnamefont {J.~E.}\ \bibnamefont
  {Bernhard}}, \ and\ \bibinfo {author} {\bibfnamefont {S.~A.}\ \bibnamefont
  {Bass}},\ }\href {\doibase 10.1103/PhysRevC.101.024911} {\bibfield  {journal}
  {\bibinfo  {journal} {Phys. Rev. C}\ }\textbf {\bibinfo {volume} {101}},\
  \bibinfo {pages} {024911} (\bibinfo {year} {2020})},\ \Eprint
  {http://arxiv.org/abs/1808.02106} {arXiv:1808.02106 [nucl-th]} \BibitemShut
  {NoStop}%
\bibitem [{\citenamefont {Bernhard}\ \emph {et~al.}(2019)\citenamefont
  {Bernhard}, \citenamefont {Moreland},\ and\ \citenamefont
  {Bass}}]{Bernhard:2019bmu}%
  \BibitemOpen
  \bibfield  {author} {\bibinfo {author} {\bibfnamefont {J.~E.}\ \bibnamefont
  {Bernhard}}, \bibinfo {author} {\bibfnamefont {J.~S.}\ \bibnamefont
  {Moreland}}, \ and\ \bibinfo {author} {\bibfnamefont {S.~A.}\ \bibnamefont
  {Bass}},\ }\href {\doibase 10.1038/s41567-019-0611-8} {\bibfield  {journal}
  {\bibinfo  {journal} {Nature Phys.}\ }\textbf {\bibinfo {volume} {15}},\
  \bibinfo {pages} {1113} (\bibinfo {year} {2019})}\BibitemShut {NoStop}%
\bibitem [{\citenamefont {Everett}\ \emph
  {et~al.}(2021{\natexlab{a}})\citenamefont {Everett} \emph
  {et~al.}}]{JETSCAPE:2020shq}%
  \BibitemOpen
  \bibfield  {author} {\bibinfo {author} {\bibfnamefont {D.}~\bibnamefont
  {Everett}} \emph {et~al.} (\bibinfo {collaboration} {JETSCAPE}),\ }\href
  {\doibase 10.1103/PhysRevLett.126.242301} {\bibfield  {journal} {\bibinfo
  {journal} {Phys. Rev. Lett.}\ }\textbf {\bibinfo {volume} {126}},\ \bibinfo
  {pages} {242301} (\bibinfo {year} {2021}{\natexlab{a}})},\ \Eprint
  {http://arxiv.org/abs/2010.03928} {arXiv:2010.03928 [hep-ph]} \BibitemShut
  {NoStop}%
\bibitem [{\citenamefont {Nijs}\ \emph
  {et~al.}(2021{\natexlab{a}})\citenamefont {Nijs}, \citenamefont {van~der
  Schee}, \citenamefont {G\"ursoy},\ and\ \citenamefont
  {Snellings}}]{Nijs:2020ors}%
  \BibitemOpen
  \bibfield  {author} {\bibinfo {author} {\bibfnamefont {G.}~\bibnamefont
  {Nijs}}, \bibinfo {author} {\bibfnamefont {W.}~\bibnamefont {van~der Schee}},
  \bibinfo {author} {\bibfnamefont {U.}~\bibnamefont {G\"ursoy}}, \ and\
  \bibinfo {author} {\bibfnamefont {R.}~\bibnamefont {Snellings}},\ }\href
  {\doibase 10.1103/PhysRevLett.126.202301} {\bibfield  {journal} {\bibinfo
  {journal} {Phys. Rev. Lett.}\ }\textbf {\bibinfo {volume} {126}},\ \bibinfo
  {pages} {202301} (\bibinfo {year} {2021}{\natexlab{a}})},\ \Eprint
  {http://arxiv.org/abs/2010.15130} {arXiv:2010.15130 [nucl-th]} \BibitemShut
  {NoStop}%
\bibitem [{\citenamefont {Nijs}\ \emph
  {et~al.}(2021{\natexlab{b}})\citenamefont {Nijs}, \citenamefont {van~der
  Schee}, \citenamefont {G\"ursoy},\ and\ \citenamefont
  {Snellings}}]{Nijs:2020roc}%
  \BibitemOpen
  \bibfield  {author} {\bibinfo {author} {\bibfnamefont {G.}~\bibnamefont
  {Nijs}}, \bibinfo {author} {\bibfnamefont {W.}~\bibnamefont {van~der Schee}},
  \bibinfo {author} {\bibfnamefont {U.}~\bibnamefont {G\"ursoy}}, \ and\
  \bibinfo {author} {\bibfnamefont {R.}~\bibnamefont {Snellings}},\ }\href
  {\doibase 10.1103/PhysRevC.103.054909} {\bibfield  {journal} {\bibinfo
  {journal} {Phys. Rev. C}\ }\textbf {\bibinfo {volume} {103}},\ \bibinfo
  {pages} {054909} (\bibinfo {year} {2021}{\natexlab{b}})},\ \Eprint
  {http://arxiv.org/abs/2010.15134} {arXiv:2010.15134 [nucl-th]} \BibitemShut
  {NoStop}%
\bibitem [{\citenamefont {Everett}\ \emph
  {et~al.}(2021{\natexlab{b}})\citenamefont {Everett} \emph
  {et~al.}}]{JETSCAPE:2020mzn}%
  \BibitemOpen
  \bibfield  {author} {\bibinfo {author} {\bibfnamefont {D.}~\bibnamefont
  {Everett}} \emph {et~al.} (\bibinfo {collaboration} {JETSCAPE}),\ }\href
  {\doibase 10.1103/PhysRevC.103.054904} {\bibfield  {journal} {\bibinfo
  {journal} {Phys. Rev. C}\ }\textbf {\bibinfo {volume} {103}},\ \bibinfo
  {pages} {054904} (\bibinfo {year} {2021}{\natexlab{b}})},\ \Eprint
  {http://arxiv.org/abs/2011.01430} {arXiv:2011.01430 [hep-ph]} \BibitemShut
  {NoStop}%
\bibitem [{\citenamefont {Heffernan}\ \emph {et~al.}(2023)\citenamefont
  {Heffernan}, \citenamefont {Gale}, \citenamefont {Jeon},\ and\ \citenamefont
  {Paquet}}]{Heffernan_CGC}%
  \BibitemOpen
  \bibfield  {author} {\bibinfo {author} {\bibfnamefont {M.~R.}\ \bibnamefont
  {Heffernan}}, \bibinfo {author} {\bibfnamefont {C.}~\bibnamefont {Gale}},
  \bibinfo {author} {\bibfnamefont {S.}~\bibnamefont {Jeon}}, \ and\ \bibinfo
  {author} {\bibfnamefont {J.-F.}\ \bibnamefont {Paquet}},\ }\href {\doibase
  10.48550/ARXIV.2302.09478} {\enquote {\bibinfo {title} {Bayesian
  quantification of strongly-interacting matter with color glass condensate
  initial conditions},}\ } (\bibinfo {year} {2023})\BibitemShut {NoStop}%
\bibitem [{\citenamefont {Liu}\ \emph {et~al.}(2015)\citenamefont {Liu},
  \citenamefont {Shen},\ and\ \citenamefont {Heinz}}]{Liu:2015nwa}%
  \BibitemOpen
  \bibfield  {author} {\bibinfo {author} {\bibfnamefont {J.}~\bibnamefont
  {Liu}}, \bibinfo {author} {\bibfnamefont {C.}~\bibnamefont {Shen}}, \ and\
  \bibinfo {author} {\bibfnamefont {U.}~\bibnamefont {Heinz}},\ }\href
  {\doibase 10.1103/PhysRevC.91.064906} {\bibfield  {journal} {\bibinfo
  {journal} {Phys. Rev. C}\ }\textbf {\bibinfo {volume} {91}},\ \bibinfo
  {pages} {064906} (\bibinfo {year} {2015})},\ \bibinfo {note} {[Erratum:
  Phys.Rev.C 92, 049904(E)]},\ \Eprint {http://arxiv.org/abs/1504.02160}
  {arXiv:1504.02160 [nucl-th]} \BibitemShut {NoStop}%
\bibitem [{\citenamefont {Nunes~da Silva}\ \emph {et~al.}(2021)\citenamefont
  {Nunes~da Silva}, \citenamefont {Chinellato}, \citenamefont {Hippert},
  \citenamefont {Serenone}, \citenamefont {Takahashi}, \citenamefont {Denicol},
  \citenamefont {Luzum},\ and\ \citenamefont {Noronha}}]{NunesdaSilva:2020bfs}%
  \BibitemOpen
  \bibfield  {author} {\bibinfo {author} {\bibfnamefont {T.}~\bibnamefont
  {Nunes~da Silva}}, \bibinfo {author} {\bibfnamefont {D.}~\bibnamefont
  {Chinellato}}, \bibinfo {author} {\bibfnamefont {M.}~\bibnamefont {Hippert}},
  \bibinfo {author} {\bibfnamefont {W.}~\bibnamefont {Serenone}}, \bibinfo
  {author} {\bibfnamefont {J.}~\bibnamefont {Takahashi}}, \bibinfo {author}
  {\bibfnamefont {G.~S.}\ \bibnamefont {Denicol}}, \bibinfo {author}
  {\bibfnamefont {M.}~\bibnamefont {Luzum}}, \ and\ \bibinfo {author}
  {\bibfnamefont {J.}~\bibnamefont {Noronha}},\ }\href {\doibase
  10.1103/PhysRevC.103.054906} {\bibfield  {journal} {\bibinfo  {journal}
  {Phys. Rev. C}\ }\textbf {\bibinfo {volume} {103}},\ \bibinfo {pages}
  {054906} (\bibinfo {year} {2021})},\ \Eprint
  {http://arxiv.org/abs/2006.02324} {arXiv:2006.02324 [nucl-th]} \BibitemShut
  {NoStop}%
\bibitem [{\citenamefont {McNelis}\ \emph
  {et~al.}(2021{\natexlab{a}})\citenamefont {McNelis}, \citenamefont {Bazow},\
  and\ \citenamefont {Heinz}}]{McNelis_2021}%
  \BibitemOpen
  \bibfield  {author} {\bibinfo {author} {\bibfnamefont {M.}~\bibnamefont
  {McNelis}}, \bibinfo {author} {\bibfnamefont {D.}~\bibnamefont {Bazow}}, \
  and\ \bibinfo {author} {\bibfnamefont {U.}~\bibnamefont {Heinz}},\ }\href
  {\doibase 10.1016/j.cpc.2021.108077} {\bibfield  {journal} {\bibinfo
  {journal} {Comput. Phys. Comm.}\ }\textbf {\bibinfo {volume} {267}},\
  \bibinfo {pages} {108077} (\bibinfo {year} {2021}{\natexlab{a}})}\BibitemShut
  {NoStop}%
\bibitem [{\citenamefont {McNelis}\ \emph {et~al.}(2018)\citenamefont
  {McNelis}, \citenamefont {Bazow},\ and\ \citenamefont
  {Heinz}}]{mcnelis20183+}%
  \BibitemOpen
  \bibfield  {author} {\bibinfo {author} {\bibfnamefont {M.}~\bibnamefont
  {McNelis}}, \bibinfo {author} {\bibfnamefont {D.}~\bibnamefont {Bazow}}, \
  and\ \bibinfo {author} {\bibfnamefont {U.}~\bibnamefont {Heinz}},\ }\href
  {\doibase 10.1103/PhysRevC.97.054912} {\bibfield  {journal} {\bibinfo
  {journal} {Phys. Rev. C}\ }\textbf {\bibinfo {volume} {97}},\ \bibinfo
  {pages} {054912} (\bibinfo {year} {2018})}\BibitemShut {NoStop}%
\bibitem [{\citenamefont {Martinez}\ and\ \citenamefont
  {Strickland}(2010)}]{Martinez:2010sc}%
  \BibitemOpen
  \bibfield  {author} {\bibinfo {author} {\bibfnamefont {M.}~\bibnamefont
  {Martinez}}\ and\ \bibinfo {author} {\bibfnamefont {M.}~\bibnamefont
  {Strickland}},\ }\href {\doibase 10.1016/j.nuclphysa.2010.08.011} {\bibfield
  {journal} {\bibinfo  {journal} {Nucl. Phys.}\ }\textbf {\bibinfo {volume}
  {A848}},\ \bibinfo {pages} {183} (\bibinfo {year} {2010})},\ \Eprint
  {http://arxiv.org/abs/1007.0889} {arXiv:1007.0889 [nucl-th]} \BibitemShut
  {NoStop}%
\bibitem [{\citenamefont {Florkowski}\ and\ \citenamefont
  {Ryblewski}(2011)}]{florkowski2011highly}%
  \BibitemOpen
  \bibfield  {author} {\bibinfo {author} {\bibfnamefont {W.}~\bibnamefont
  {Florkowski}}\ and\ \bibinfo {author} {\bibfnamefont {R.}~\bibnamefont
  {Ryblewski}},\ }\href {\doibase 10.1103/PhysRevC.83.034907} {\bibfield
  {journal} {\bibinfo  {journal} {Phys. Rev. C}\ }\textbf {\bibinfo {volume}
  {83}},\ \bibinfo {pages} {034907} (\bibinfo {year} {2011})}\BibitemShut
  {NoStop}%
\bibitem [{\citenamefont {Alqahtani}\ \emph
  {et~al.}(2017{\natexlab{a}})\citenamefont {Alqahtani}, \citenamefont
  {Nopoush}, \citenamefont {Ryblewski},\ and\ \citenamefont
  {Strickland}}]{alqahtani20173+}%
  \BibitemOpen
  \bibfield  {author} {\bibinfo {author} {\bibfnamefont {M.}~\bibnamefont
  {Alqahtani}}, \bibinfo {author} {\bibfnamefont {M.}~\bibnamefont {Nopoush}},
  \bibinfo {author} {\bibfnamefont {R.}~\bibnamefont {Ryblewski}}, \ and\
  \bibinfo {author} {\bibfnamefont {M.}~\bibnamefont {Strickland}},\ }\href
  {\doibase 10.1103/PhysRevLett.119.042301} {\bibfield  {journal} {\bibinfo
  {journal} {Phys. Rev. Lett.}\ }\textbf {\bibinfo {volume} {119}},\ \bibinfo
  {pages} {042301} (\bibinfo {year} {2017}{\natexlab{a}})}\BibitemShut
  {NoStop}%
\bibitem [{\citenamefont {Alqahtani}\ \emph
  {et~al.}(2017{\natexlab{b}})\citenamefont {Alqahtani}, \citenamefont
  {Nopoush}, \citenamefont {Ryblewski},\ and\ \citenamefont
  {Strickland}}]{alqahtani2017anisotropic}%
  \BibitemOpen
  \bibfield  {author} {\bibinfo {author} {\bibfnamefont {M.}~\bibnamefont
  {Alqahtani}}, \bibinfo {author} {\bibfnamefont {M.}~\bibnamefont {Nopoush}},
  \bibinfo {author} {\bibfnamefont {R.}~\bibnamefont {Ryblewski}}, \ and\
  \bibinfo {author} {\bibfnamefont {M.}~\bibnamefont {Strickland}},\ }\href
  {\doibase 10.1103/PhysRevC.96.044910} {\bibfield  {journal} {\bibinfo
  {journal} {Phys. Rev. C}\ }\textbf {\bibinfo {volume} {96}},\ \bibinfo
  {pages} {044910} (\bibinfo {year} {2017}{\natexlab{b}})}\BibitemShut
  {NoStop}%
\bibitem [{\citenamefont {Almaalol}\ \emph {et~al.}(2019)\citenamefont
  {Almaalol}, \citenamefont {Alqahtani},\ and\ \citenamefont
  {Strickland}}]{almaalol2019anisotropic}%
  \BibitemOpen
  \bibfield  {author} {\bibinfo {author} {\bibfnamefont {D.}~\bibnamefont
  {Almaalol}}, \bibinfo {author} {\bibfnamefont {M.}~\bibnamefont {Alqahtani}},
  \ and\ \bibinfo {author} {\bibfnamefont {M.}~\bibnamefont {Strickland}},\
  }\href {\doibase 10.1103/PhysRevC.99.044902} {\bibfield  {journal} {\bibinfo
  {journal} {Phys. Rev. C}\ }\textbf {\bibinfo {volume} {99}},\ \bibinfo
  {pages} {044902} (\bibinfo {year} {2019})}\BibitemShut {NoStop}%
\bibitem [{\citenamefont {Alqahtani}\ and\ \citenamefont
  {Strickland}(2021)}]{alqahtani2021bulk}%
  \BibitemOpen
  \bibfield  {author} {\bibinfo {author} {\bibfnamefont {M.}~\bibnamefont
  {Alqahtani}}\ and\ \bibinfo {author} {\bibfnamefont {M.}~\bibnamefont
  {Strickland}},\ }\href {\doibase 10.1140/epjc/s10052-021-09832-z} {\bibfield
  {journal} {\bibinfo  {journal} {The European Physical Journal C}\ }\textbf
  {\bibinfo {volume} {81}},\ \bibinfo {pages} {1022} (\bibinfo {year}
  {2021})}\BibitemShut {NoStop}%
\bibitem [{\citenamefont {Bazow}\ \emph {et~al.}(2014)\citenamefont {Bazow},
  \citenamefont {Heinz},\ and\ \citenamefont {Strickland}}]{bazow2014second}%
  \BibitemOpen
  \bibfield  {author} {\bibinfo {author} {\bibfnamefont {D.}~\bibnamefont
  {Bazow}}, \bibinfo {author} {\bibfnamefont {U.}~\bibnamefont {Heinz}}, \ and\
  \bibinfo {author} {\bibfnamefont {M.}~\bibnamefont {Strickland}},\ }\href
  {\doibase 10.1103/PhysRevC.90.054910} {\bibfield  {journal} {\bibinfo
  {journal} {Phys. Rev. C}\ }\textbf {\bibinfo {volume} {90}},\ \bibinfo
  {pages} {054910} (\bibinfo {year} {2014})}\BibitemShut {NoStop}%
\bibitem [{git()}]{gitcode}%
  \BibitemOpen
  \href@noop {} {}\bibinfo {howpublished}
  {\url{https://github.com/danOSU/Bayesian_parameter_inferece_for_VAH}}\BibitemShut
  {NoStop}%
\bibitem [{\citenamefont {Moreland}\ \emph {et~al.}(2015)\citenamefont
  {Moreland}, \citenamefont {Bernhard},\ and\ \citenamefont
  {Bass}}]{Moreland:2014oya}%
  \BibitemOpen
  \bibfield  {author} {\bibinfo {author} {\bibfnamefont {J.~S.}\ \bibnamefont
  {Moreland}}, \bibinfo {author} {\bibfnamefont {J.~E.}\ \bibnamefont
  {Bernhard}}, \ and\ \bibinfo {author} {\bibfnamefont {S.~A.}\ \bibnamefont
  {Bass}},\ }\href {\doibase 10.1103/PhysRevC.92.011901} {\bibfield  {journal}
  {\bibinfo  {journal} {Phys. Rev. C}\ }\textbf {\bibinfo {volume} {92}},\
  \bibinfo {pages} {011901(R)} (\bibinfo {year} {2015})}\BibitemShut {NoStop}%
\bibitem [{tre()}]{trento_code}%
  \BibitemOpen
  \href@noop {} {}\bibinfo {howpublished}
  {\url{https://github.com/Duke-QCD/trento.git}}\BibitemShut {NoStop}%
\bibitem [{\citenamefont {Romatschke}\ and\ \citenamefont
  {Romatschke}(2019)}]{Romatschke:2017ejr}%
  \BibitemOpen
  \bibfield  {author} {\bibinfo {author} {\bibfnamefont {P.}~\bibnamefont
  {Romatschke}}\ and\ \bibinfo {author} {\bibfnamefont {U.}~\bibnamefont
  {Romatschke}},\ }\href {\doibase 10.1017/9781108651998} {\emph {\bibinfo
  {title} {{Relativistic fluid dynamics in and out of equilibrium}}}},\
  Cambridge Monographs on Mathematical Physics\ (\bibinfo  {publisher}
  {Cambridge University Press},\ \bibinfo {year} {2019})\ \Eprint
  {http://arxiv.org/abs/1712.05815} {arXiv:1712.05815 [nucl-th]} \BibitemShut
  {NoStop}%
\bibitem [{\citenamefont {Kolb}\ and\ \citenamefont
  {Heinz}(2004)}]{Kolb:2003dz}%
  \BibitemOpen
  \bibfield  {author} {\bibinfo {author} {\bibfnamefont {P.~F.}\ \bibnamefont
  {Kolb}}\ and\ \bibinfo {author} {\bibfnamefont {U.}~\bibnamefont {Heinz}},\
  }in\ \href {\doibase 10.1142/9789812795533_0010} {\emph {\bibinfo {booktitle}
  {Quark-Gluon Plasma 3}}},\ \bibinfo {editor} {edited by\ \bibinfo {editor}
  {\bibfnamefont {R.~C.}\ \bibnamefont {Hwa}}\ and\ \bibinfo {editor}
  {\bibfnamefont {X.-N.}\ \bibnamefont {Wang}}}\ (\bibinfo  {publisher} {World
  Scientific},\ \bibinfo {address} {Singapore},\ \bibinfo {year} {2004})\ pp.\
  \bibinfo {pages} {634--714},\ \Eprint {http://arxiv.org/abs/nucl-th/0305084}
  {arXiv:nucl-th/0305084} \BibitemShut {NoStop}%
\bibitem [{\citenamefont {Heinz}(2010)}]{Heinz:2009xj}%
  \BibitemOpen
  \bibfield  {author} {\bibinfo {author} {\bibfnamefont {U.}~\bibnamefont
  {Heinz}},\ }\href {\doibase 10.1007/978-3-642-01539-7_9} {\bibfield
  {journal} {\bibinfo  {journal} {Landolt-Bornstein}\ }\textbf {\bibinfo
  {volume} {23}},\ \bibinfo {pages} {240} (\bibinfo {year} {2010})},\ \Eprint
  {http://arxiv.org/abs/0901.4355} {arXiv:0901.4355 [nucl-th]} \BibitemShut
  {NoStop}%
\bibitem [{\citenamefont {Heinz}\ and\ \citenamefont
  {Snellings}(2013)}]{Heinz:2013th}%
  \BibitemOpen
  \bibfield  {author} {\bibinfo {author} {\bibfnamefont {U.}~\bibnamefont
  {Heinz}}\ and\ \bibinfo {author} {\bibfnamefont {R.}~\bibnamefont
  {Snellings}},\ }\href {\doibase 10.1146/annurev-nucl-102212-170540}
  {\bibfield  {journal} {\bibinfo  {journal} {Ann. Rev. Nucl. Part. Sci.}\
  }\textbf {\bibinfo {volume} {63}},\ \bibinfo {pages} {123} (\bibinfo {year}
  {2013})},\ \Eprint {http://arxiv.org/abs/1301.2826} {arXiv:1301.2826
  [nucl-th]} \BibitemShut {NoStop}%
\bibitem [{\citenamefont {Gale}\ \emph {et~al.}(2013)\citenamefont {Gale},
  \citenamefont {Jeon},\ and\ \citenamefont {Schenke}}]{Gale}%
  \BibitemOpen
  \bibfield  {author} {\bibinfo {author} {\bibfnamefont {C.}~\bibnamefont
  {Gale}}, \bibinfo {author} {\bibfnamefont {S.}~\bibnamefont {Jeon}}, \ and\
  \bibinfo {author} {\bibfnamefont {B.}~\bibnamefont {Schenke}},\ }\href
  {\doibase 10.1142/S0217751X13400113} {\bibfield  {journal} {\bibinfo
  {journal} {Int. J. Mod. Phys. A}\ }\textbf {\bibinfo {volume} {28}},\
  \bibinfo {pages} {1340011} (\bibinfo {year} {2013})},\ \Eprint
  {http://arxiv.org/abs/https://doi.org/10.1142/S0217751X13400113}
  {https://doi.org/10.1142/S0217751X13400113} \BibitemShut {NoStop}%
\bibitem [{\citenamefont {Jeon}\ and\ \citenamefont
  {Heinz}(2015)}]{Jeon:2015dfa}%
  \BibitemOpen
  \bibfield  {author} {\bibinfo {author} {\bibfnamefont {S.}~\bibnamefont
  {Jeon}}\ and\ \bibinfo {author} {\bibfnamefont {U.}~\bibnamefont {Heinz}},\
  }\href {\doibase 10.1142/S0218301315300106} {\bibfield  {journal} {\bibinfo
  {journal} {Int. J. Mod. Phys. E}\ }\textbf {\bibinfo {volume} {24}},\
  \bibinfo {pages} {1530010} (\bibinfo {year} {2015})},\ \Eprint
  {http://arxiv.org/abs/1503.03931} {arXiv:1503.03931 [hep-ph]} \BibitemShut
  {NoStop}%
\bibitem [{\citenamefont {Müller}(1967)}]{Muller:1967}%
  \BibitemOpen
  \bibfield  {author} {\bibinfo {author} {\bibfnamefont {I.}~\bibnamefont
  {Müller}},\ }\href {\doibase https://doi.org/10.1007/BF01326412} {\bibfield
  {journal} {\bibinfo  {journal} {Phys. Z}\ }\textbf {\bibinfo {volume}
  {198}},\ \bibinfo {pages} {329–344} (\bibinfo {year} {1967})}\BibitemShut
  {NoStop}%
\bibitem [{\citenamefont {Israel}(1976)}]{Israel:1976tn}%
  \BibitemOpen
  \bibfield  {author} {\bibinfo {author} {\bibfnamefont {W.}~\bibnamefont
  {Israel}},\ }\href {\doibase 10.1016/0003-4916(76)90064-6} {\bibfield
  {journal} {\bibinfo  {journal} {Annals Phys.}\ }\textbf {\bibinfo {volume}
  {100}},\ \bibinfo {pages} {310} (\bibinfo {year} {1976})}\BibitemShut
  {NoStop}%
\bibitem [{\citenamefont {Israel}\ and\ \citenamefont
  {Stewart}(1979)}]{Israel:1979wp}%
  \BibitemOpen
  \bibfield  {author} {\bibinfo {author} {\bibfnamefont {W.}~\bibnamefont
  {Israel}}\ and\ \bibinfo {author} {\bibfnamefont {J.~M.}\ \bibnamefont
  {Stewart}},\ }\href {\doibase 10.1016/0003-4916(79)90130-1} {\bibfield
  {journal} {\bibinfo  {journal} {Annals Phys.}\ }\textbf {\bibinfo {volume}
  {118}},\ \bibinfo {pages} {341} (\bibinfo {year} {1979})}\BibitemShut
  {NoStop}%
\bibitem [{\citenamefont {Baym}(1984)}]{Baym:1984np}%
  \BibitemOpen
  \bibfield  {author} {\bibinfo {author} {\bibfnamefont {G.}~\bibnamefont
  {Baym}},\ }\href {\doibase 10.1016/0370-2693(84)91863-X} {\bibfield
  {journal} {\bibinfo  {journal} {Phys. Lett. B}\ }\textbf {\bibinfo {volume}
  {138}},\ \bibinfo {pages} {18} (\bibinfo {year} {1984})}\BibitemShut
  {NoStop}%
\bibitem [{\citenamefont {Broniowski}\ \emph {et~al.}(2009)\citenamefont
  {Broniowski}, \citenamefont {Florkowski}, \citenamefont {Chojnacki},\ and\
  \citenamefont {Kisiel}}]{Broniowski:2008qk}%
  \BibitemOpen
  \bibfield  {author} {\bibinfo {author} {\bibfnamefont {W.}~\bibnamefont
  {Broniowski}}, \bibinfo {author} {\bibfnamefont {W.}~\bibnamefont
  {Florkowski}}, \bibinfo {author} {\bibfnamefont {M.}~\bibnamefont
  {Chojnacki}}, \ and\ \bibinfo {author} {\bibfnamefont {A.}~\bibnamefont
  {Kisiel}},\ }\href {\doibase 10.1103/PhysRevC.80.034902} {\bibfield
  {journal} {\bibinfo  {journal} {Phys. Rev.}\ }\textbf {\bibinfo {volume}
  {C80}},\ \bibinfo {pages} {034902} (\bibinfo {year} {2009})},\ \Eprint
  {http://arxiv.org/abs/0812.3393} {arXiv:0812.3393 [nucl-th]} \BibitemShut
  {NoStop}%
\bibitem [{\citenamefont {Romatschke}(2015)}]{Romatschke:2015dha}%
  \BibitemOpen
  \bibfield  {author} {\bibinfo {author} {\bibfnamefont {P.}~\bibnamefont
  {Romatschke}},\ }\href {\doibase 10.1140/epjc/s10052-015-3646-8} {\bibfield
  {journal} {\bibinfo  {journal} {Eur. Phys. J. C}\ }\textbf {\bibinfo {volume}
  {75}},\ \bibinfo {pages} {429} (\bibinfo {year} {2015})},\ \Eprint
  {http://arxiv.org/abs/1504.02529} {arXiv:1504.02529 [nucl-th]} \BibitemShut
  {NoStop}%
\bibitem [{\citenamefont {Liu}(2015)}]{Liu-thesis}%
  \BibitemOpen
  \bibfield  {author} {\bibinfo {author} {\bibfnamefont {J.}~\bibnamefont
  {Liu}},\ }\emph {\bibinfo {title} {Pre-equilibrium evolution effects}},\
  \href {http://rave.ohiolink.edu/etdc/view?acc_num=osu1449185522} {Ph.D.
  thesis},\ \bibinfo  {school} {Ohio State University} (\bibinfo {year}
  {2015})\BibitemShut {NoStop}%
\bibitem [{\citenamefont {Chattopadhyay}\ \emph {et~al.}(2022)\citenamefont
  {Chattopadhyay}, \citenamefont {Jaiswal}, \citenamefont {Du}, \citenamefont
  {Heinz},\ and\ \citenamefont {Pal}}]{Chattopadhyay:2021ive}%
  \BibitemOpen
  \bibfield  {author} {\bibinfo {author} {\bibfnamefont {C.}~\bibnamefont
  {Chattopadhyay}}, \bibinfo {author} {\bibfnamefont {S.}~\bibnamefont
  {Jaiswal}}, \bibinfo {author} {\bibfnamefont {L.}~\bibnamefont {Du}},
  \bibinfo {author} {\bibfnamefont {U.}~\bibnamefont {Heinz}}, \ and\ \bibinfo
  {author} {\bibfnamefont {S.}~\bibnamefont {Pal}},\ }\href {\doibase
  10.1016/j.physletb.2021.136820} {\bibfield  {journal} {\bibinfo  {journal}
  {Phys. Lett. B}\ }\textbf {\bibinfo {volume} {824}},\ \bibinfo {pages}
  {136820} (\bibinfo {year} {2022})},\ \Eprint
  {http://arxiv.org/abs/2107.05500} {arXiv:2107.05500 [nucl-th]} \BibitemShut
  {NoStop}%
\bibitem [{\citenamefont {Jaiswal}\ \emph {et~al.}(2022)\citenamefont
  {Jaiswal}, \citenamefont {Chattopadhyay}, \citenamefont {Du}, \citenamefont
  {Heinz},\ and\ \citenamefont {Pal}}]{Jaiswal:2021uvv}%
  \BibitemOpen
  \bibfield  {author} {\bibinfo {author} {\bibfnamefont {S.}~\bibnamefont
  {Jaiswal}}, \bibinfo {author} {\bibfnamefont {C.}~\bibnamefont
  {Chattopadhyay}}, \bibinfo {author} {\bibfnamefont {L.}~\bibnamefont {Du}},
  \bibinfo {author} {\bibfnamefont {U.}~\bibnamefont {Heinz}}, \ and\ \bibinfo
  {author} {\bibfnamefont {S.}~\bibnamefont {Pal}},\ }\href {\doibase
  10.1103/PhysRevC.105.024911} {\bibfield  {journal} {\bibinfo  {journal}
  {Phys. Rev. C}\ }\textbf {\bibinfo {volume} {105}},\ \bibinfo {pages}
  {024911} (\bibinfo {year} {2022})},\ \Eprint
  {http://arxiv.org/abs/2107.10248} {arXiv:2107.10248 [hep-ph]} \BibitemShut
  {NoStop}%
\bibitem [{\citenamefont {Anishetty}\ \emph {et~al.}(1980)\citenamefont
  {Anishetty}, \citenamefont {Koehler},\ and\ \citenamefont
  {McLerran}}]{PhysRevD.22.2793}%
  \BibitemOpen
  \bibfield  {author} {\bibinfo {author} {\bibfnamefont {R.}~\bibnamefont
  {Anishetty}}, \bibinfo {author} {\bibfnamefont {P.}~\bibnamefont {Koehler}},
  \ and\ \bibinfo {author} {\bibfnamefont {L.}~\bibnamefont {McLerran}},\
  }\href {\doibase 10.1103/PhysRevD.22.2793} {\bibfield  {journal} {\bibinfo
  {journal} {Phys. Rev. D}\ }\textbf {\bibinfo {volume} {22}},\ \bibinfo
  {pages} {2793} (\bibinfo {year} {1980})}\BibitemShut {NoStop}%
\bibitem [{\citenamefont {Bjorken}(1983)}]{PhysRevD.27.140}%
  \BibitemOpen
  \bibfield  {author} {\bibinfo {author} {\bibfnamefont {J.~D.}\ \bibnamefont
  {Bjorken}},\ }\href {\doibase 10.1103/PhysRevD.27.140} {\bibfield  {journal}
  {\bibinfo  {journal} {Phys. Rev. D}\ }\textbf {\bibinfo {volume} {27}},\
  \bibinfo {pages} {140} (\bibinfo {year} {1983})}\BibitemShut {NoStop}%
\bibitem [{\citenamefont {Molnar}\ \emph {et~al.}(2016)\citenamefont {Molnar},
  \citenamefont {Niemi},\ and\ \citenamefont {Rischke}}]{Molnar:2016vvu}%
  \BibitemOpen
  \bibfield  {author} {\bibinfo {author} {\bibfnamefont {E.}~\bibnamefont
  {Molnar}}, \bibinfo {author} {\bibfnamefont {H.}~\bibnamefont {Niemi}}, \
  and\ \bibinfo {author} {\bibfnamefont {D.~H.}\ \bibnamefont {Rischke}},\
  }\href {\doibase 10.1103/PhysRevD.93.114025} {\bibfield  {journal} {\bibinfo
  {journal} {Phys. Rev.}\ }\textbf {\bibinfo {volume} {D93}},\ \bibinfo {pages}
  {114025} (\bibinfo {year} {2016})},\ \Eprint
  {http://arxiv.org/abs/1602.00573} {arXiv:1602.00573 [nucl-th]} \BibitemShut
  {NoStop}%
\bibitem [{\citenamefont {Bazavov}\ \emph {et~al.}(2018)\citenamefont
  {Bazavov}, \citenamefont {Petreczky},\ and\ \citenamefont
  {Weber}}]{PhysRevD.97.014510}%
  \BibitemOpen
  \bibfield  {author} {\bibinfo {author} {\bibfnamefont {A.}~\bibnamefont
  {Bazavov}}, \bibinfo {author} {\bibfnamefont {P.}~\bibnamefont {Petreczky}},
  \ and\ \bibinfo {author} {\bibfnamefont {J.~H.}\ \bibnamefont {Weber}},\
  }\href {\doibase 10.1103/PhysRevD.97.014510} {\bibfield  {journal} {\bibinfo
  {journal} {Phys. Rev. D}\ }\textbf {\bibinfo {volume} {97}},\ \bibinfo
  {pages} {014510} (\bibinfo {year} {2018})}\BibitemShut {NoStop}%
\bibitem [{\citenamefont {Alqahtani}\ \emph
  {et~al.}(2017{\natexlab{c}})\citenamefont {Alqahtani}, \citenamefont
  {Nopoush},\ and\ \citenamefont {Strickland}}]{Alqahtani:2016rth}%
  \BibitemOpen
  \bibfield  {author} {\bibinfo {author} {\bibfnamefont {M.}~\bibnamefont
  {Alqahtani}}, \bibinfo {author} {\bibfnamefont {M.}~\bibnamefont {Nopoush}},
  \ and\ \bibinfo {author} {\bibfnamefont {M.}~\bibnamefont {Strickland}},\
  }\href {\doibase 10.1103/PhysRevC.95.034906} {\bibfield  {journal} {\bibinfo
  {journal} {Phys. Rev. C}\ }\textbf {\bibinfo {volume} {95}},\ \bibinfo
  {pages} {034906} (\bibinfo {year} {2017}{\natexlab{c}})},\ \Eprint
  {http://arxiv.org/abs/1605.02101} {arXiv:1605.02101 [nucl-th]} \BibitemShut
  {NoStop}%
\bibitem [{\citenamefont {Tinti}\ \emph {et~al.}(2017)\citenamefont {Tinti},
  \citenamefont {Jaiswal},\ and\ \citenamefont
  {Ryblewski}}]{PhysRevD.95.054007}%
  \BibitemOpen
  \bibfield  {author} {\bibinfo {author} {\bibfnamefont {L.}~\bibnamefont
  {Tinti}}, \bibinfo {author} {\bibfnamefont {A.}~\bibnamefont {Jaiswal}}, \
  and\ \bibinfo {author} {\bibfnamefont {R.}~\bibnamefont {Ryblewski}},\ }\href
  {\doibase 10.1103/PhysRevD.95.054007} {\bibfield  {journal} {\bibinfo
  {journal} {Phys. Rev. D}\ }\textbf {\bibinfo {volume} {95}},\ \bibinfo
  {pages} {054007} (\bibinfo {year} {2017})}\BibitemShut {NoStop}%
\bibitem [{\citenamefont {Song}\ \emph
  {et~al.}(2011{\natexlab{a}})\citenamefont {Song}, \citenamefont {Bass},
  \citenamefont {Heinz}, \citenamefont {Hirano},\ and\ \citenamefont
  {Shen}}]{Song:2011hk}%
  \BibitemOpen
  \bibfield  {author} {\bibinfo {author} {\bibfnamefont {H.}~\bibnamefont
  {Song}}, \bibinfo {author} {\bibfnamefont {S.~A.}\ \bibnamefont {Bass}},
  \bibinfo {author} {\bibfnamefont {U.}~\bibnamefont {Heinz}}, \bibinfo
  {author} {\bibfnamefont {T.}~\bibnamefont {Hirano}}, \ and\ \bibinfo {author}
  {\bibfnamefont {C.}~\bibnamefont {Shen}},\ }\href {\doibase
  10.1103/PhysRevC.83.054910} {\bibfield  {journal} {\bibinfo  {journal} {Phys.
  Rev. C}\ }\textbf {\bibinfo {volume} {83}},\ \bibinfo {pages} {054910}
  (\bibinfo {year} {2011}{\natexlab{a}})},\ \bibinfo {note} {[Erratum:
  Phys.Rev.C 86, 059903 (2012)]},\ \Eprint {http://arxiv.org/abs/1101.4638}
  {arXiv:1101.4638 [nucl-th]} \BibitemShut {NoStop}%
\bibitem [{sma()}]{smash_code}%
  \BibitemOpen
  \href@noop {} {}\bibinfo {howpublished}
  {\url{https://github.com/smash-transport/smash}}\BibitemShut {NoStop}%
\bibitem [{\citenamefont {Huovinen}\ and\ \citenamefont
  {Petersen}(2012)}]{Huovinen:2012is}%
  \BibitemOpen
  \bibfield  {author} {\bibinfo {author} {\bibfnamefont {P.}~\bibnamefont
  {Huovinen}}\ and\ \bibinfo {author} {\bibfnamefont {H.}~\bibnamefont
  {Petersen}},\ }\href {\doibase 10.1140/epja/i2012-12171-9} {\bibfield
  {journal} {\bibinfo  {journal} {Eur. Phys. J.}\ }\textbf {\bibinfo {volume}
  {A48}},\ \bibinfo {pages} {171} (\bibinfo {year} {2012})},\ \Eprint
  {http://arxiv.org/abs/1206.3371} {arXiv:1206.3371 [nucl-th]} \BibitemShut
  {NoStop}%
\bibitem [{\citenamefont {McNelis}\ \emph
  {et~al.}(2021{\natexlab{b}})\citenamefont {McNelis}, \citenamefont
  {Everett},\ and\ \citenamefont {Heinz}}]{McNelis:2019auj}%
  \BibitemOpen
  \bibfield  {author} {\bibinfo {author} {\bibfnamefont {M.}~\bibnamefont
  {McNelis}}, \bibinfo {author} {\bibfnamefont {D.}~\bibnamefont {Everett}}, \
  and\ \bibinfo {author} {\bibfnamefont {U.}~\bibnamefont {Heinz}},\ }\href
  {\doibase 10.1016/j.cpc.2020.107604} {\bibfield  {journal} {\bibinfo
  {journal} {Comput. Phys. Commun.}\ }\textbf {\bibinfo {volume} {258}},\
  \bibinfo {pages} {107604} (\bibinfo {year} {2021}{\natexlab{b}})},\ \Eprint
  {http://arxiv.org/abs/1912.08271} {arXiv:1912.08271 [nucl-th]} \BibitemShut
  {NoStop}%
\bibitem [{is3()}]{is3d_code}%
  \BibitemOpen
  \href@noop {} {}\bibinfo {howpublished}
  {\url{https://github.com/derekeverett/iS3D}}\BibitemShut {NoStop}%
\bibitem [{\citenamefont {Cooper}\ and\ \citenamefont
  {Frye}(1974)}]{Cooper:1974mv}%
  \BibitemOpen
  \bibfield  {author} {\bibinfo {author} {\bibfnamefont {F.}~\bibnamefont
  {Cooper}}\ and\ \bibinfo {author} {\bibfnamefont {G.}~\bibnamefont {Frye}},\
  }\href {\doibase 10.1103/PhysRevD.10.186} {\bibfield  {journal} {\bibinfo
  {journal} {Phys. Rev. D}\ }\textbf {\bibinfo {volume} {10}},\ \bibinfo
  {pages} {186} (\bibinfo {year} {1974})}\BibitemShut {NoStop}%
\bibitem [{\citenamefont {McNelis}\ and\ \citenamefont
  {Heinz}(2021)}]{PhysRevC.103.064903}%
  \BibitemOpen
  \bibfield  {author} {\bibinfo {author} {\bibfnamefont {M.}~\bibnamefont
  {McNelis}}\ and\ \bibinfo {author} {\bibfnamefont {U.}~\bibnamefont
  {Heinz}},\ }\href {\doibase 10.1103/PhysRevC.103.064903} {\bibfield
  {journal} {\bibinfo  {journal} {Phys. Rev. C}\ }\textbf {\bibinfo {volume}
  {103}},\ \bibinfo {pages} {064903} (\bibinfo {year} {2021})}\BibitemShut
  {NoStop}%
\bibitem [{\citenamefont {Everett}\ \emph
  {et~al.}(2021{\natexlab{c}})\citenamefont {Everett}, \citenamefont
  {Chattopadhyay},\ and\ \citenamefont {Heinz}}]{Everett:2021ulz}%
  \BibitemOpen
  \bibfield  {author} {\bibinfo {author} {\bibfnamefont {D.}~\bibnamefont
  {Everett}}, \bibinfo {author} {\bibfnamefont {C.}~\bibnamefont
  {Chattopadhyay}}, \ and\ \bibinfo {author} {\bibfnamefont {U.}~\bibnamefont
  {Heinz}},\ }\href {\doibase 10.1103/PhysRevC.103.064902} {\bibfield
  {journal} {\bibinfo  {journal} {Phys. Rev. C}\ }\textbf {\bibinfo {volume}
  {103}},\ \bibinfo {pages} {064902} (\bibinfo {year} {2021}{\natexlab{c}})},\
  \Eprint {http://arxiv.org/abs/2101.01130} {arXiv:2101.01130 [hep-ph]}
  \BibitemShut {NoStop}%
\bibitem [{\citenamefont {Phillips}\ \emph {et~al.}(2021)\citenamefont
  {Phillips} \emph {et~al.}}]{Phillips:2020dmw}%
  \BibitemOpen
  \bibfield  {author} {\bibinfo {author} {\bibfnamefont {D.~R.}\ \bibnamefont
  {Phillips}} \emph {et~al.},\ }\href {\doibase 10.1088/1361-6471/abf1df}
  {\bibfield  {journal} {\bibinfo  {journal} {J. Phys. G}\ }\textbf {\bibinfo
  {volume} {48}},\ \bibinfo {pages} {072001} (\bibinfo {year} {2021})},\
  \Eprint {http://arxiv.org/abs/2012.07704} {arXiv:2012.07704 [nucl-th]}
  \BibitemShut {NoStop}%
\bibitem [{\citenamefont {Nonaka}\ and\ \citenamefont
  {Bass}(2007)}]{Nonaka:2006yn}%
  \BibitemOpen
  \bibfield  {author} {\bibinfo {author} {\bibfnamefont {C.}~\bibnamefont
  {Nonaka}}\ and\ \bibinfo {author} {\bibfnamefont {S.~A.}\ \bibnamefont
  {Bass}},\ }\href {\doibase 10.1103/PhysRevC.75.014902} {\bibfield  {journal}
  {\bibinfo  {journal} {Phys. Rev.}\ }\textbf {\bibinfo {volume} {C75}},\
  \bibinfo {pages} {014902} (\bibinfo {year} {2007})},\ \Eprint
  {http://arxiv.org/abs/nucl-th/0607018} {arXiv:nucl-th/0607018 [nucl-th]}
  \BibitemShut {NoStop}%
\bibitem [{\citenamefont {Hirano}\ \emph {et~al.}(2008)\citenamefont {Hirano},
  \citenamefont {Heinz}, \citenamefont {Kharzeev}, \citenamefont {Lacey},\ and\
  \citenamefont {Nara}}]{Hirano:2007ei}%
  \BibitemOpen
  \bibfield  {author} {\bibinfo {author} {\bibfnamefont {T.}~\bibnamefont
  {Hirano}}, \bibinfo {author} {\bibfnamefont {U.}~\bibnamefont {Heinz}},
  \bibinfo {author} {\bibfnamefont {D.}~\bibnamefont {Kharzeev}}, \bibinfo
  {author} {\bibfnamefont {R.}~\bibnamefont {Lacey}}, \ and\ \bibinfo {author}
  {\bibfnamefont {Y.}~\bibnamefont {Nara}},\ }\href {\doibase
  10.1103/PhysRevC.77.044909} {\bibfield  {journal} {\bibinfo  {journal} {Phys.
  Rev.}\ }\textbf {\bibinfo {volume} {C77}},\ \bibinfo {pages} {044909}
  (\bibinfo {year} {2008})},\ \Eprint {http://arxiv.org/abs/0710.5795}
  {arXiv:0710.5795 [nucl-th]} \BibitemShut {NoStop}%
\bibitem [{\citenamefont {Petersen}\ \emph {et~al.}(2008)\citenamefont
  {Petersen}, \citenamefont {Steinheimer}, \citenamefont {Burau}, \citenamefont
  {Bleicher},\ and\ \citenamefont {Stocker}}]{Petersen:2008dd}%
  \BibitemOpen
  \bibfield  {author} {\bibinfo {author} {\bibfnamefont {H.}~\bibnamefont
  {Petersen}}, \bibinfo {author} {\bibfnamefont {J.}~\bibnamefont
  {Steinheimer}}, \bibinfo {author} {\bibfnamefont {G.}~\bibnamefont {Burau}},
  \bibinfo {author} {\bibfnamefont {M.}~\bibnamefont {Bleicher}}, \ and\
  \bibinfo {author} {\bibfnamefont {H.}~\bibnamefont {Stocker}},\ }\href
  {\doibase 10.1103/PhysRevC.78.044901} {\bibfield  {journal} {\bibinfo
  {journal} {Phys. Rev.}\ }\textbf {\bibinfo {volume} {C78}},\ \bibinfo {pages}
  {044901} (\bibinfo {year} {2008})},\ \Eprint {http://arxiv.org/abs/0806.1695}
  {arXiv:0806.1695 [nucl-th]} \BibitemShut {NoStop}%
\bibitem [{\citenamefont {Song}\ \emph
  {et~al.}(2011{\natexlab{b}})\citenamefont {Song}, \citenamefont {Bass},\ and\
  \citenamefont {Heinz}}]{Song:2010aq}%
  \BibitemOpen
  \bibfield  {author} {\bibinfo {author} {\bibfnamefont {H.}~\bibnamefont
  {Song}}, \bibinfo {author} {\bibfnamefont {S.~A.}\ \bibnamefont {Bass}}, \
  and\ \bibinfo {author} {\bibfnamefont {U.}~\bibnamefont {Heinz}},\ }\href
  {\doibase 10.1103/PhysRevC.83.024912} {\bibfield  {journal} {\bibinfo
  {journal} {Phys. Rev.}\ }\textbf {\bibinfo {volume} {C83}},\ \bibinfo {pages}
  {024912} (\bibinfo {year} {2011}{\natexlab{b}})},\ \Eprint
  {http://arxiv.org/abs/1012.0555} {arXiv:1012.0555 [nucl-th]} \BibitemShut
  {NoStop}%
\bibitem [{\citenamefont {Heinz}\ \emph {et~al.}(2012)\citenamefont {Heinz},
  \citenamefont {Shen},\ and\ \citenamefont {Song}}]{Heinz:2011kt}%
  \BibitemOpen
  \bibfield  {author} {\bibinfo {author} {\bibfnamefont {U.}~\bibnamefont
  {Heinz}}, \bibinfo {author} {\bibfnamefont {C.}~\bibnamefont {Shen}}, \ and\
  \bibinfo {author} {\bibfnamefont {H.}~\bibnamefont {Song}},\ }\href {\doibase
  10.1063/1.3700674} {\bibfield  {journal} {\bibinfo  {journal} {AIP Conf.
  Proc.}\ }\textbf {\bibinfo {volume} {1441}},\ \bibinfo {pages} {766}
  (\bibinfo {year} {2012})},\ \Eprint {http://arxiv.org/abs/1108.5323}
  {arXiv:1108.5323 [nucl-th]} \BibitemShut {NoStop}%
\bibitem [{\citenamefont {Song}\ \emph {et~al.}(2014)\citenamefont {Song},
  \citenamefont {Bass},\ and\ \citenamefont {Heinz}}]{Song:2013qma}%
  \BibitemOpen
  \bibfield  {author} {\bibinfo {author} {\bibfnamefont {H.}~\bibnamefont
  {Song}}, \bibinfo {author} {\bibfnamefont {S.~A.}\ \bibnamefont {Bass}}, \
  and\ \bibinfo {author} {\bibfnamefont {U.}~\bibnamefont {Heinz}},\ }\href
  {\doibase 10.1103/PhysRevC.89.034919} {\bibfield  {journal} {\bibinfo
  {journal} {Phys. Rev. C}\ }\textbf {\bibinfo {volume} {89}},\ \bibinfo
  {pages} {034919} (\bibinfo {year} {2014})}\BibitemShut {NoStop}%
\bibitem [{\citenamefont {Zhu}\ \emph {et~al.}(2015)\citenamefont {Zhu},
  \citenamefont {Meng}, \citenamefont {Song},\ and\ \citenamefont
  {Liu}}]{Zhu:2015dfa}%
  \BibitemOpen
  \bibfield  {author} {\bibinfo {author} {\bibfnamefont {X.}~\bibnamefont
  {Zhu}}, \bibinfo {author} {\bibfnamefont {F.}~\bibnamefont {Meng}}, \bibinfo
  {author} {\bibfnamefont {H.}~\bibnamefont {Song}}, \ and\ \bibinfo {author}
  {\bibfnamefont {Y.-X.}\ \bibnamefont {Liu}},\ }\href {\doibase
  10.1103/PhysRevC.91.034904} {\bibfield  {journal} {\bibinfo  {journal} {Phys.
  Rev.}\ }\textbf {\bibinfo {volume} {C91}},\ \bibinfo {pages} {034904}
  (\bibinfo {year} {2015})},\ \Eprint {http://arxiv.org/abs/1501.03286}
  {arXiv:1501.03286 [nucl-th]} \BibitemShut {NoStop}%
\bibitem [{\citenamefont {Ryu}\ \emph {et~al.}(2018)\citenamefont {Ryu},
  \citenamefont {Paquet}, \citenamefont {Shen}, \citenamefont {Denicol},
  \citenamefont {Schenke}, \citenamefont {Jeon},\ and\ \citenamefont
  {Gale}}]{Ryu:2017qzn}%
  \BibitemOpen
  \bibfield  {author} {\bibinfo {author} {\bibfnamefont {S.}~\bibnamefont
  {Ryu}}, \bibinfo {author} {\bibfnamefont {J.-F.}\ \bibnamefont {Paquet}},
  \bibinfo {author} {\bibfnamefont {C.}~\bibnamefont {Shen}}, \bibinfo {author}
  {\bibfnamefont {G.}~\bibnamefont {Denicol}}, \bibinfo {author} {\bibfnamefont
  {B.}~\bibnamefont {Schenke}}, \bibinfo {author} {\bibfnamefont
  {S.}~\bibnamefont {Jeon}}, \ and\ \bibinfo {author} {\bibfnamefont
  {C.}~\bibnamefont {Gale}},\ }\href {\doibase 10.1103/PhysRevC.97.034910}
  {\bibfield  {journal} {\bibinfo  {journal} {Phys. Rev.}\ }\textbf {\bibinfo
  {volume} {C97}},\ \bibinfo {pages} {034910} (\bibinfo {year} {2018})},\
  \Eprint {http://arxiv.org/abs/1704.04216} {arXiv:1704.04216 [nucl-th]}
  \BibitemShut {NoStop}%
\bibitem [{\citenamefont {Aamodt}\ \emph
  {et~al.}(2011{\natexlab{a}})\citenamefont {Aamodt} \emph
  {et~al.}}]{Aamodt:2010cz}%
  \BibitemOpen
  \bibfield  {author} {\bibinfo {author} {\bibfnamefont {K.}~\bibnamefont
  {Aamodt}} \emph {et~al.} (\bibinfo {collaboration} {ALICE}),\ }\href
  {\doibase 10.1103/PhysRevLett.106.032301} {\bibfield  {journal} {\bibinfo
  {journal} {Phys. Rev. Lett.}\ }\textbf {\bibinfo {volume} {106}},\ \bibinfo
  {pages} {032301} (\bibinfo {year} {2011}{\natexlab{a}})},\ \Eprint
  {http://arxiv.org/abs/1012.1657} {arXiv:1012.1657 [nucl-ex]} \BibitemShut
  {NoStop}%
\bibitem [{\citenamefont {Adam}\ \emph {et~al.}(2016)\citenamefont {Adam} \emph
  {et~al.}}]{Adam:2016thv}%
  \BibitemOpen
  \bibfield  {author} {\bibinfo {author} {\bibfnamefont {J.}~\bibnamefont
  {Adam}} \emph {et~al.} (\bibinfo {collaboration} {ALICE}),\ }\href {\doibase
  10.1103/PhysRevC.94.034903} {\bibfield  {journal} {\bibinfo  {journal} {Phys.
  Rev.}\ }\textbf {\bibinfo {volume} {C94}},\ \bibinfo {pages} {034903}
  (\bibinfo {year} {2016})},\ \Eprint {http://arxiv.org/abs/1603.04775}
  {arXiv:1603.04775 [nucl-ex]} \BibitemShut {NoStop}%
\bibitem [{\citenamefont {Abelev}\ \emph {et~al.}(2013)\citenamefont {Abelev}
  \emph {et~al.}}]{Abelev:2013vea}%
  \BibitemOpen
  \bibfield  {author} {\bibinfo {author} {\bibfnamefont {B.}~\bibnamefont
  {Abelev}} \emph {et~al.} (\bibinfo {collaboration} {ALICE}),\ }\href
  {\doibase 10.1103/PhysRevC.88.044910} {\bibfield  {journal} {\bibinfo
  {journal} {Phys. Rev.}\ }\textbf {\bibinfo {volume} {C88}},\ \bibinfo {pages}
  {044910} (\bibinfo {year} {2013})},\ \Eprint {http://arxiv.org/abs/1303.0737}
  {arXiv:1303.0737 [hep-ex]} \BibitemShut {NoStop}%
\bibitem [{\citenamefont {Aamodt}\ \emph
  {et~al.}(2011{\natexlab{b}})\citenamefont {Aamodt} \emph
  {et~al.}}]{ALICE:2011ab}%
  \BibitemOpen
  \bibfield  {author} {\bibinfo {author} {\bibfnamefont {K.}~\bibnamefont
  {Aamodt}} \emph {et~al.} (\bibinfo {collaboration} {ALICE}),\ }\href
  {\doibase 10.1103/PhysRevLett.107.032301} {\bibfield  {journal} {\bibinfo
  {journal} {Phys. Rev. Lett.}\ }\textbf {\bibinfo {volume} {107}},\ \bibinfo
  {pages} {032301} (\bibinfo {year} {2011}{\natexlab{b}})},\ \Eprint
  {http://arxiv.org/abs/1105.3865} {arXiv:1105.3865 [nucl-ex]} \BibitemShut
  {NoStop}%
\bibitem [{\citenamefont {Sivia}\ and\ \citenamefont {Skilling}(2006)}]{Sivia}%
  \BibitemOpen
  \bibfield  {author} {\bibinfo {author} {\bibfnamefont {D.}~\bibnamefont
  {Sivia}}\ and\ \bibinfo {author} {\bibfnamefont {J.}~\bibnamefont
  {Skilling}},\ }\href@noop {} {\emph {\bibinfo {title} {Data analysis: a
  Bayesian tutorial}}}\ (\bibinfo  {publisher} {OUP Oxford},\ \bibinfo {year}
  {2006})\BibitemShut {NoStop}%
\bibitem [{\citenamefont {Gelman}\ \emph {et~al.}(2004)\citenamefont {Gelman},
  \citenamefont {Carlin}, \citenamefont {Stern},\ and\ \citenamefont
  {Rubin}}]{Gelman2004}%
  \BibitemOpen
  \bibfield  {author} {\bibinfo {author} {\bibfnamefont {A.}~\bibnamefont
  {Gelman}}, \bibinfo {author} {\bibfnamefont {J.~B.}\ \bibnamefont {Carlin}},
  \bibinfo {author} {\bibfnamefont {H.~S.}\ \bibnamefont {Stern}}, \ and\
  \bibinfo {author} {\bibfnamefont {D.~B.}\ \bibnamefont {Rubin}},\ }\href@noop
  {} {\emph {\bibinfo {title} {Bayesian Data Analysis}}},\ \bibinfo {edition}
  {2nd}\ ed.\ (\bibinfo  {publisher} {Chapman and Hall/CRC},\ \bibinfo {year}
  {2004})\BibitemShut {NoStop}%
\bibitem [{\citenamefont {Trotta}(2008{\natexlab{a}})}]{Trotta:2008qt}%
  \BibitemOpen
  \bibfield  {author} {\bibinfo {author} {\bibfnamefont {R.}~\bibnamefont
  {Trotta}},\ }\href {\doibase 10.1080/00107510802066753} {\bibfield  {journal}
  {\bibinfo  {journal} {Contemp. Phys.}\ }\textbf {\bibinfo {volume} {49}},\
  \bibinfo {pages} {71} (\bibinfo {year} {2008}{\natexlab{a}})},\ \Eprint
  {http://arxiv.org/abs/0803.4089} {arXiv:0803.4089 [astro-ph]} \BibitemShut
  {NoStop}%
\bibitem [{\citenamefont {Kennedy}\ and\ \citenamefont {O'Hagan}(2001)}]{BMC}%
  \BibitemOpen
  \bibfield  {author} {\bibinfo {author} {\bibfnamefont {M.~C.}\ \bibnamefont
  {Kennedy}}\ and\ \bibinfo {author} {\bibfnamefont {A.}~\bibnamefont
  {O'Hagan}},\ }\href {\doibase 10.1111/1467-9868.00294} {\bibfield  {journal}
  {\bibinfo  {journal} {J. Roy. Statist. Soc., Series B (Statistical
  Methodology)}\ }\textbf {\bibinfo {volume} {63}},\ \bibinfo {pages} {425}
  (\bibinfo {year} {2001})}\BibitemShut {NoStop}%
\bibitem [{\citenamefont {Higdon}\ \emph {et~al.}(2004)\citenamefont {Higdon},
  \citenamefont {Kennedy}, \citenamefont {Cavendish}, \citenamefont {Cafeo},\
  and\ \citenamefont {Ryne}}]{Higdon2004}%
  \BibitemOpen
  \bibfield  {author} {\bibinfo {author} {\bibfnamefont {D.}~\bibnamefont
  {Higdon}}, \bibinfo {author} {\bibfnamefont {M.}~\bibnamefont {Kennedy}},
  \bibinfo {author} {\bibfnamefont {J.~C.}\ \bibnamefont {Cavendish}}, \bibinfo
  {author} {\bibfnamefont {J.~A.}\ \bibnamefont {Cafeo}}, \ and\ \bibinfo
  {author} {\bibfnamefont {R.~D.}\ \bibnamefont {Ryne}},\ }\href {\doibase
  10.1137/S1064827503426693} {\bibfield  {journal} {\bibinfo  {journal} {SIAM
  Journal on Scientific Computing}\ }\textbf {\bibinfo {volume} {26}},\
  \bibinfo {pages} {448} (\bibinfo {year} {2004})}\BibitemShut {NoStop}%
\bibitem [{\citenamefont {Gramacy}(2020)}]{gramacy2020surrogates}%
  \BibitemOpen
  \bibfield  {author} {\bibinfo {author} {\bibfnamefont {R.~B.}\ \bibnamefont
  {Gramacy}},\ }\enquote {\bibinfo {title} {Surrogates: {G}aussian process
  modeling, design and optimization for the applied sciences},}\ \ (\bibinfo
  {publisher} {Chapman Hall/CRC},\ \bibinfo {address} {Boca Raton, Florida},\
  \bibinfo {year} {2020})\ \bibinfo {note}
  {\url{http://bobby.gramacy.com/surrogates/}}\BibitemShut {NoStop}%
\bibitem [{\citenamefont {Rasmussen}(2004)}]{Rasmussen2004}%
  \BibitemOpen
  \bibfield  {author} {\bibinfo {author} {\bibfnamefont {C.~E.}\ \bibnamefont
  {Rasmussen}},\ }\enquote {\bibinfo {title} {Gaussian processes in machine
  learning},}\ in\ \href {\doibase 10.1007/978-3-540-28650-9_4} {\emph
  {\bibinfo {booktitle} {Advanced Lectures on Machine Learning}}},\ \bibinfo
  {editor} {edited by\ \bibinfo {editor} {\bibfnamefont {O.}~\bibnamefont
  {Bousquet}}, \bibinfo {editor} {\bibfnamefont {U.}~\bibnamefont {von
  Luxburg}}, \ and\ \bibinfo {editor} {\bibfnamefont {G.}~\bibnamefont
  {R{\"a}tsch}}}\ (\bibinfo  {publisher} {Springer Berlin Heidelberg},\
  \bibinfo {address} {Berlin, Heidelberg},\ \bibinfo {year} {2004})\ pp.\
  \bibinfo {pages} {63--71}\BibitemShut {NoStop}%
\bibitem [{\citenamefont {Joseph}\ \emph {et~al.}(2019)\citenamefont {Joseph},
  \citenamefont {Wang}, \citenamefont {Gu}, \citenamefont {Lyu},\ and\
  \citenamefont {Tuo}}]{MED}%
  \BibitemOpen
  \bibfield  {author} {\bibinfo {author} {\bibfnamefont {V.~R.}\ \bibnamefont
  {Joseph}}, \bibinfo {author} {\bibfnamefont {D.}~\bibnamefont {Wang}},
  \bibinfo {author} {\bibfnamefont {L.}~\bibnamefont {Gu}}, \bibinfo {author}
  {\bibfnamefont {S.}~\bibnamefont {Lyu}}, \ and\ \bibinfo {author}
  {\bibfnamefont {R.}~\bibnamefont {Tuo}},\ }\href {\doibase
  10.1080/00401706.2018.1552203} {\bibfield  {journal} {\bibinfo  {journal}
  {Technometrics}\ }\textbf {\bibinfo {volume} {61}},\ \bibinfo {pages} {297}
  (\bibinfo {year} {2019})}\BibitemShut {NoStop}%
\bibitem [{\citenamefont {Iman}\ \emph {et~al.}(1980)\citenamefont {Iman},
  \citenamefont {Davenport},\ and\ \citenamefont {Zeigler}}]{LHS}%
  \BibitemOpen
  \bibfield  {author} {\bibinfo {author} {\bibfnamefont {R.~L.}\ \bibnamefont
  {Iman}}, \bibinfo {author} {\bibfnamefont {J.~M.}\ \bibnamefont {Davenport}},
  \ and\ \bibinfo {author} {\bibfnamefont {D.~K.}\ \bibnamefont {Zeigler}},\
  }\href@noop {} {\emph {\bibinfo {title} {Latin hypercube sampling (program
  user's guide)}}},\ \bibinfo {type} {Tech. Rep.}\ (\bibinfo  {institution}
  {Sandia Labs., Albuquerque, NM (USA)},\ \bibinfo {year} {1980})\BibitemShut
  {NoStop}%
\bibitem [{\citenamefont {Joseph}\ \emph {et~al.}(2015)\citenamefont {Joseph},
  \citenamefont {Dasgupta}, \citenamefont {Tuo},\ and\ \citenamefont
  {Wu}}]{Joseph2015}%
  \BibitemOpen
  \bibfield  {author} {\bibinfo {author} {\bibfnamefont {V.~R.}\ \bibnamefont
  {Joseph}}, \bibinfo {author} {\bibfnamefont {T.}~\bibnamefont {Dasgupta}},
  \bibinfo {author} {\bibfnamefont {R.}~\bibnamefont {Tuo}}, \ and\ \bibinfo
  {author} {\bibfnamefont {C.~F.~J.}\ \bibnamefont {Wu}},\ }\href {\doibase
  10.1080/00401706.2014.881749} {\bibfield  {journal} {\bibinfo  {journal}
  {Technometrics}\ }\textbf {\bibinfo {volume} {57}},\ \bibinfo {pages} {64}
  (\bibinfo {year} {2015})}\BibitemShut {NoStop}%
\bibitem [{\citenamefont {Higdon}\ \emph {et~al.}(2008)\citenamefont {Higdon},
  \citenamefont {Gattiker}, \citenamefont {Williams},\ and\ \citenamefont
  {Rightley}}]{Higdon2008}%
  \BibitemOpen
  \bibfield  {author} {\bibinfo {author} {\bibfnamefont {D.}~\bibnamefont
  {Higdon}}, \bibinfo {author} {\bibfnamefont {J.}~\bibnamefont {Gattiker}},
  \bibinfo {author} {\bibfnamefont {B.}~\bibnamefont {Williams}}, \ and\
  \bibinfo {author} {\bibfnamefont {M.}~\bibnamefont {Rightley}},\ }\href
  {http://www.jstor.org/stable/27640080} {\bibfield  {journal} {\bibinfo
  {journal} {Journal of the American Statistical Association}\ }\textbf
  {\bibinfo {volume} {103}},\ \bibinfo {pages} {570} (\bibinfo {year}
  {2008})}\BibitemShut {NoStop}%
\bibitem [{\citenamefont {Ramsay}\ and\ \citenamefont
  {Silverman}(1997)}]{Ramsay97functionaldata}%
  \BibitemOpen
  \bibfield  {author} {\bibinfo {author} {\bibfnamefont {J.}~\bibnamefont
  {Ramsay}}\ and\ \bibinfo {author} {\bibfnamefont {B.}~\bibnamefont
  {Silverman}},\ }\href@noop {} {\emph {\bibinfo {title} {Functional Data
  Analysis}}}\ (\bibinfo  {publisher} {Springer},\ \bibinfo {year}
  {1997})\BibitemShut {NoStop}%
\bibitem [{\citenamefont {Ankenman}\ \emph {et~al.}(2009)\citenamefont
  {Ankenman}, \citenamefont {Nelson},\ and\ \citenamefont
  {Staum}}]{Ankenman2009}%
  \BibitemOpen
  \bibfield  {author} {\bibinfo {author} {\bibfnamefont {B.}~\bibnamefont
  {Ankenman}}, \bibinfo {author} {\bibfnamefont {B.~L.}\ \bibnamefont
  {Nelson}}, \ and\ \bibinfo {author} {\bibfnamefont {J.}~\bibnamefont
  {Staum}},\ }\href {\doibase 10.1287/opre.1090.0754} {\bibfield  {journal}
  {\bibinfo  {journal} {Operations Research}\ }\textbf {\bibinfo {volume}
  {58}},\ \bibinfo {pages} {371} (\bibinfo {year} {2009})}\BibitemShut
  {NoStop}%
\bibitem [{\citenamefont {Santner}\ \emph {et~al.}(2003)\citenamefont
  {Santner}, \citenamefont {Williams}, \citenamefont {Notz},\ and\
  \citenamefont {Williams}}]{santner2003design}%
  \BibitemOpen
  \bibfield  {author} {\bibinfo {author} {\bibfnamefont {T.~J.}\ \bibnamefont
  {Santner}}, \bibinfo {author} {\bibfnamefont {B.~J.}\ \bibnamefont
  {Williams}}, \bibinfo {author} {\bibfnamefont {W.~I.}\ \bibnamefont {Notz}},
  \ and\ \bibinfo {author} {\bibfnamefont {B.~J.}\ \bibnamefont {Williams}},\
  }\href@noop {} {\emph {\bibinfo {title} {The Design and Analysis of Computer
  Experiments}}}\ (\bibinfo  {publisher} {Springer},\ \bibinfo {year}
  {2003})\BibitemShut {NoStop}%
\bibitem [{\citenamefont {Abelev}\ \emph {et~al.}(2014)\citenamefont {Abelev}
  \emph {et~al.}}]{Abelev:2014ckr}%
  \BibitemOpen
  \bibfield  {author} {\bibinfo {author} {\bibfnamefont {B.~B.}\ \bibnamefont
  {Abelev}} \emph {et~al.} (\bibinfo {collaboration} {ALICE}),\ }\href
  {\doibase 10.1140/epjc/s10052-014-3077-y} {\bibfield  {journal} {\bibinfo
  {journal} {Eur. Phys. J.}\ }\textbf {\bibinfo {volume} {C74}},\ \bibinfo
  {pages} {3077} (\bibinfo {year} {2014})},\ \Eprint
  {http://arxiv.org/abs/1407.5530} {arXiv:1407.5530 [nucl-ex]} \BibitemShut
  {NoStop}%
\bibitem [{\citenamefont {Peters}(2008)}]{MCMC}%
  \BibitemOpen
  \bibfield  {author} {\bibinfo {author} {\bibfnamefont {G.}~\bibnamefont
  {Peters}},\ }\href {\doibase https://doi.org/10.1002/sim.3240} {\bibfield
  {journal} {\bibinfo  {journal} {Statistics in Medicine}\ }\textbf {\bibinfo
  {volume} {27}},\ \bibinfo {pages} {3213} (\bibinfo {year} {2008})},\ \Eprint
  {http://arxiv.org/abs/https://onlinelibra-ry.wiley.com/doi/pdf/10.1002/sim.3240}
  {https://onlinelibra-ry.wiley.com/doi/pdf/10.1002/sim.3240} \BibitemShut
  {NoStop}%
\bibitem [{\citenamefont {Trotta}(2008{\natexlab{b}})}]{Trotta_2008}%
  \BibitemOpen
  \bibfield  {author} {\bibinfo {author} {\bibfnamefont {R.}~\bibnamefont
  {Trotta}},\ }\href {\doibase 10.1080/00107510802066753} {\bibfield  {journal}
  {\bibinfo  {journal} {Contemporary Physics}\ }\textbf {\bibinfo {volume}
  {49}},\ \bibinfo {pages} {71–104} (\bibinfo {year}
  {2008}{\natexlab{b}})}\BibitemShut {NoStop}%
\bibitem [{\citenamefont {Vousden}\ \emph {et~al.}(2015)\citenamefont
  {Vousden}, \citenamefont {Farr},\ and\ \citenamefont
  {Mandel}}]{Vousden_2015}%
  \BibitemOpen
  \bibfield  {author} {\bibinfo {author} {\bibfnamefont {W.~D.}\ \bibnamefont
  {Vousden}}, \bibinfo {author} {\bibfnamefont {W.~M.}\ \bibnamefont {Farr}}, \
  and\ \bibinfo {author} {\bibfnamefont {I.}~\bibnamefont {Mandel}},\ }\href
  {\doibase 10.1093/mnras/stv2422} {\bibfield  {journal} {\bibinfo  {journal}
  {Monthly Notices of the Royal Astronomical Society}\ }\textbf {\bibinfo
  {volume} {455}},\ \bibinfo {pages} {1919} (\bibinfo {year}
  {2015})}\BibitemShut {NoStop}%
\bibitem [{\citenamefont {Foreman-Mackey}\ \emph {et~al.}(2013)\citenamefont
  {Foreman-Mackey}, \citenamefont {Hogg}, \citenamefont {Lang},\ and\
  \citenamefont {Goodman}}]{Foreman_Mackey_2013}%
  \BibitemOpen
  \bibfield  {author} {\bibinfo {author} {\bibfnamefont {D.}~\bibnamefont
  {Foreman-Mackey}}, \bibinfo {author} {\bibfnamefont {D.~W.}\ \bibnamefont
  {Hogg}}, \bibinfo {author} {\bibfnamefont {D.}~\bibnamefont {Lang}}, \ and\
  \bibinfo {author} {\bibfnamefont {J.}~\bibnamefont {Goodman}},\ }\href
  {\doibase 10.1086/670067} {\bibfield  {journal} {\bibinfo  {journal}
  {Publications of the Astronomical Society of the Pacific}\ }\textbf {\bibinfo
  {volume} {125}},\ \bibinfo {pages} {306} (\bibinfo {year}
  {2013})}\BibitemShut {NoStop}%
\bibitem [{pte()}]{ptemcee_code}%
  \BibitemOpen
  \href@noop {} {}\bibinfo {howpublished}
  {\url{https://github.com/willvousden/ptemcee}}\BibitemShut {NoStop}%
\bibitem [{\citenamefont {Florkowski}\ \emph {et~al.}(2018)\citenamefont
  {Florkowski}, \citenamefont {Heller},\ and\ \citenamefont
  {Spalinski}}]{Florkowski:2017olj}%
  \BibitemOpen
  \bibfield  {author} {\bibinfo {author} {\bibfnamefont {W.}~\bibnamefont
  {Florkowski}}, \bibinfo {author} {\bibfnamefont {M.~P.}\ \bibnamefont
  {Heller}}, \ and\ \bibinfo {author} {\bibfnamefont {M.}~\bibnamefont
  {Spalinski}},\ }\href {\doibase 10.1088/1361-6633/aaa091} {\bibfield
  {journal} {\bibinfo  {journal} {Rept. Prog. Phys.}\ }\textbf {\bibinfo
  {volume} {81}},\ \bibinfo {pages} {046001} (\bibinfo {year} {2018})},\
  \Eprint {http://arxiv.org/abs/1707.02282} {arXiv:1707.02282 [hep-ph]}
  \BibitemShut {NoStop}%
\bibitem [{\citenamefont {Jaiswal}\ \emph {et~al.}(2019)\citenamefont
  {Jaiswal}, \citenamefont {Chattopadhyay}, \citenamefont {Jaiswal},
  \citenamefont {Pal},\ and\ \citenamefont {Heinz}}]{Jaiswal:2019cju}%
  \BibitemOpen
  \bibfield  {author} {\bibinfo {author} {\bibfnamefont {S.}~\bibnamefont
  {Jaiswal}}, \bibinfo {author} {\bibfnamefont {C.}~\bibnamefont
  {Chattopadhyay}}, \bibinfo {author} {\bibfnamefont {A.}~\bibnamefont
  {Jaiswal}}, \bibinfo {author} {\bibfnamefont {S.}~\bibnamefont {Pal}}, \ and\
  \bibinfo {author} {\bibfnamefont {U.}~\bibnamefont {Heinz}},\ }\href
  {\doibase 10.1103/PhysRevC.100.034901} {\bibfield  {journal} {\bibinfo
  {journal} {Phys. Rev. C}\ }\textbf {\bibinfo {volume} {100}},\ \bibinfo
  {pages} {034901} (\bibinfo {year} {2019})},\ \Eprint
  {http://arxiv.org/abs/1907.07965} {arXiv:1907.07965 [nucl-th]} \BibitemShut
  {NoStop}%
\bibitem [{\citenamefont {Kurkela}\ \emph {et~al.}(2020)\citenamefont
  {Kurkela}, \citenamefont {van~der Schee}, \citenamefont {Wiedemann},\ and\
  \citenamefont {Wu}}]{Kurkela:2019set}%
  \BibitemOpen
  \bibfield  {author} {\bibinfo {author} {\bibfnamefont {A.}~\bibnamefont
  {Kurkela}}, \bibinfo {author} {\bibfnamefont {W.}~\bibnamefont {van~der
  Schee}}, \bibinfo {author} {\bibfnamefont {U.~A.}\ \bibnamefont {Wiedemann}},
  \ and\ \bibinfo {author} {\bibfnamefont {B.}~\bibnamefont {Wu}},\ }\href
  {\doibase 10.1103/PhysRevLett.124.102301} {\bibfield  {journal} {\bibinfo
  {journal} {Phys. Rev. Lett.}\ }\textbf {\bibinfo {volume} {124}},\ \bibinfo
  {pages} {102301} (\bibinfo {year} {2020})},\ \Eprint
  {http://arxiv.org/abs/1907.08101} {arXiv:1907.08101 [hep-ph]} \BibitemShut
  {NoStop}%
\bibitem [{\citenamefont {Liyanage}()}]{vah_tool}%
  \BibitemOpen
  \bibfield  {author} {\bibinfo {author} {\bibfnamefont {D.}~\bibnamefont
  {Liyanage}},\ }\href@noop {} {\enquote {\bibinfo {title} {Online emulators
  for vah},}\ }\bibinfo {howpublished}
  {\url{https://danosu-visualization-vah-streamlit-widget-wq49dw.streamlit.app}}\BibitemShut
  {NoStop}%
\bibitem [{\citenamefont {Gelis}\ \emph {et~al.}(2010)\citenamefont {Gelis},
  \citenamefont {Iancu}, \citenamefont {Jalilian-Marian},\ and\ \citenamefont
  {Venugopalan}}]{Gelis:2010nm}%
  \BibitemOpen
  \bibfield  {author} {\bibinfo {author} {\bibfnamefont {F.}~\bibnamefont
  {Gelis}}, \bibinfo {author} {\bibfnamefont {E.}~\bibnamefont {Iancu}},
  \bibinfo {author} {\bibfnamefont {J.}~\bibnamefont {Jalilian-Marian}}, \ and\
  \bibinfo {author} {\bibfnamefont {R.}~\bibnamefont {Venugopalan}},\ }\href
  {\doibase 10.1146/annurev.nucl.010909.083629} {\bibfield  {journal} {\bibinfo
   {journal} {Ann. Rev. Nucl. Part. Sci.}\ }\textbf {\bibinfo {volume} {60}},\
  \bibinfo {pages} {463} (\bibinfo {year} {2010})},\ \Eprint
  {http://arxiv.org/abs/1002.0333} {arXiv:1002.0333 [hep-ph]} \BibitemShut
  {NoStop}%
\bibitem [{\citenamefont {Epelbaum}\ and\ \citenamefont
  {Gelis}(2013)}]{Epelbaum:2013ekf}%
  \BibitemOpen
  \bibfield  {author} {\bibinfo {author} {\bibfnamefont {T.}~\bibnamefont
  {Epelbaum}}\ and\ \bibinfo {author} {\bibfnamefont {F.}~\bibnamefont
  {Gelis}},\ }\href {\doibase 10.1103/PhysRevLett.111.232301} {\bibfield
  {journal} {\bibinfo  {journal} {Phys. Rev. Lett.}\ }\textbf {\bibinfo
  {volume} {111}},\ \bibinfo {pages} {232301} (\bibinfo {year} {2013})},\
  \Eprint {http://arxiv.org/abs/1307.2214} {arXiv:1307.2214 [hep-ph]}
  \BibitemShut {NoStop}%
\bibitem [{\citenamefont {Nijs}\ and\ \citenamefont {van~der
  Schee}(2022{\natexlab{a}})}]{Nijs:2021clz}%
  \BibitemOpen
  \bibfield  {author} {\bibinfo {author} {\bibfnamefont {G.}~\bibnamefont
  {Nijs}}\ and\ \bibinfo {author} {\bibfnamefont {W.}~\bibnamefont {van~der
  Schee}},\ }\href {\doibase 10.1103/PhysRevC.106.044903} {\bibfield  {journal}
  {\bibinfo  {journal} {Phys. Rev. C}\ }\textbf {\bibinfo {volume} {106}},\
  \bibinfo {pages} {044903} (\bibinfo {year} {2022}{\natexlab{a}})},\ \Eprint
  {http://arxiv.org/abs/2110.13153} {arXiv:2110.13153 [nucl-th]} \BibitemShut
  {NoStop}%
\bibitem [{\citenamefont {Khachatryan}\ \emph {et~al.}(2016)\citenamefont
  {Khachatryan} \emph {et~al.}}]{CMS:2015nfb}%
  \BibitemOpen
  \bibfield  {author} {\bibinfo {author} {\bibfnamefont {V.}~\bibnamefont
  {Khachatryan}} \emph {et~al.} (\bibinfo {collaboration} {CMS}),\ }\href
  {\doibase 10.1016/j.physletb.2016.06.027} {\bibfield  {journal} {\bibinfo
  {journal} {Phys. Lett. B}\ }\textbf {\bibinfo {volume} {759}},\ \bibinfo
  {pages} {641} (\bibinfo {year} {2016})},\ \Eprint
  {http://arxiv.org/abs/1509.03893} {arXiv:1509.03893 [hep-ex]} \BibitemShut
  {NoStop}%
\bibitem [{\citenamefont
  {{A}{L}{I}{C}{E}-Collaboration}(2022)}]{ALICE:2022xir}%
  \BibitemOpen
  \bibfield  {author} {\bibinfo {author} {\bibnamefont
  {{A}{L}{I}{C}{E}-Collaboration}},\ }\href@noop {} {\  (\bibinfo {year}
  {2022})},\ \Eprint {http://arxiv.org/abs/2204.10148} {arXiv:2204.10148
  [nucl-ex]} \BibitemShut {NoStop}%
\bibitem [{\citenamefont {Nijs}\ and\ \citenamefont {van~der
  Schee}(2022{\natexlab{b}})}]{Nijs:2022rme}%
  \BibitemOpen
  \bibfield  {author} {\bibinfo {author} {\bibfnamefont {G.}~\bibnamefont
  {Nijs}}\ and\ \bibinfo {author} {\bibfnamefont {W.}~\bibnamefont {van~der
  Schee}},\ }\href {\doibase 10.1103/PhysRevLett.129.232301} {\bibfield
  {journal} {\bibinfo  {journal} {Phys. Rev. Lett.}\ }\textbf {\bibinfo
  {volume} {129}},\ \bibinfo {pages} {232301} (\bibinfo {year}
  {2022}{\natexlab{b}})},\ \Eprint {http://arxiv.org/abs/2206.13522}
  {arXiv:2206.13522 [nucl-th]} \BibitemShut {NoStop}%
\bibitem [{\citenamefont {Sobol'}(1990)}]{sobol1990sensitivity}%
  \BibitemOpen
  \bibfield  {author} {\bibinfo {author} {\bibfnamefont {I.~M.}\ \bibnamefont
  {Sobol'}},\ }\href@noop {} {\bibfield  {journal} {\bibinfo  {journal}
  {Matematicheskoe modelirovanie}\ }\textbf {\bibinfo {volume} {2}},\ \bibinfo
  {pages} {112} (\bibinfo {year} {1990})}\BibitemShut {NoStop}%
\bibitem [{\citenamefont {Liyanage}\ \emph {et~al.}(2022)\citenamefont
  {Liyanage}, \citenamefont {Ji}, \citenamefont {Everett}, \citenamefont
  {Heffernan}, \citenamefont {Heinz}, \citenamefont {Mak},\ and\ \citenamefont
  {Paquet}}]{dan_tl}%
  \BibitemOpen
  \bibfield  {author} {\bibinfo {author} {\bibfnamefont {D.}~\bibnamefont
  {Liyanage}}, \bibinfo {author} {\bibfnamefont {Y.}~\bibnamefont {Ji}},
  \bibinfo {author} {\bibfnamefont {D.}~\bibnamefont {Everett}}, \bibinfo
  {author} {\bibfnamefont {M.}~\bibnamefont {Heffernan}}, \bibinfo {author}
  {\bibfnamefont {U.}~\bibnamefont {Heinz}}, \bibinfo {author} {\bibfnamefont
  {S.}~\bibnamefont {Mak}}, \ and\ \bibinfo {author} {\bibfnamefont {J.-F.}\
  \bibnamefont {Paquet}},\ }\href {\doibase 10.1103/PhysRevC.105.034910}
  {\bibfield  {journal} {\bibinfo  {journal} {Phys. Rev. C}\ }\textbf {\bibinfo
  {volume} {105}},\ \bibinfo {pages} {034910} (\bibinfo {year}
  {2022})}\BibitemShut {NoStop}%
\bibitem [{\citenamefont {Jacques}\ \emph {et~al.}(2006)\citenamefont
  {Jacques}, \citenamefont {Lavergne},\ and\ \citenamefont
  {Devictor}}]{jacques2006sensitivity}%
  \BibitemOpen
  \bibfield  {author} {\bibinfo {author} {\bibfnamefont {J.}~\bibnamefont
  {Jacques}}, \bibinfo {author} {\bibfnamefont {C.}~\bibnamefont {Lavergne}}, \
  and\ \bibinfo {author} {\bibfnamefont {N.}~\bibnamefont {Devictor}},\
  }\href@noop {} {\bibfield  {journal} {\bibinfo  {journal} {Reliability
  Engineering \& System Safety}\ }\textbf {\bibinfo {volume} {91}},\ \bibinfo
  {pages} {1126} (\bibinfo {year} {2006})}\BibitemShut {NoStop}%
\bibitem [{\citenamefont {McNelis}(2021)}]{McNelis_thesis}%
  \BibitemOpen
  \bibfield  {author} {\bibinfo {author} {\bibfnamefont {M.}~\bibnamefont
  {McNelis}},\ }\emph {\bibinfo {title} {Far-from-equilibrium hydrodynamic
  simulations}},\ \href {\doibase 10.48550/ARXIV.2105.06007} {Ph.D. thesis},\
  \bibinfo  {school} {The Ohio State University} (\bibinfo {year} {2021}),\
  \Eprint {http://arxiv.org/abs/2105.06007} {arXiv:2105.06007 [hep-ph]}
  \BibitemShut {NoStop}%
\bibitem [{\citenamefont {Storn}\ and\ \citenamefont
  {Price}(1997)}]{diff_evolution}%
  \BibitemOpen
  \bibfield  {author} {\bibinfo {author} {\bibfnamefont {R.}~\bibnamefont
  {Storn}}\ and\ \bibinfo {author} {\bibfnamefont {K.}~\bibnamefont {Price}},\
  }\href {\doibase 10.1023/A:1008202821328} {\bibfield  {journal} {\bibinfo
  {journal} {Journal of Global Optimization}\ }\textbf {\bibinfo {volume}
  {11}},\ \bibinfo {pages} {341} (\bibinfo {year} {1997})}\BibitemShut
  {NoStop}%
\bibitem [{\citenamefont {Virtanen}\ \emph {et~al.}(2020)\citenamefont
  {Virtanen}, \citenamefont {Gommers}, \citenamefont {Oliphant}, \citenamefont
  {Haberland}, \citenamefont {Reddy} \emph {et~al.}}]{2020SciPy-NMeth}%
  \BibitemOpen
  \bibfield  {author} {\bibinfo {author} {\bibfnamefont {P.}~\bibnamefont
  {Virtanen}}, \bibinfo {author} {\bibfnamefont {R.}~\bibnamefont {Gommers}},
  \bibinfo {author} {\bibfnamefont {T.~E.}\ \bibnamefont {Oliphant}}, \bibinfo
  {author} {\bibfnamefont {M.}~\bibnamefont {Haberland}}, \bibinfo {author}
  {\bibfnamefont {T.}~\bibnamefont {Reddy}},  \emph {et~al.},\ }\href {\doibase
  10.1038/s41592-019-0686-2} {\bibfield  {journal} {\bibinfo  {journal} {Nature
  Methods}\ }\textbf {\bibinfo {volume} {17}},\ \bibinfo {pages} {261}
  (\bibinfo {year} {2020})}\BibitemShut {NoStop}%
\end{thebibliography}%

\end{document}